\documentclass[useAMS,usenatbib]{mn2e}
\usepackage{graphicx}
\usepackage{graphics,float}                
\usepackage{epsf}
\usepackage{amsmath}                                
\usepackage{amssymb}
\usepackage{picinpar}                               
\usepackage{relsize,setspace,subfigure}
\usepackage{tabularx}
\usepackage[innercaption]{sidecap}                  
\usepackage[figuresright]{rotating}
\usepackage{url}

\title{Future Evolution of Bound Superclusters in an Accelerating Universe}
\author[Araya-Melo et al.]{\parbox{\textwidth}{Pablo A. Araya-Melo$^{1,2,3}$, 
    Andreas Reisenegger$^4$, Andr\'es Meza$^5$, Rien van de Weygaert$^1$, 
    Rolando D\"{u}nner$^4$, and Hern\'an Quintana$^4$}
\vspace*{4pt} \\
$^1$Kapteyn Astronomical Institute, University of Groningen, P.O.
Box 800, 9700 AV Groningen, The Netherlands\\
$^2$Korea Institute for Advanced Study, Dongdaemon-gu, Seoul 130-722, Korea\\
$^3$Jacobs University Bremen, Campus Ring 1, 28759 Bremen, Germany\\
$^4$Departamento de Astronom\'ia y Astrof\'isica, Facultad de F\'isica,
P. Universidad Cat\'olica de Chile, Casilla 306, Santiago 22, Chile\\
$^5$Departamento de Ciencias Fisicas, Facultad de Ingenieria, Universidad
Andres Bello, Santiago, Chile}
\begin{document}

\date{\today}

\pagerange{\pageref{firstpage}--\pageref{lastpage}} \pubyear{2007}

\maketitle

\label{firstpage}

\begin{abstract}
The evolution of marginally bound supercluster-like objects in an accelerating 
$\Lambda$CDM Universe is followed, by means of cosmological simulations, from 
the present time to an expansion factor $a=100$. The objects are identified on 
the basis of the binding density criterion introduced by \cite{dunner06}. 
Superclusters are identified with the ones whose mass 
${\rm M}>10^{15}h^{-1}$M$_{\odot}$, the most massive one with 
${\rm M}\sim8\times10^{15}h^{-1}$M$_{\odot}$, comparable to the Shapley 
supercluster. The spatial distribution of the superclusters remains essentially
the same after the present epoch, reflecting the halting growth of the Cosmic 
Web as $\Lambda$ gets to dominate the expansion of the Universe. The same trend
can be seen in the stagnation of the development of the mass function of 
virialized haloes and bound objects. The situation is considerably different 
when looking at the internal evolution, quantified in terms of their shape, 
compactness and density profile, and substructure in terms of their 
multiplicity function. We find a continuing evolution from a wide range of 
triaxial shapes at $a=1$ to almost perfect spherical shapes at $a=100$. We also
find a systematic trend towards a higher concentration. Meanwhile, we see their
substructure gradually disappearing, as the surrounding subclumps fall in and 
merge to form one coherent, virialized system.
\end{abstract}


\section{Introduction}
The evidence for an accelerated expansion of the Universe has established the 
dominant presence of a `dark energy' component. In the present cosmological 
paradigm, the Universe entered into an accelerating phase at $z\approx 0.7$. 
Observational evidence points towards a dark energy component which behaves 
like Einstein's cosmological constant. As long as the matter density in the 
Universe dominated over that of dark energy, the gravitational growth of matter
concentrations resulted in the emergence of ever larger structures. Once dark 
energy came to dominate the dynamics of the Universe and the Universe got into 
accelerated expansion, structure formation came to a halt 
\citep{peebles80,heath77}. With the present-day Universe having reached this 
stage, the largest identifiable objects that will ever populate our Universe 
may be the ones that we observe in the process of formation at the present 
cosmological time. While no larger objects will emerge, these sufficiently 
overdense and bound patches will not be much affected by the global cosmic 
acceleration. They will remain bound and evolve as if they are {\it island 
universes}: they turn into isolated evolving regions 
\citep{chiueh02,nagamine03,busha03,dunner06}. 

While clusters of galaxies are the most massive and most recently 
collapsed and virialized structures, the present-day superclusters are arguably
the largest bound but not yet fully evolved objects in our Universe. In our 
accelerating Universe, we may assume they are the objects that ultimately will 
turn into island universes. A large range of observational studies, mostly 
based on optically or X-ray selected samples, show that clusters are strongly 
clustered and group together in large supercluster complexes 
\citep[see e.g.][]{oort1983,bahcall1988,einasto1994,einasto2001,quintana2000}. 
These superclusters, the largest structures identifiable in the present 
Universe, are enormous structures comprising a few up to dozens of rich 
clusters of galaxies, a large number of more modestly sized clumps, and 
thousands of galaxies spread between these density concentrations. 

In this study, we aim at contrasting the large-scale evolution of structure in
an accelerated Universe with that of the internal evolution of bound objects. 
In order to infer what will be the largest bound regions in our Universe, 
the {\it island universes}, we study the mass function of bound objects. 
The abundance or mass function of superclusters serves as a good indicator of 
the growth of structure of a cosmological model. While large-scale structure 
formation comes to a halt, this will manifest itself in the asymptotic behaviour
of the supercluster mass function. Meanwhile, the internal evolution of the 
superclusters continues as they contract and collapse into the largest 
virialized entities the Universe will ever contain. 

We address three aspects of the continuing internal evolution of bound regions:
their shape, density profile and internal substructure in terms of their 
cluster multiplicity. The shape of supercluster regions is one of the most 
sensitive probes of their evolutionary stage. We know that superclusters in the
present-day Universe are mostly flattened or elongated structures, usually 
identified with the most prominent filaments and sheets in the galaxy 
distribution \citep[e.g.][]{plionis92,sathya1998,basilakos01,sheth2003,
einasto2007c}. The Pisces-Perseus supercluster chain is a particularly 
well-known example of a strongly elongated filament 
\citep[see e.g.][]{hayngiov1986}. The distribution of shapes of bound 
structures is a combination of at least two factors. One is the shape of the 
proto-supercluster in the initial density field. The second factor is the 
evolutionary state of the bound structure. We know that the gravitational 
collapse of cosmic overdensities --- whose progenitors in the primordial 
density perturbation field will never be spherical 
\citep{peacheav1985,bbks} --- proceeds in a distinctly anisotropic fashion via 
flattened and elongated configurations towards a final more compact triaxial 
virialized state \citep[see eg.][]{zeldovich1970,icke73,whitesilk1979,
eistloeb1995,bondmyers1996,sathya1996,desjacques2008,weybond2008}. 

The collapse of the superclusters will also result in a continuous sharpening 
of the internal mass distribution, reflected in the steepening of their 
density profile. While they are in the process of collapse, internal 
substructure of constituent clusters remains recognizable. While the subclumps 
merge into an ever more massive central concentration the supercluster 
substructure gradually fades, resulting in an increasingly uniform mass 
distribution. The evolving and decreasing level of substructure will be 
followed in terms of the evolving supercluster multiplicity function, i.e. the 
number of cluster-sized clumps within the supercluster region. 

Representing moderate density enhancements on scale of tens of Mpc, in the 
present Universe superclusters are still expanding with the Hubble flow, 
although at a slightly decelerated rate, or have just started contracting. 
Because these structures have not yet fully formed, virialized and clearly 
separated from each other, it is difficult to identify them unambiguously. In 
most studies, superclusters have been defined by more or less arbitrary 
criteria, mostly on the basis of a grouping and/or percolation algorithm 
\citep[see e.g][]{oort1983,bahcall1988,einasto1994,einasto2001,quintana2000}. 
This introduces the need for a user-specified percolation radius. 
\citet[hereafter Paper I]{dunner06} attempted to define a more physically based
criterion, identifying superclusters with the biggest gravitationally bound 
structures that will be able to form in our Universe. On the basis of this, 
they worked out a lower density limit for gravitationally bound structures. 
This limit is based on the density contrast that a spherical shell needs to 
enclose to remain bound to a spherically symmetric overdensity. 

We use this spherical density criterion to identify bound structures in a large
cosmological box. In this we follow the work of \cite{chiueh02} and 
\cite{dunner06}. \cite{chiueh02} numerically solved the spherical collapse 
model equations for self-consistent growing mode perturbations in order to 
obtain a theoretical criterion for the mean density enclosed in the outer 
gravitationally bound shell. The resulting density criterion was evaluated by 
\cite{dunner06} on the basis of numerical simulations. They generalized it by 
deriving the analytical solution which also forms the basis of the current 
study, and in \cite{dunner07} extended the criterion to limits for bound 
structures in redshift space. 

Various authors have addressed the future evolution of cosmic structure  
\citep{chiueh02,busha03,nagamine03,dunner06,dunner07,hoffman07,busha07}. The 
internal evolution of the density and velocity structures of bound objects was 
followed by \cite{busha03}, with \cite{busha07} focusing on the effects of 
small-scale structure on the formation of dark matter halos in two different 
cosmologies. \cite{nagamine03} and \cite{hoffman07} specifically focused on the
evolution of the Local Universe. \cite{nagamine03} found that the Local Group
will get detached from the rest of the Universe, and that its physical distance
to other systems will increase exponentially. \cite{hoffman07} investigated
the dependence on dark matter and dark energy by contrasting $\Lambda$CDM and 
OCDM models. They concluded that the evolution of structure in comoving 
coordinates at long term is determined mainly by the matter density rather 
than by the dark energy. A key point of attention in \cite{nagamine03}, 
\cite{hoffman07} and \cite{busha07} was the mass function of objects in their 
simulations, on which they all agree that it hardly changes after the 
current cosmic epoch. 

This paper is the first in a series addressing the future evolution of 
structure in FRW Universes. In an accompanying publication we will specifically 
look at the influence of dark matter and the cosmological constant on the 
emerging (super)cluster population, based on the work described in 
\cite{arayamelo08}. The present study concentrates on the details of this 
evolution in a standard flat $\Lambda$CDM Universe. 

This paper is organized as follows. In section \ref{sec:scdef}, we present a
review of the spherical collapse model, including a derivation of the critical 
overdensity for a structure to remain bound. Section
\ref{sec:simulation} describes the simulation and the group finder algorithm 
that we employ when determining the mass functions. This is followed in 
sect.~\ref{sec:evolution} by a a qualitative description of the evolution, 
including a case study of the evolution of some typical bound mass clumps 
from the present epoch to $a=100$. The lack of evolution in their spatial 
distribution in the same time interval is studied in sect.~\ref{sec:spatial}. 
Section \ref{sec:ps_mf} presents the mass functions of the bound structures at 
$a=1$ and $a=100$ and a comparison with the ones obtained by the 
Press-Schechter formalism and its variants. The evolution of the shapes of the 
structures is studied in section \ref{sec:shapes}. In section 
\ref{sec:dens_prof} we look into the mass distribution and density profiles. 
Section \ref{sec:multi} presents the stark changes in the supercluster 
multiplicity function. In sect.~\ref{sec:shapley} its results are combined with
those on the supercluster mass functions obtained in section~\ref{sec:ps_mf} to 
relate our findings to the presence and abundance of monster supercluster 
complexes like the Shapley and Horologium-Reticulum supercluster. Finally, in 
section~\ref{sec:conclusions}, we discuss our findings and draw conclusions on 
various issues addressed by our study.

\section{Supercluster Collapse Model}
\label{sec:scdef}
The present study is based on a physical criterion for the definition of 
superclusters proposed by \cite{dunner06}, namely that they are the largest 
bound (though not yet virialized) structures in the Universe. The practical 
implementation of this definition suggested in that paper is through an 
(approximate) density threshold for regions that are gravitationally bound. 
Given the anisotropic nature of the collapse and the tidal influence of the 
cosmic surroundings \citep{zeldovich1970,icke73,bondmyers1996,sheth99,
desjacques2008,weybond2008} this may only yield a rough approximation. The work
by \cite{sheth2002} did show that a density threshold does depend on shape and 
environment. One could imagine a variety of alternative physical definitions 
for superclusters, also relating to the assumption that they are the largest 
bound structures in the Universe which just have commenced to condense out of 
the cosmic background. One particular criterion would be to invoke the 
corresponding velocity field and identify them with bound regions that have 
turned around and started to contract, on the way towards complete collapse and
virialization.  

Here we follow \cite{dunner06} and assume that a global density threshold 
criterion, in combination with a few extra assumptions, assures a reasonably 
accurate identification process. This has indeed been demonstrated in the same 
study in a comparison of the criterion with the outcome of numerical 
simulations. 

\cite{dunner06} derive an analytical density threshold criterion for bound 
regions in a Universe with dark energy. We summarize this criterion and 
derivation in the subsections below. In addition to the density threshold, we 
assure that a given bound region has started to materialize as a recognizable 
entity by including the additional requirement of the bound regions to have a 
virialized core. In a final step we group the overlapping identified spherical 
bound supercluster objects in order to outline a region that in the 
observational reality would be recognized as a supercluster. 

The spherical density criterion described below forms a key ingredient of our 
supercluster definition and for the identification procedure for singling out 
bound spherical regions associated with clusters (sect.~\ref{sec:boundhalo}).
\subsection{Spherical Collapse Model}
\label{sec:scm}
The spherical collapse model \citep{gunn72,lilje91,lahav91} describes the 
evolution of a spherically symmetric mass density perturbation in an expanding 
Universe \citep[also see][]{mota04}. Its great virtue is the ability to 
completely follow the nonlinear evolution of a collapsing shell, as the 
dynamics is fully and solely determined by the (constant) mass interior to the 
shell. Even though the gravitational collapse of generic cosmological 
structures tends to be highly inhomogeneous and anisotropic, the spherical 
model has proven to provide a surprisingly accurate description of the more 
complex reality. Even in situations where it is not able to provide accurate 
quantitative predictions it may be used as a good reference for interpretation 
of results. 

We consider a flat FRW Universe with a cosmological constant $\Lambda$. 
The mass density parameter at the current epoch is $\Omega_{m,0}$, 
while $\Omega_{\Lambda,0}$ is the present value of the cosmological 
density parameter. The Hubble parameter at the current epoch is $H_0$.

A mass shell with a (physical) radius $r(t)$ at time $t$ encloses a fixed 
mass $M$. The starting point of our derivation is the energy $E$ per unit mass 
of the shell, which satisfies the equation \citep{peebles84}:
\begin{equation}
\label{eq:lambda}
E=\frac{1}{2}\left(\frac{dr}{dt}\right)^{2}-\frac{GM}{r}
-\frac{\Lambda r^{2}}{6}=\mathrm{constant}\,.
\end{equation}
By introducing the dimensionless variables
\begin{align}
\tilde{r}&\,=\,\left(\frac{\Lambda}{3GM}\right)^{\frac{1}{3}}r
\label{eq:radio}\,, \\
\tilde{t}&\,=\,\left(\frac{\Lambda}{3}\right)^{\frac{1}{2}}t
\label{eq:aditiempo}\,,\\
\tilde{E}&\,=\,\left(\frac{G^{2}M^{2}\Lambda}{3}\right)^{-\frac{1}{3}}E\,.
\label{eq:energydims}
\end{align}
the energy equation can be recast into the simpler dimensionless form
\begin{equation}
\tilde{E}\,=\,\frac{1}{2}\left(\frac{d\tilde{r}}{d\tilde{t}}\right)^{2}-
\frac{1}{\tilde{r}}-\frac{\tilde{r}^{2}}{2}\,.
\label{eq:energy}
\end{equation}

\subsection{Critical shell and turnaround radius}
\label{sec:critshell}
To delineate a bound regions around a spherical mass concentration we have to 
identify the critical shell. It is the shell which separates the regions that 
will expand forever and the ones that will at some moment in time turn 
around and fall in onto the core of the region. 

In a Universe with a cosmological constant, the critical shell is the one that 
delimits the region of gravitational attraction and the region of repulsion. 
This translates into the radius for which the (dimensionless) potential energy 
${\tilde V}$,
\begin{equation}
\tilde{V}=-\frac{1}{\tilde{r}}-\frac{\tilde{r}^{2}}{2}
\label{eq:potential}
\end{equation}
\noindent is maximized. The maximum of this potential occurs at 
$\tilde{r}^{\ast}=1$. The critical shell (indicated by the subscript $'cs'$) is
the shell with the maximum possible energy to remain attached to the spherical 
mass concentration, 
\begin{equation}
{\tilde E}^{\ast}\,=\,\tilde{V}({\tilde r}^{\ast}=1)\,=\,-\frac{3}{2}\,.
\end{equation}
\noindent The evolution of the critical shell's radius 
$\tilde r_{\rm cs}({\tilde t})$ can be inferred by integrating the energy 
equation (\ref{eq:energy}) from ${\tilde t}=0$ ($\tilde r=0$) until epoch 
$\tilde t$, for the energy ${\tilde E}={\tilde E}^{\ast}=-\frac{3}{2}$ of the 
critical shell. In a flat Universe with a cosmological constant, the vacuum 
energy density parameter $\Omega_\Lambda$ is a monotonically increasing 
function of the age $\tilde t$ of the Universe \citep{peebles80}, 
\begin{equation}
\Omega_{\Lambda}({\tilde t})\equiv\frac{\Lambda}{3H^{2}}=\tanh^{2}
\left(\frac{3\tilde{t}}{2}\right)\,.
\label{eq:omglambda}
\end{equation}
The evolution of the critical shell $\tilde r_{\rm cs}({\tilde t})$ may 
therefore be expressed in terms of $\Omega_{\Lambda}({\tilde t})$, 
\begin{equation}
\Omega_{\Lambda}(\tilde{t})=\left[\frac{\chi(\tilde{r})-1}
{\chi(\tilde{r})+1}\right]^{2}\,,
\label{eq:omegalambda}
\end{equation}
where the variable $\chi(\tilde{r})$ is given by
\begin{equation}
\chi(\tilde{r})=\left[\frac{1+2\tilde{r}+
\sqrt{3\tilde{r}(\tilde{r}+2)}}{1+2\tilde{r}-
\sqrt{3\tilde{r}(\tilde{r}+2)}}\right]^{\frac{\sqrt{3}}{2}}\times\left(1
+\tilde{r}+\sqrt{\tilde{r}(\tilde{r}+2)}\right)^{-3}\,.
\label{eq:tilder}
\end{equation}
\noindent Evaluation of this expression shows that the critical shell 
radius $\tilde r_{\rm cs}$ is an increasing function of $\Omega_{\Lambda}$. The 
shell reaches its maximum - turnaround - radius $r_{\rm max}$ as 
$\Omega_{\Lambda}\to 1$ and $t\to\infty$. The maximum radius is reached when 
$V({\tilde r}_{\rm max})\,=\,{\tilde E}^{\ast}=-\frac{3}{2}$, i.e. when
\begin{equation}
r_{\rm max}\,=\,\left({\displaystyle 3GM \over \displaystyle \Lambda}\right)^{1/3}\,,
\end{equation}
\noindent so that the normalized radius can be interpreted as \\
$\tilde r\,=\,r/r_{\rm max}$. 

For the cosmology at hand ($\Omega_{m,0}=0.3$, $\Omega_{\Lambda,0}=0.7$) the 
critical shell currently has a dimensionless radius $\tilde{r}_{0}=0.84$, i.e. 
it has a value 84\% of its maximum radius. 
\subsection{Conditions for a critical shell}
\label{sec:densthr}
For the translation of the radius $\tilde r_{\rm cs}$ of a bound object into a 
density criterion, we evaluate the average mass density $\bar{\rho}_{s}$ 
enclosed by a given shell,
\begin{equation}
\bar{\rho}_{s}\,=\,\frac{3M}{4\pi r^{3}}\,.
\label{eq:densesf}
\end{equation}
in terms of the critical density $\rho_{c}=3H^{2}/8\pi G$. The shell is bound 
if
\begin{equation}
\frac{\bar{\rho}_{s}}{\rho_{c}}\,\ge\,\frac{\rho_{cs}}{\rho_{c}}\,=\,
\frac{2\Omega_{\Lambda}}{\tilde{r}_{cs}^{3}}\,=\,2.36\,.
\label{eq:ratio}
\end{equation}
\noindent The value 2.36 corresponds to a Universe with 
$\Omega_{\Lambda,0}=0.7$. Note that for $\Omega_{\Lambda}=1$ ($t\to\infty$) the 
critical shell's density is $\rho_{cs}/\rho_{c}=2$. 

The corresponding density excess $\delta$ of the spherical mass concentration 
with respect to the global cosmic background $\rho_u(t)$ 
($=\rho_{c,0}\,\Omega_{m,0}/{a^3}$, with $\rho_{c,0}$ the critical density at 
the present epoch), may be inferred from the expression
\begin{equation}
1+\delta\,\equiv\,\frac{\rho}{\rho_{u}}=\frac{2\Omega_{\Lambda 0}}{\Omega_{m,0}}
\left(\frac{a}{\tilde{r}}\right)^3\,.
\label{eq:deltamasuno}
\end{equation}
On the basis of this equation we find that a critically bound shell at the 
present epoch has a density excess in the order of $\delta_{cs}\approx 6.9$. 
\subsection{Bound Object: \\ \ \ \ \ \ \ \ linearly extrapolated density threshold}
\label{sec:deltab}
To be able to identify the primordial regions that correspond to bound, 
collapsing and/or virialized objects at any arbitrary redshift $z$, we need the
value of the corresponding linearly extrapolated densities. 

According to gravitational instability theory \citep{peebles80}, in the linear 
regime the density excess $\delta(a)$ grows as 
\begin{equation}
\delta(a)\,=\,D(a)\,\delta_0
\label{eq:denlin}
\end{equation}
where $D(a)$ is the linear density growth factor (growing mode). In a FRW 
Universe with matter and a cosmological constant, it can be computed from 
\citep{heath77,peebles80}
\begin{align}
D(a)\,\equiv\,a g(a)\,=\,\frac{5\Omega_{m,0}H_{0}^{2}}{2}H(a)\int_{0}^{a}
\frac{da'}{a'^{3}H(a')^{3}}\nonumber\\
\end{align}
where $g(a)$ is the growth with respect to that in an Einstein-de Sitter 
Universe ($D(a)$ is normalized such that $D(a)\approx a$ for $a \rightarrow 
0$). 

\smallskip
\noindent In order to find the linear density excess $\delta_0$ for the 
critically bound shell, one should evaluate its evolution at early epochs 
($a\ll1$). At these early times -- when density perturbations are still very 
small, $\delta \ll1$ -- the linearly extrapolated density excess 
(Eqn.~\ref{eq:denlin}) represents a good approximation for the (real) density 
of the object (Eqn.~\ref{eq:ratio}). 

Using the fact that the early Universe is very close to an Einstein-de Sitter 
Universe and expands accordingly, $a(t)\propto t^{2/3}$, we may infer that the 
early density excess $\delta$ of the bound sphere and its dimensionless radius 
$\tilde{r}$ is (see appendix~\ref{app:rdelta}) 
\begin{equation}
{\delta}(t)\,=\,\frac{9}{10}\,{\tilde r}(t)\,.
\end{equation}
Using the approximate evolution of ${\tilde r}(a)\propto a $ implied by 
Eqn.~(\ref{eq:deltamasuno}), for early times ($1+\delta \approx 1$) we find 
\begin{equation}
{\delta}(t)\,=\,\frac{9}{10}\,\left(\frac{2\Omega_{\Lambda,0}}
{\Omega_{m,0}}\right)^{1/3}\,a(t)\,.
\end{equation}

\smallskip
\noindent The above leads us directly to the value of the linear density excess 
$\delta_0$, as at the early Einstein-de Sitter phase 
$\delta(t) \approx a(t) \delta_0$,
\begin{equation}
\delta_{0}\,=\,\frac{\delta}{a}\,=\,\frac{9}{10}
\left(\frac{2\Omega_{\Lambda,0}}{\Omega_{m,0}}\right)^{1/3}\,\approx\,1.504\,.
\label{eq:deltabnd}
\end{equation}
The corresponding linearly extrapolated density excess for marginally 
bound structures at the present epoch is
\begin{equation}
\delta_b(a=1)\,=\,1.17\,,
\label{eq:deltabnd0}
\end{equation}
where we have used the approximation for $g(a)=D(a)/a$ \citep{carroll92} 
\begin{equation}
\begin{split}
g(a)&\approx\frac{5}{2}\Omega_{m}(a)\times\\&\left[
\Omega_{m}(a)^{4/7}-\Omega_{\Lambda}(a)
+\left(1+\frac{\Omega_{m}(a)}{2}\right)
\left(1+\frac{\Omega_{\Lambda}(a)}{70}\right)\right]^{-1}
\end{split}
\label{eq:growth}\,
\end{equation}
These objects are due to reach turnaround at $a\rightarrow\infty$. 

\subsection{Tests of the Spherical Binding Criterion}
Reality is always far more complex than a simple analytical criterion is 
liable to cover. In addition to distinct anisotropies, internal inhomogeneities
and velocity dispersions will influence the viability of the derived spherical 
binding criterion. By means of N-body simulations \cite{dunner06} tested the 
binding density criterion $\rho_{cs}/\rho_{c}=2.36$ (for the current epoch) and
the criterion involving the mass enclosed within the radius $r_{\rm cs}$. On 
average 72\% of the mass enclosed within $r_{\rm cs}$ is indeed gravitationally
bound to the structure. At the same time it was found that a mere 0.3\% of the 
mass bound to the object is not enclosed within this radius. 
\section{The Computer Simulation}
\label{sec:simulation}
\begin{figure*}
\vskip -0.25truecm
\mbox{\hskip -0.5truecm\includegraphics[width=0.49\textwidth]{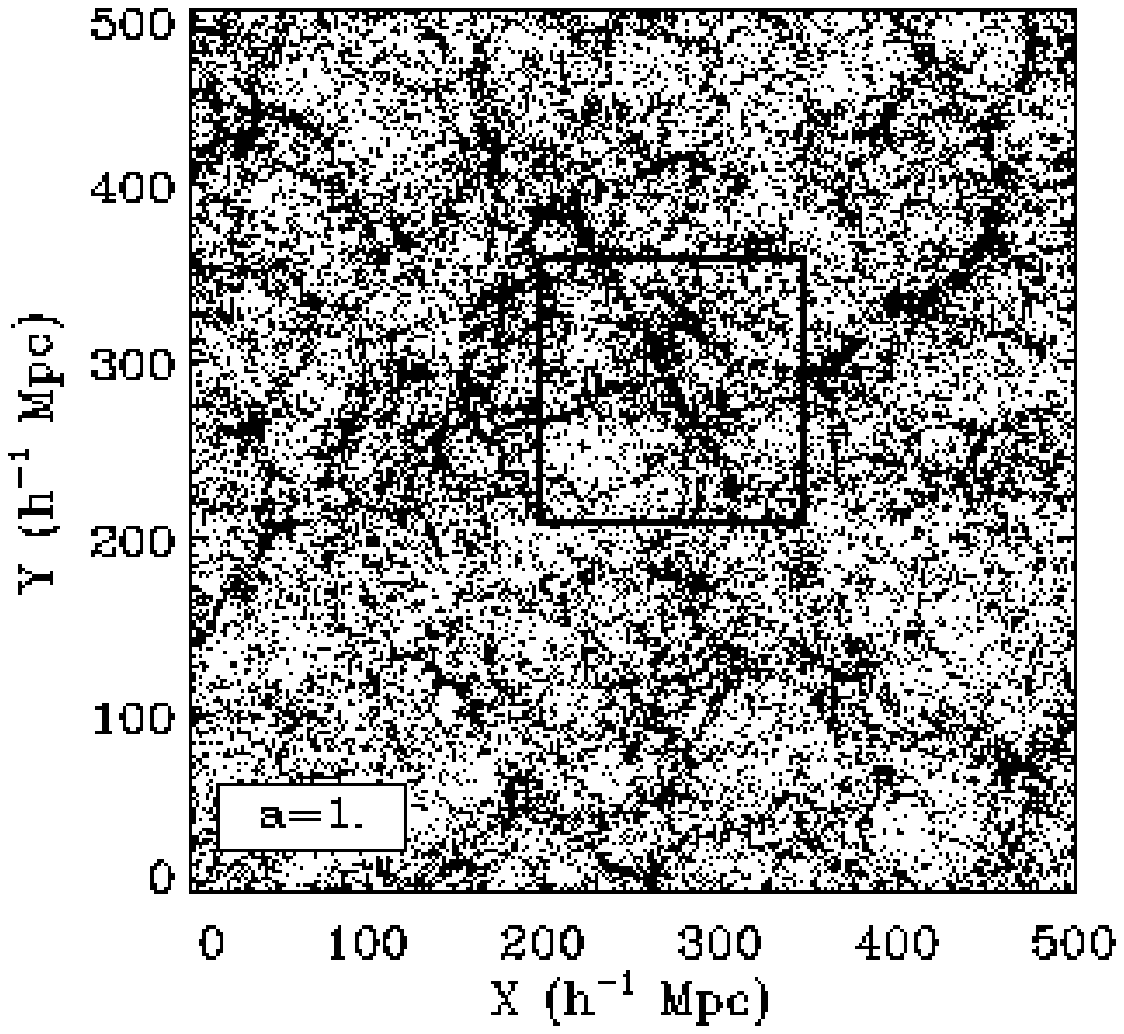}}
\includegraphics[width=0.49\textwidth]{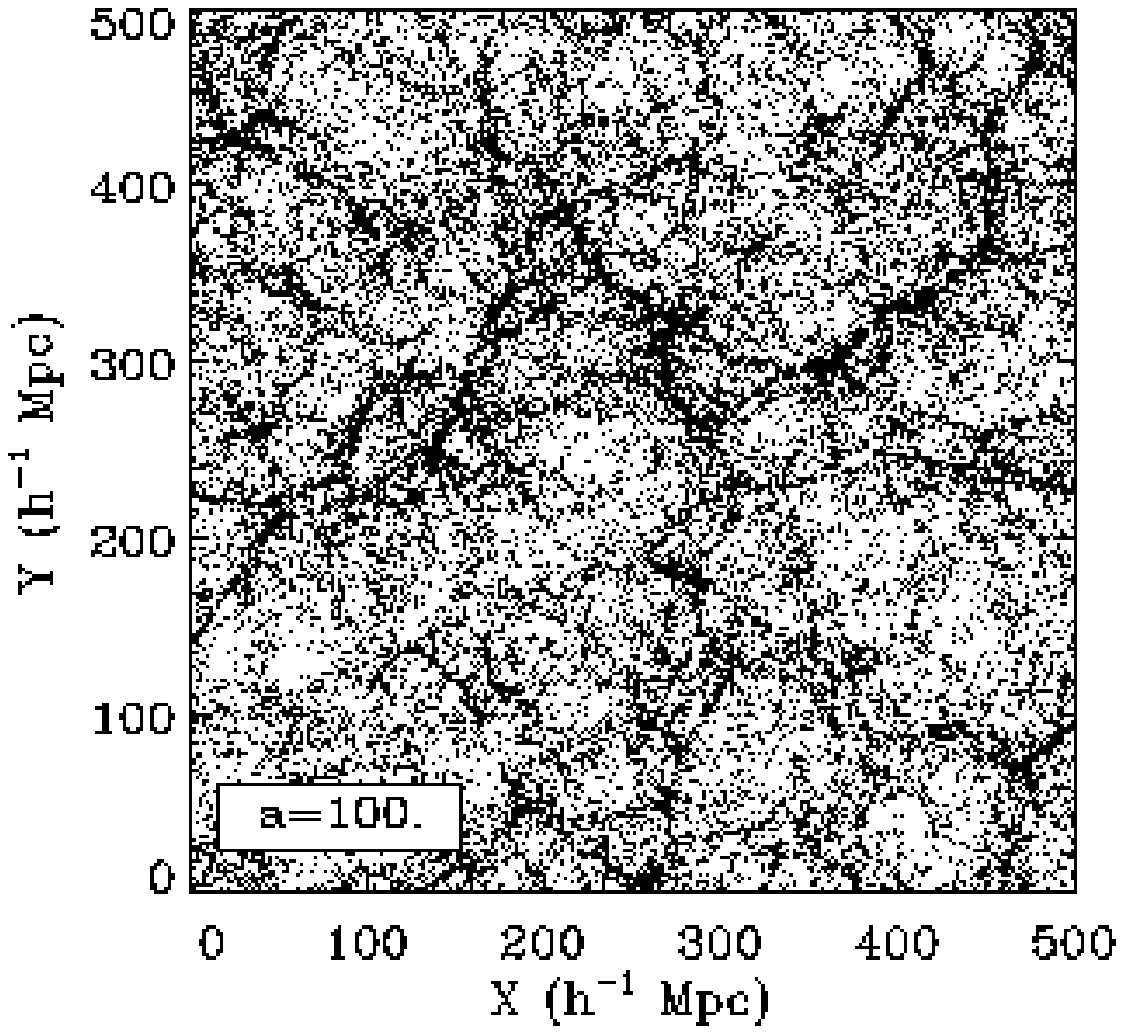}
\vskip -0.25truecm
\mbox{\hskip -0.5truecm\includegraphics[width=0.49\textwidth]{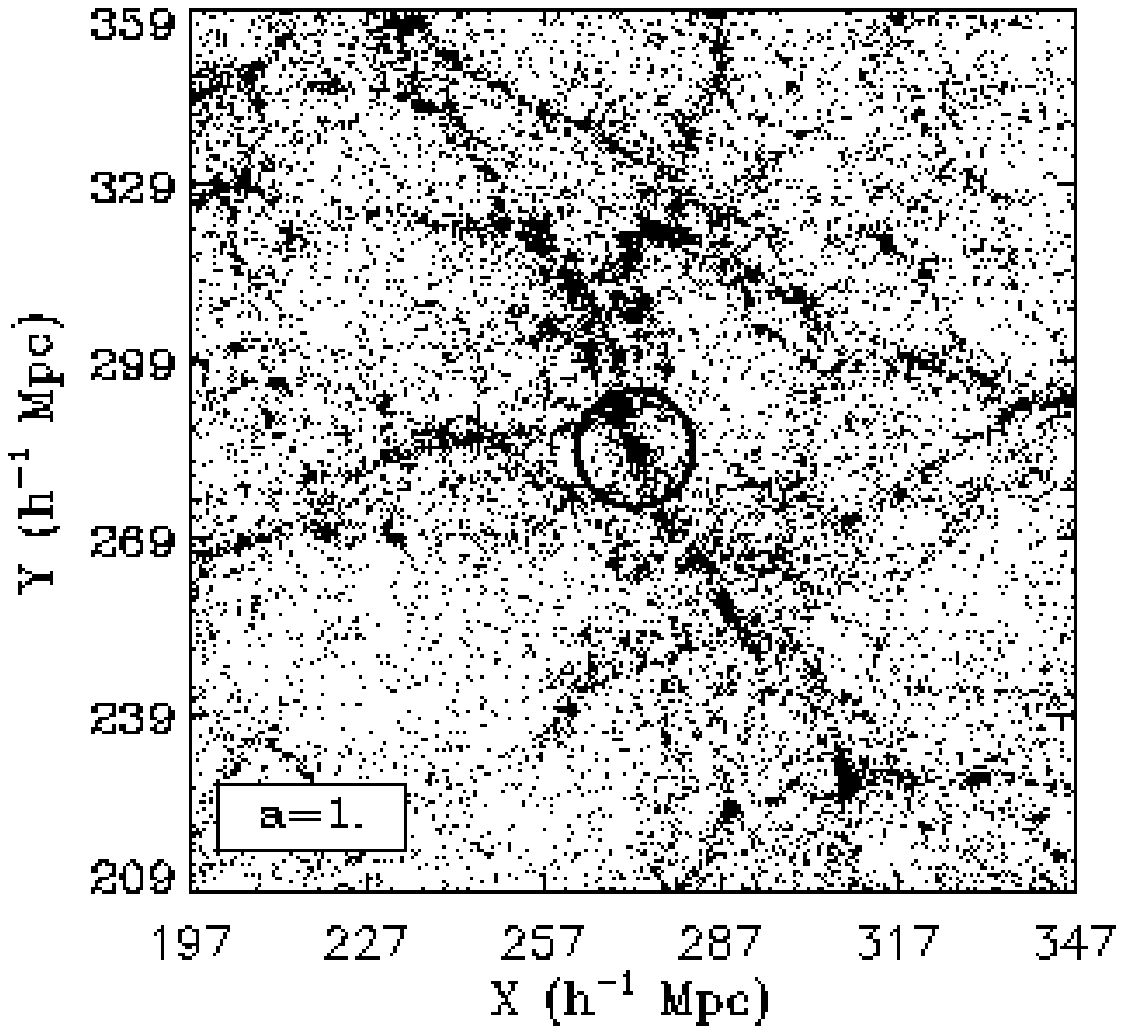}}
\includegraphics[width=0.49\textwidth]{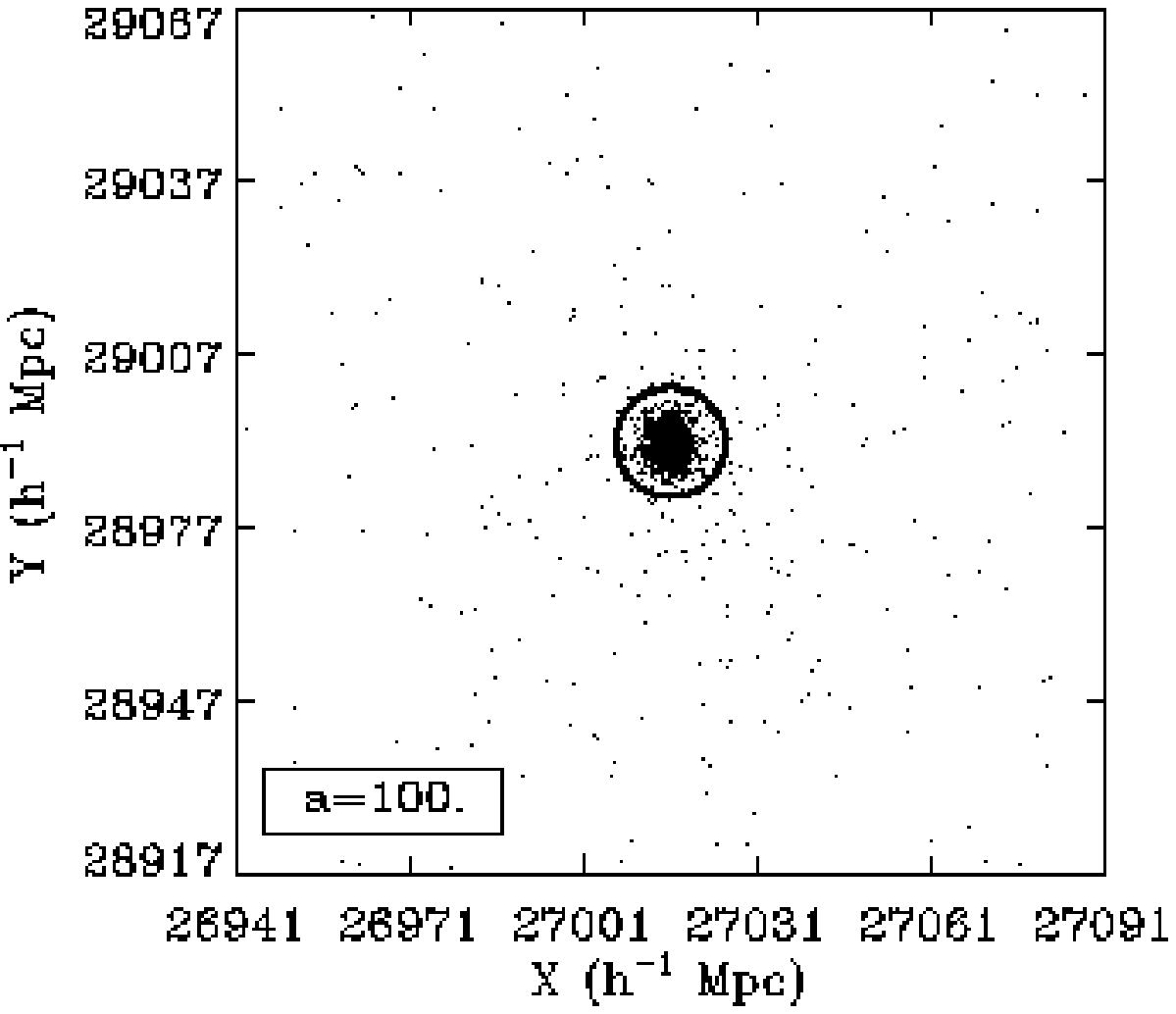}
\vskip -0.25truecm
\caption{\small{Simulated mass distribution at $a=1$ (left column) and 
$a=100$ (right column). The top panels show the particle distribution 
in a 500$h^{-1}$Mpc box, in a $30h^{-1}$Mpc thick slice projected along the $z$-axis. Top left: mass 
distribution at $a=1$. Top right: mass distribution at $a=100$.
Lower left: zoom-in on the $a=1$ mass distribution within $50h^{-1}$Mpc box 
indicated by the square in the top lefthand panel (similar $30h^{-1}$Mpc 
thick slice along $z$ direction). Lower right: region of same $50h^{-1}$Mpc 
{\it physical} size as the lower left panel, at $a=100$, around the same bound 
supercluster.}}
\label{fig:particles}
\end{figure*}

We simulate a standard flat $\Lambda$CDM Universe with cosmological parameters 
$\Omega_{m,0}=0.3$, $\Omega_{\Lambda,0}=0.7$, and $h=0.7$, where the 
Hubble parameter is given by $H_{0}=100h$ km s$^{-1}$Mpc$^{-1}$. The 
normalization of the power spectrum is $\sigma_{8}=1$. In order to have a 
large sample of bound objects, the simulation box has a side length of 
500$h^{-1}$ Mpc and contains 512$^{3}$ dark matter particles of mass 
$m_{dm}=7.75\times10^{10}h^{-1}$M$_{\odot}$ (see Fig.~\ref{fig:particles}). 
The initial conditions are  
generated at expansion factor $a=0.02$ (redshift $z=49$), and  evolved 
until $a=100$ using  the massive parallel tree N-Body/SPH code GADGET-2 
\citep{springel05}. The Plummer-equivalent softening was set at 
$\epsilon_{Pl}=20$ $h^{-1}$kpc in physical units from $a=1/3$ to $a=100$, while
it was taken to be fixed in comoving units at higher redshift. Given the mass 
resolution and the size of the box, our simulation allows us to reliably 
identify massive superclusters with $\sim 80,000$ particles. The simulation 
was performed on the Beowulf Cluster at the University of Groningen.

We took snapshots at five different timesteps: starting at the present 
time ($a=1$), we studied the mass distribution at $a=2$, $a=5$, $a=10$, to 
ultimately end up in the far future at $a=100$. The latter was taken as 
a representative epoch at which the internal evolution of all bound objects 
appeared to have been completed. 

\subsection{Identification of bound objects and superclusters}
\label{sec:boundid}
We identify and extract groups and objects in the simulation on 
the basis of a random subsample of $256^3$ particles ($1/8$ of the total 
particle number). For the identification of bound structures we apply a 
four-step procedure. In the first step, we find all virialized halos in the
simulations with more than 50 particles. Subsequently, we incorporate the 
surrounding spherical region bound to these halos. We then join
the bound spheres that overlap each other into single bound objects. 
Finally, among these objects we select the most massive ones,  
the superclusters in the simulation volume. 

\subsubsection{Mass range}
A first practical issue is the minimal amount of particles we 
deem necessary to accept a halo/object detection as significant. 
We choose a minimum of 50 particles, corresponding to a mass cut of 
$M\ge 3.1\times 10^{13}h^{-1}$M$_{\odot}$. This means our cluster and 
bound object sample has an implicit bias in not containing any objects with a 
mass less than the mass limit. This is not a problem for the most massive 
objects, but may produce an incomplete sample for lower mass objects.

\subsubsection{HOP and virialized halo finding}
\label{sec:virhalo}
In order to find groups of particles present in our simulation, we use HOP 
\citep{eisenstein98}. This algorithm first assigns a density estimate
at every particle position by smoothing the density field with an SPH-like
kernel using the $n_{dens}$ nearest neighbors of a given particle. In our case,
we use $n_{dens}=64$. Subsequently, particles are linked by associating each 
particle to the densest particle from the list of its $n_{hop}$ closest 
neighbors. We use $n_{hop}=16$. The process is repeated until it reaches the 
particle that is its own densest neighbor. 

The algorithm associates all particles to their local maxima. This procedure 
often causes groups to fragment. To correct this, groups are merged if the 
bridge between them exceeds some chosen density thresholds. Three density 
thresholds are defined as follows \citep{cohn01}:
\begin{itemize}
\item[-] $\delta_{out}$: 
\item[] \ \ \ the required density for a particle to be in a group.
\item[-] $\delta_{saddle}$: 
\item[] \ \ \ the minimum boundary density between two groups 
\item[] \ \ \ so that they may have merged. 
\item[-] $\delta_{peak}$: 
\item[] \ \ \ the minimum central density for a group to be
\item[] \ \ \ independently viable.
\end{itemize}
We follow the criterion of \cite{eisenstein98}: 
$\delta_{outer}$:$\delta_{saddle}$:$\delta_{peak}$=1:2.5:3. The value of 
$\delta_{peak}$ is associated with that of the corresponding density 
$\Delta_{vir}(a)$ of the virialized core of the bound regions.

\subsubsection{Virial Density Value}
To determine the value of the virial density $\Delta_{vir}(a)$ in the HOP 
formalism we resort to the dynamical evolution of spherical top-hat 
perturbation. The value of $\Delta_{vir}$ is obtained from the solution to the 
collapse of a spherical top-hat perturbation under the assumption that the 
object has just virialized. Its value is $\Delta_{vir}=18\pi^{2}$ for an 
Einstein-de Sitter Universe. For the cosmology described here, at $a=1$ its 
value is $\Delta_{vir}(a=1)\approx 337$. The value at later epochs, or in other
cosmologies, is obtained by solving the spherical collapse equations 
numerically \citep{gunn72,lacey93,eke1996,kitayama1996,bryan98}. An extensive 
description of this can be found in \cite{arayamelo08}.

At $a=100$, $\Omega_{m}=4.3\times 10^{-7}$. For this situation 
we resort directly to the virial theorem to determine the characteristic 
virial radius. According to the virial theorem the kinetic energy $K_{vir}$ of 
a body whose potential is of the form $V_{vir}=R^n$ is equal to 
$K_{vir}=(n/2)V_{vir}$ \citep{landau1960,lahav91}. For a general case of a 
virialized object in a Universe with matter and a cosmological constant, 
\begin{equation}
{\tilde K}_{vir}\,=\,-\frac{1}{2}\tilde{V}_{G,vir}\,+\,\tilde{V}_{\Lambda,vir}\,.
\end{equation}
Note that here we write energies in dimensionless form (see e.g. 
Eqn.~\ref{eq:energy} and Eqn.~\ref{eq:potential}). In this equation 
$\tilde{V}_{G}$ is the gravitational potential energy and $\tilde{V}_{\Lambda}$ 
is the potential energy due to the cosmological constant \citep{lahav91}. 
Hence, the total energy ${\tilde E}_{vir}={\tilde K}_{vir}+{\tilde V}_{vir}$ of 
a virialized object is 
\begin{equation}
\tilde{E}_{vir}=\frac{1}{2}{\tilde{V}_{G,vir}}+2\tilde{V}_{\Lambda,vir}
\label{eq:energy_at_vir}
\end{equation}

\noindent Because of energy conservation, the energy at maximum expansion is 
equal to the energy at virialization. This translates into the following 
equation for the dimensionless radius
\begin{equation}
\tilde{r}^{2}+\frac{1}{2\tilde{r}}=\frac{3}{2}\,.
\label{eq:cubic}
\end{equation}
This is a cubic equation with solutions $\tilde{r}\approx-1.366$, $\tilde{r}=1$ 
and $\tilde{r}\approx0.366$. The first is an unphysical solution. The solution 
$\tilde{r}=1$ is the maximum radius of the critically bound shell (see 
Eqn.~\ref{sec:densthr}). The third value, $\tilde{r}\approx0.366$, is the 
virial radius of the enclosed mass $M$. The corresponding virial density excess
of the mass clump would be $1+\Delta_{vir}=\bar{\rho}/\rho_{c}\approx40.8$. 

\subsubsection{Clusters}
Besides the definition of a sample of superclusters, we also identify 
the clusters in our simulation. Their identity is more straightforward to 
define, since we may presume they are virialized. 

Virialized halos, as identified by HOP (sect.~\ref{sec:virhalo}), with masses 
larger than $3\times10^{13}h^{-1}$M$_{\odot}$, are considered {\it clusters}. 
Note that, by definition, these clusters are identical to the bound object 
cores produced in step 1 of our supercluster finding procedure 
(sect.~\ref{sec:virhalo}). 

\begin{figure}
\vskip -0.5truecm
\includegraphics[width=0.5\textwidth]{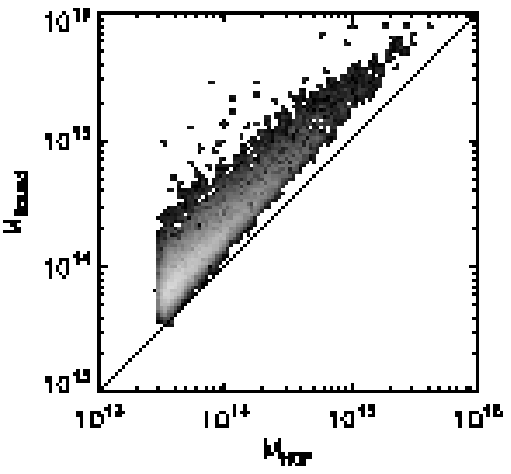}\\
\vskip -0.25truecm
\includegraphics[width=0.5\textwidth]{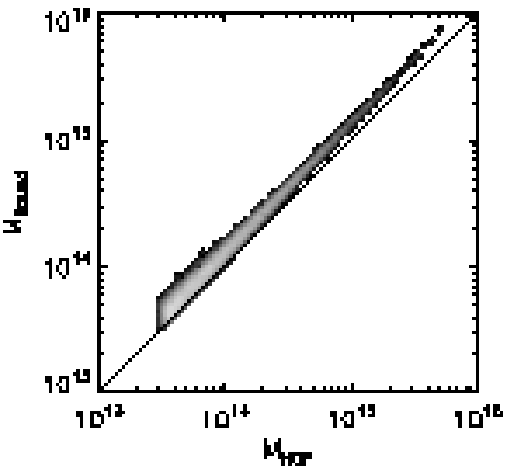}
\vskip -0.25truecm
\caption{Scatter plots of the mass of identified bound objects, M$_{\rm bound}$, 
against the mass of the virialized HOP halo that forms its core, M$_{\rm HOP}$. 
The density of points (M$_{\rm HOP}$,M$_{\rm Bound}$) in the scatter diagram is 
depicted in terms of an isodensity contour map. The highest concentration of 
points has the darkest (black) shade, gradually fading towards light colours. 
Top: $a=1$. Bottom: $a=100$}
\label{fig:scatter}
\end{figure}

\subsubsection{Bound Halo Identification}
\label{sec:boundhalo}
Once we have identified the virialized halos in our sample, we have to proceed 
by outlining the gravitationally bound region around these cores and, if 
necessary, join them together into a supercluster. In practice, we do 
this as follows.

Of an identified HOP halo, we take the location of the densest particle as a
first estimate of the center of mass. Subsequently, we grow a sphere around 
this center, with the radius being increased until the mean overdensity within
the corresponding radius reaches a value of 300$\rho_{c}$. This value is chosen
in order to find the densest core of the structure. We then calculate the 
center of mass of this sphere and repeat the process, iterating until the shift
in the center between successive iterations is less than 1\% of the radius. 

With the final center of mass, we apply the criterion of Eqn.~\ref{eq:ratio} 
for identifying the bound spherical region around the HOP core. To this end 
we determine the radius $r_{cs}$ at which the average interior density 
$\rho_{cs}/\rho_c$ reaches a value: $\rho_{cs}/\rho_c=2.36$ at $a=1$, 
$2.22$ at $a=2$, $2.06$ at $a=5$, $2.02$ at $a=10$ and $2.00$ at $a=100$ 
(see sect.~\ref{sec:densthr}). 

\subsubsection{Joining Halos: Bound Objects}
\label{sec:join}
The procedure outlined in the previous paragraph will frequently lead to
overlapping bound spheres that in reality will be bound to each other. In order
to account for this, we follow a radical prescription. If two spheres overlap,
we proceed with the most massive one and join the lower mass sphere to the high 
mass one while removing it from the list of objects.

We found that by following this procedure at $a=100$, such overlaps 
do not occur. It implies that in the far future nearly all bound objects are 
compact and isolated islands in the Universe. At much earlier epochs, and 
specifically at $a=1$, it does turn out to occur for a significant fraction of 
the bound spheres.

\subsubsection{Superclusters and Bound Objects}
Because not all bound objects would be prominent enough to be 
a supercluster, we use a mass threshold to select the superclusters 
amongst the bound objects in our sample. The mass threshold 
is chosen to be $M$$_{sc}\,=\,10^{15}h^{-1}$M$_{\odot}$, 
approximately the mass of the Local Supercluster. Although this  
value is somewhat arbitrary, and a somewhat different value might 
also have been viable, it represents a reasonable order of 
magnitude estimate. 

As a result, we reserve the name {\it bound objects} for all the 
objects that have ended up in our sample, while {\it superclusters} 
are the subset with masses higher than our supercluster mass threshold 
of $10^{15}$M$_{\odot}$. 

\subsubsection{Sample completeness}
Fig.~\ref{fig:scatter} helps us to evaluate the completeness of our object 
sample. The panels contain scatter plots of the total mass of the bound 
objects versus the HOP mass of their corresponding virialized cores.

Evidently, the lower right region is empty: HOP groups will always be less 
massive than the bound groups. There is a correlation between both masses, but 
with a high scatter. As expected, the evolution of bound objects towards fully 
virialized clumps expresses itself in a substantially stronger correlation at 
$a=100$ than at $a=1$. On the basis of these relations, we may conclude that at 
$a=1$ the sample is complete for masses greater than 
$2\times10^{14}h^{-1}$M$_{\odot}$, while at $a=100$ the sample is complete for 
masses down to $6\times10^{13}h^{-1}$M$_{\odot}$. In order to keep the samples 
comparable, we use a mass completeness threshold of 
$2\times10^{14}h^{-1}$M$_{\odot}$ at $a=1$ and $1.4\times10^{14}h^{-1}$M$_{\odot}$ 
at $a=100$ (see discussion sect.~\ref{sec:ps_mf}). 

\subsubsection{The Object Sample}
\label{sec:objectsample}
At $a=1$ HOP finds $\sim 20600$ virialized ``clusters'' with more than 50 
particles, i.e. halos with a total mass 
$M\ge 3.1\times 10^{13}h^{-1} $M$_{\odot}$. At $a=100$ it finds $\sim 18000$ 
virialized objects. These will be taken as the starting point for our 
supercluster finding procedure. They also constitute the cluster sample in our 
simulation. 

After determining the connected bound region and joining these overlapping 
bound spheres (see sec.~\ref{sec:join}), plus checking for sample completeness, 
we finally end up with a sample of $\sim 4900$ bound objects at the current 
epoch.  At the other epochs, from $a=2$ to $a=100$, this is approximately the 
same number. Of these, $\sim 535$ are superclusters (at $a=1$), while seventeen
are truely massive supercluster complexes with 
$M$$_{sc}> 5\times10^{15}h^{-1}$M$_{\odot}$. 

\section{Evolution of Bound Objects}
\label{sec:evolution}
The two top panels of Fig.~\ref{fig:particles} show a slice of 30$h^{-1}$Mpc 
width of the particle distribution projected along the $z$ axis, at $a=1$
and $a=100$. By taking a region of the same physical size at both epochs, the 
effect of the de Sitter expansion of the Universe becomes manifestly clear.

At $a=1$, the large-scale structure of the cosmic web is well
established and its morphology and character hardly change thereafter. 
The lower left panel zooms in on the square region of the top-left panel. 
Centered on a massive structure, it shows the mass distribution at $a=1$. The 
radius of the circle is that of the bound region, according to the criterion of 
Eqn.~\ref{eq:ratio}. It shows that it is well connected with the surrounding 
structure. The same object, but now at $a=100$, is depicted in the lower right 
panel (with the same \emph{physical scale}). We see that the size of the bound 
object is nearly the same at both expansion factors. While in comoving 
coordinates the accelerated expansion of the Universe results in a freezing of 
structure growth on scales much larger than the initial size of superclusters, 
in physical coordinates the separation of structures continues and grows 
exponentially in time. This results in clearly detached regions which evolve in
complete isolation: genuine {\it cosmic islands}. 

\begin{figure*}
\vskip -0.5truecm
\mbox{\hskip -8.0truecm\includegraphics[width=0.45\textwidth]{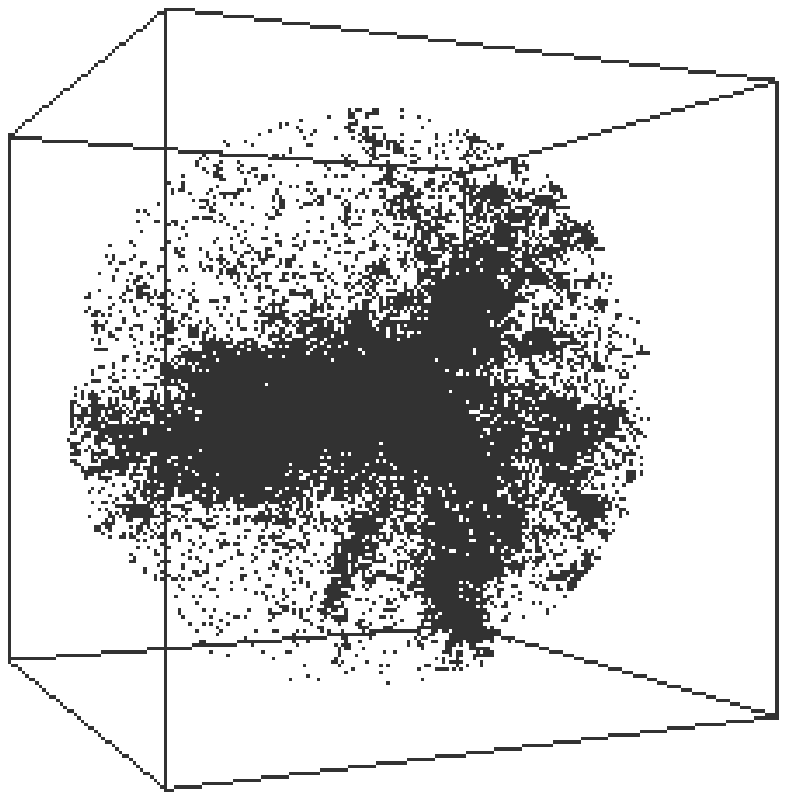}}
\vskip -4.15truecm
\mbox{\hskip 9.25truecm\includegraphics[width=0.45\textwidth]{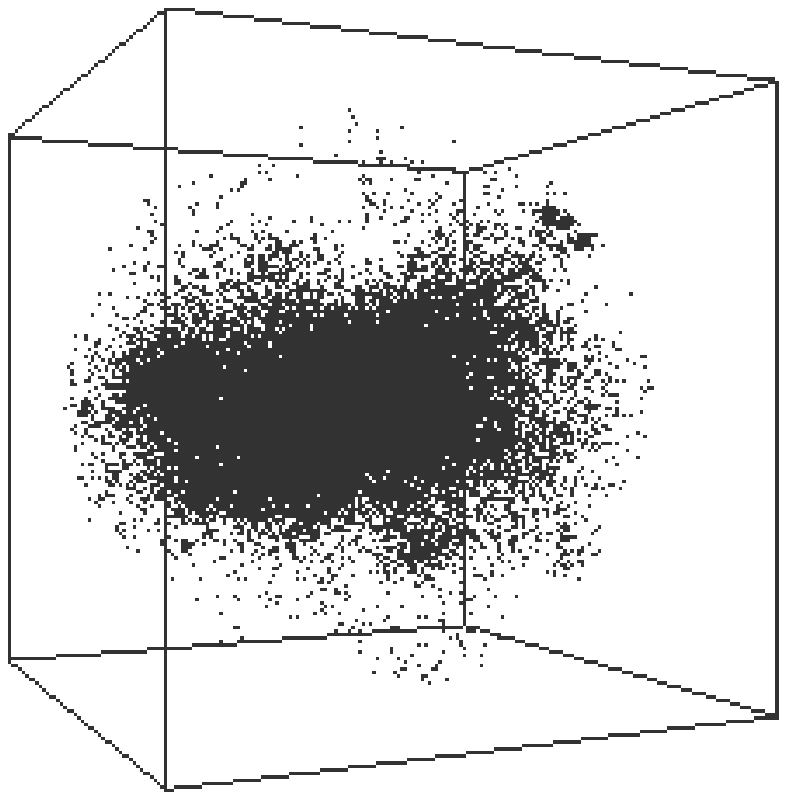}}
\vskip -4.15truecm
\mbox{\hskip -8.0truecm\includegraphics[width=0.45\textwidth]{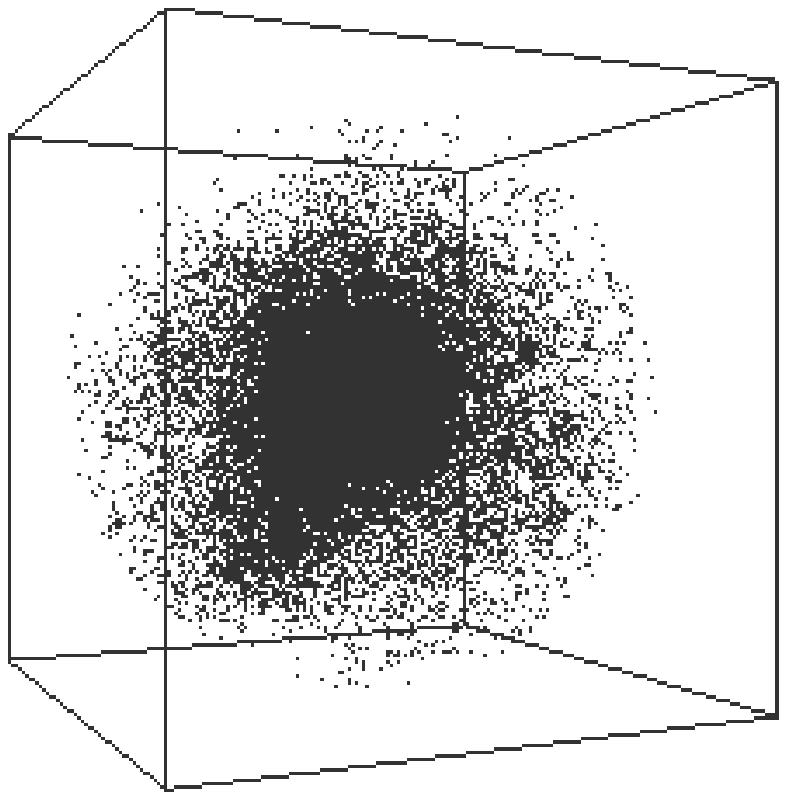}}
\vskip -4.15truecm
\mbox{\hskip 9.25truecm\includegraphics[width=0.45\textwidth]{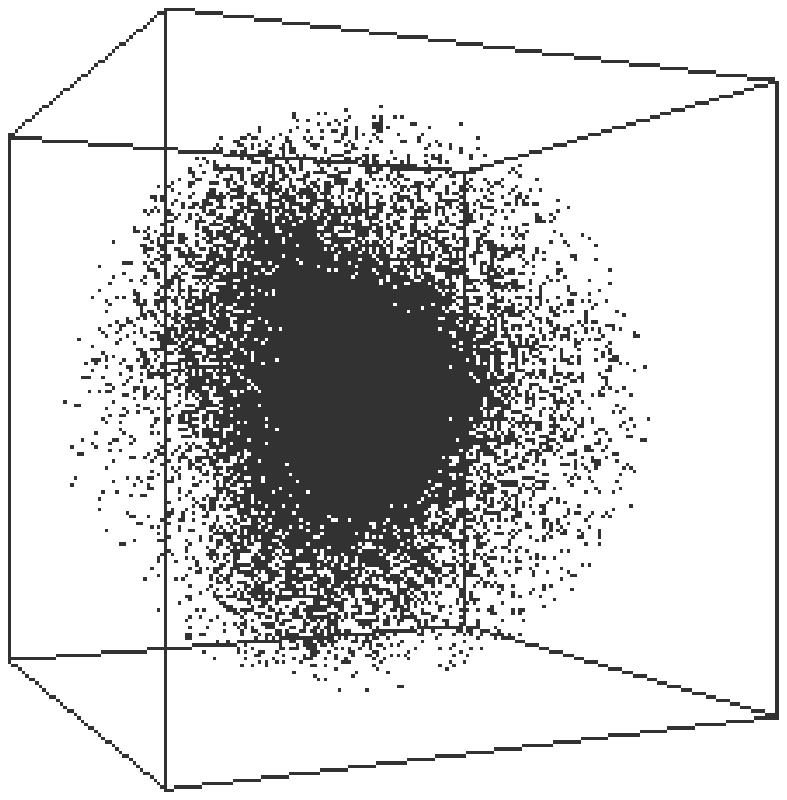}}
\vskip -4.15truecm
\mbox{\hskip -8.0truecm\includegraphics[width=0.45\textwidth]{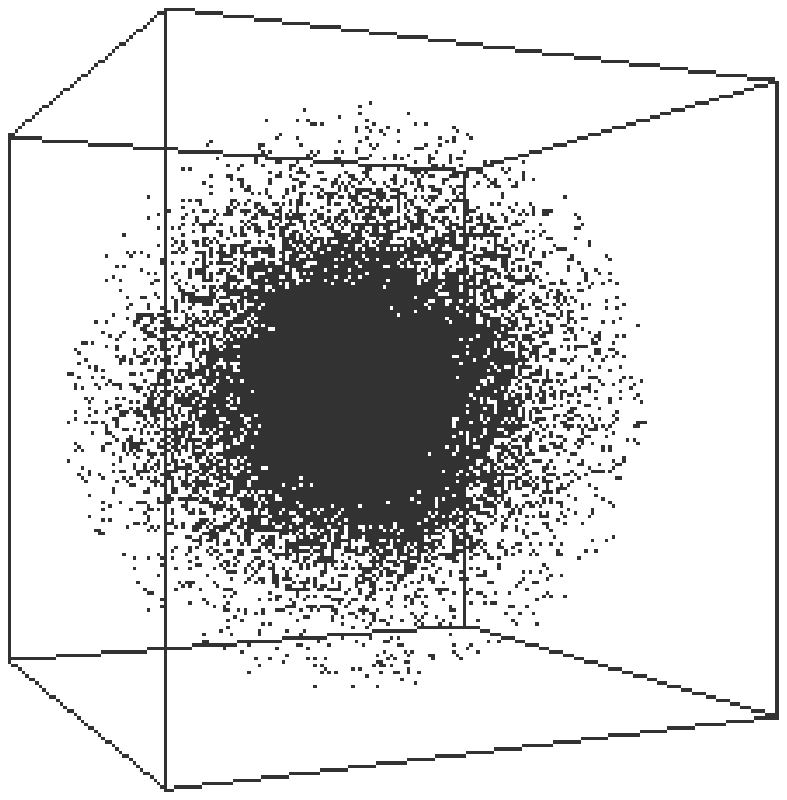}}
\caption{\small{Evolution of one of the most massive objects in our simulation, 
with a present-day mass of $M$$\sim6.8\times10^{15}h^{-1}$M$_{\odot}$, at $a=1$. 
The supercluster is shown at $a=1$, $a=2$, $a=5$, $a=10$ and $a=100$ 
(zigzagging from top left to bottom left). The size of the box is always 
$14h^{-1}\hbox{Mpc}$ in physical coordinates.}}
\label{fig:cumulos_a0}
\end{figure*}

\begin{figure*}
\includegraphics[width=0.195\textwidth]{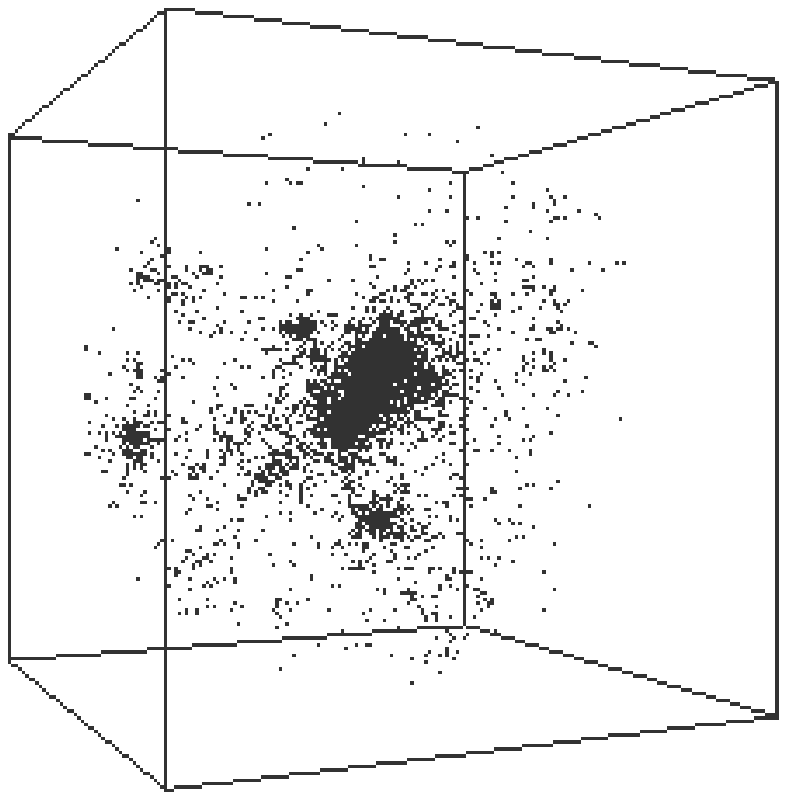}
\includegraphics[width=0.195\textwidth]{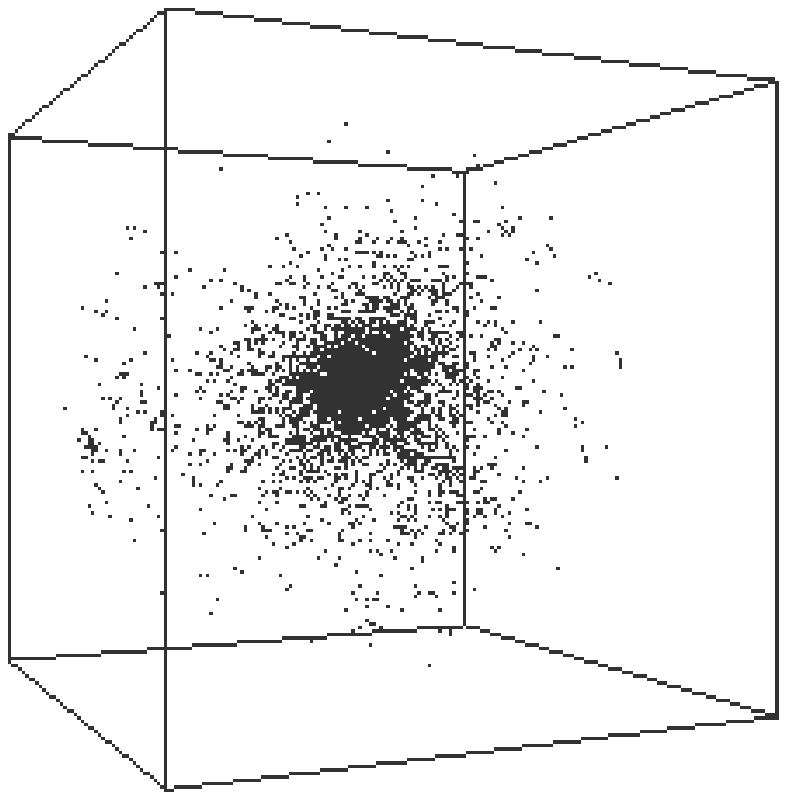}
\includegraphics[width=0.195\textwidth]{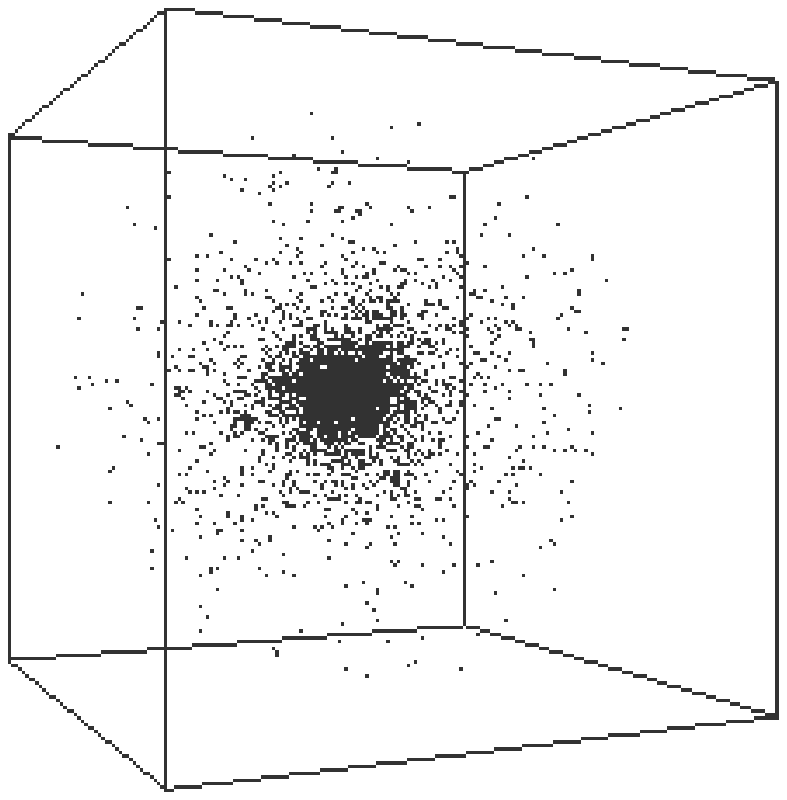}
\includegraphics[width=0.195\textwidth]{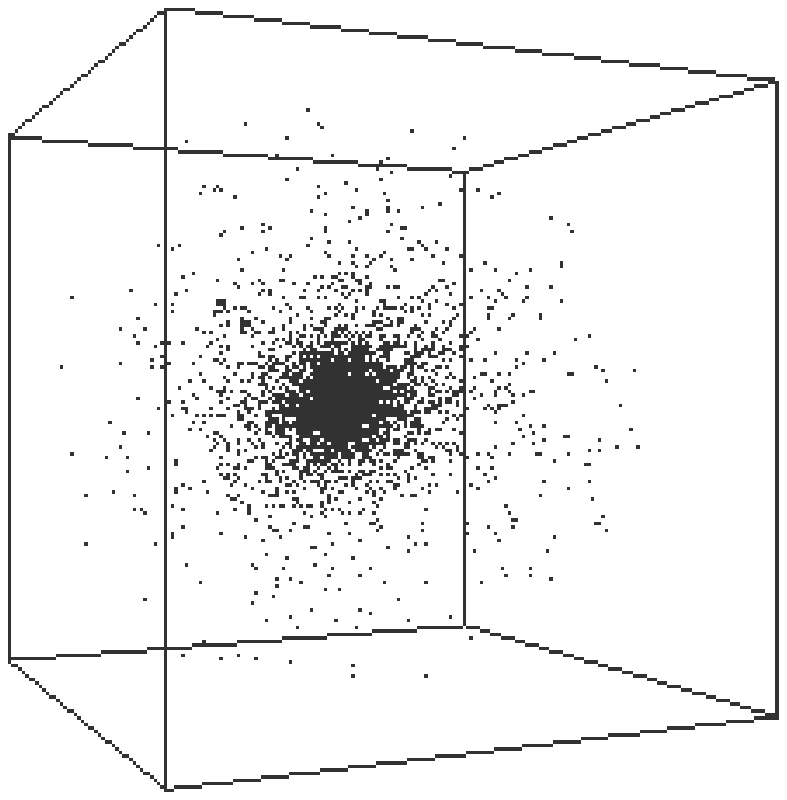}
\includegraphics[width=0.195\textwidth]{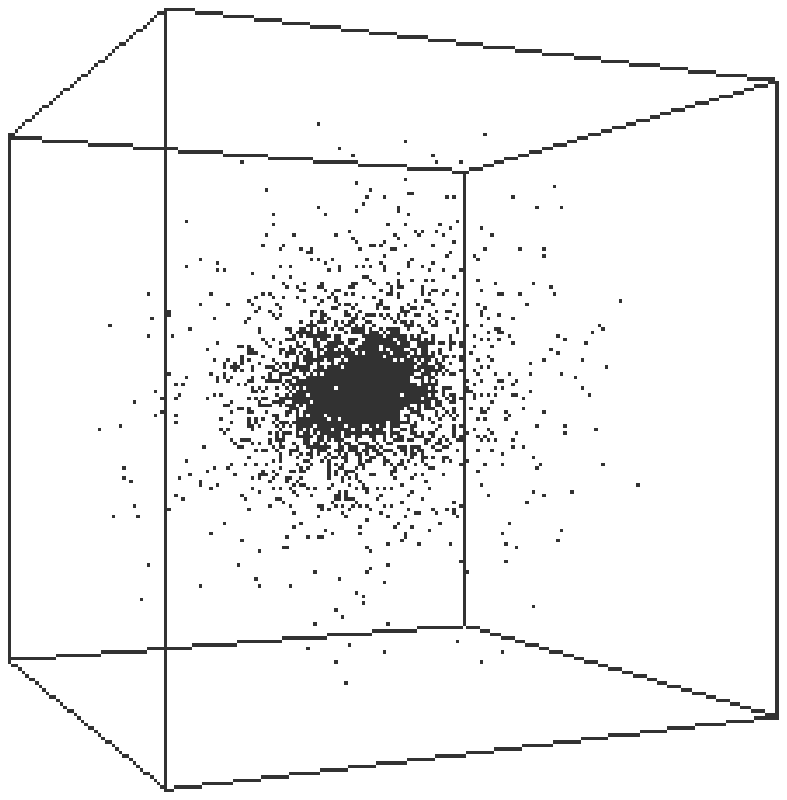}
\includegraphics[width=0.195\textwidth]{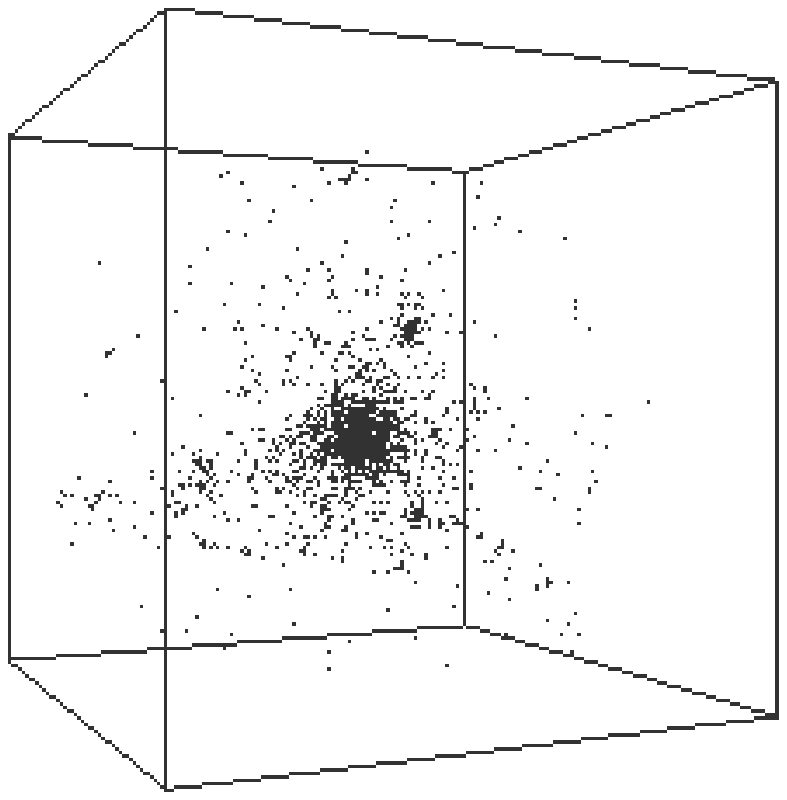}
\includegraphics[width=0.195\textwidth]{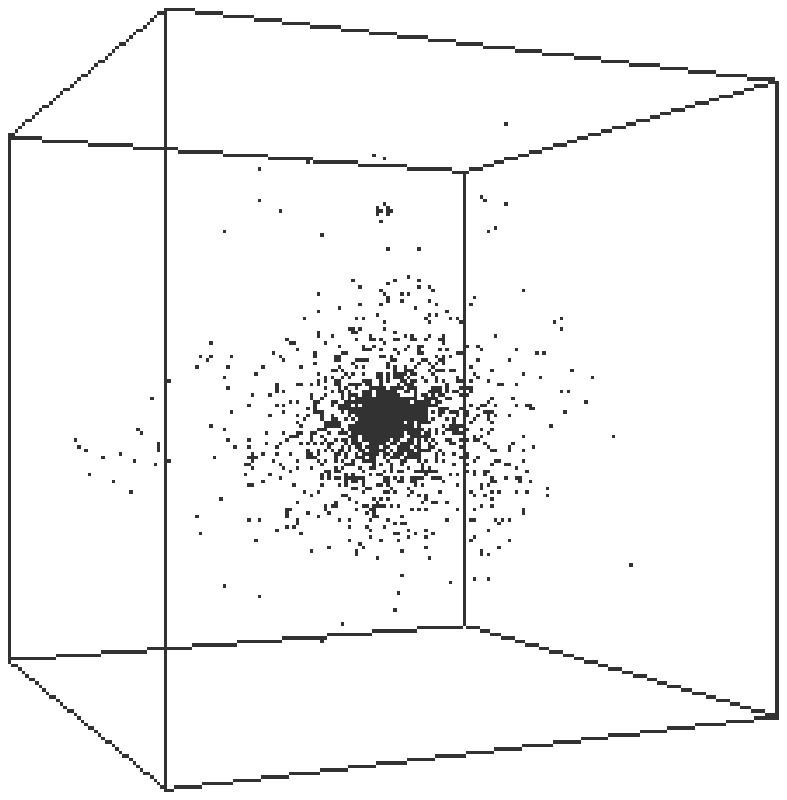}
\includegraphics[width=0.195\textwidth]{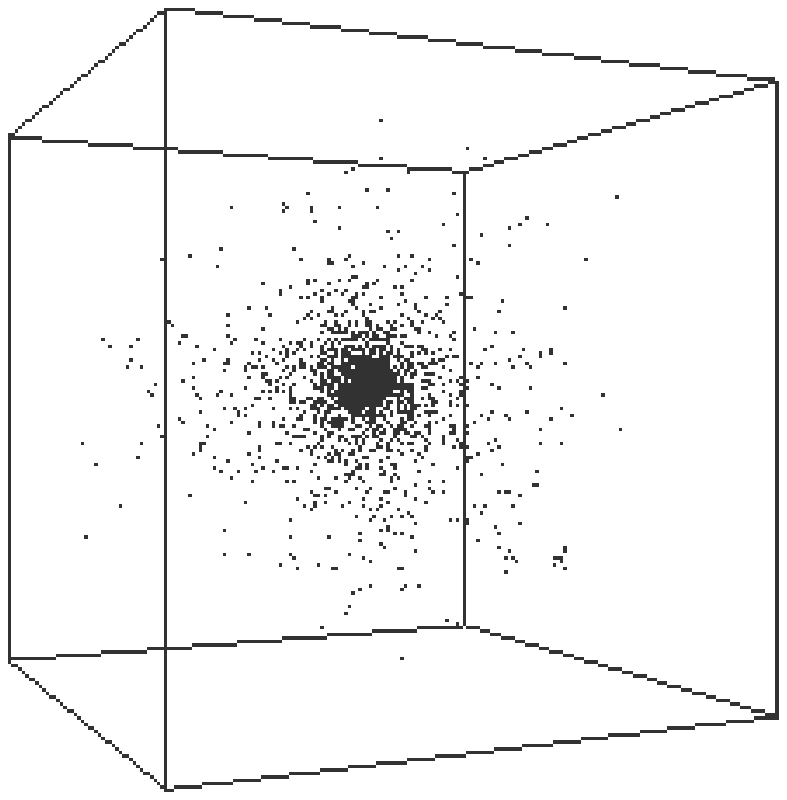}
\includegraphics[width=0.195\textwidth]{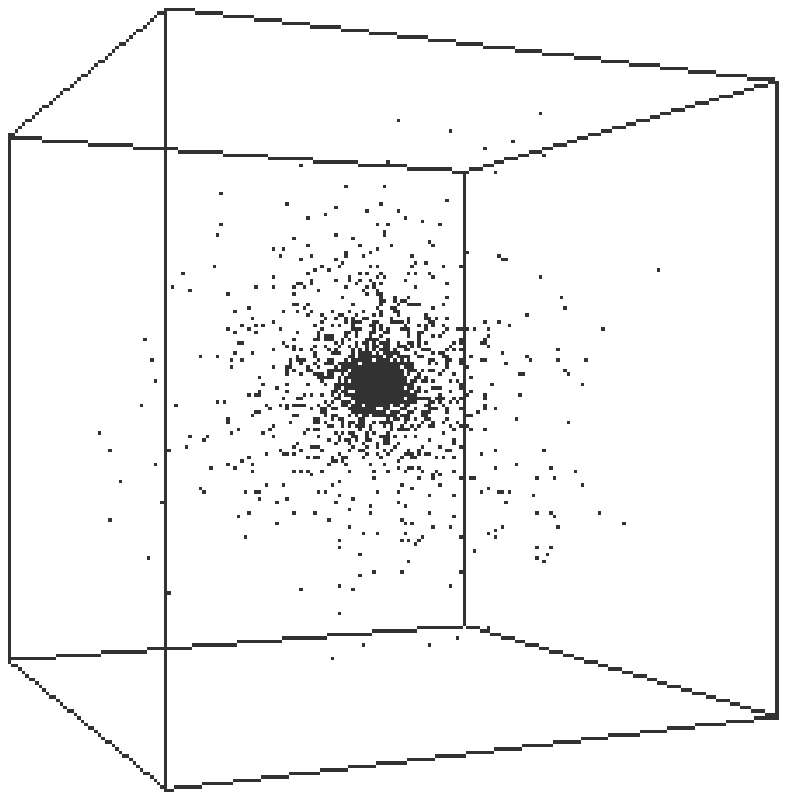}
\includegraphics[width=0.195\textwidth]{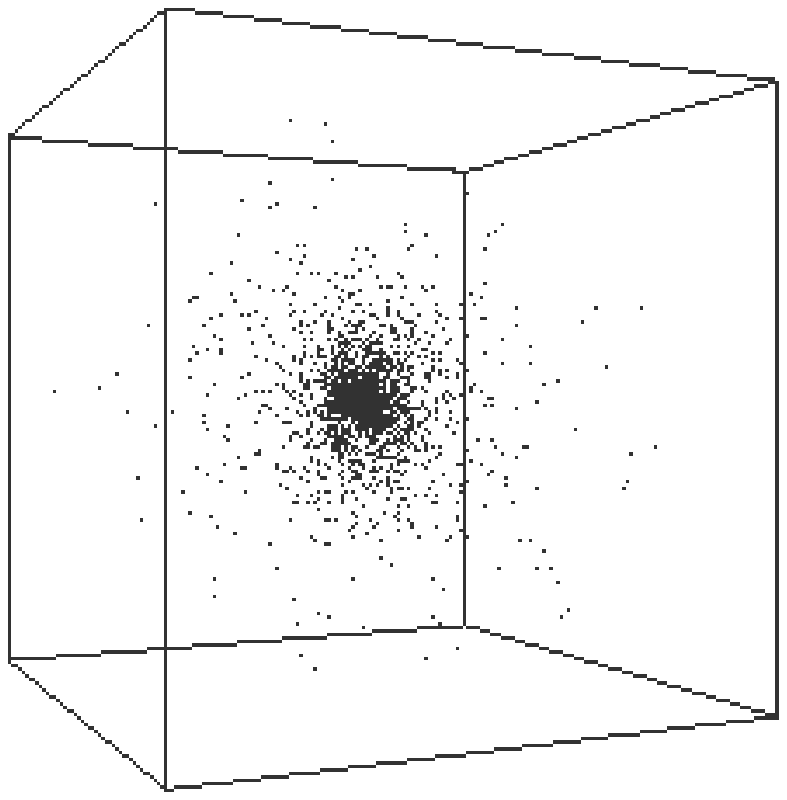}
\caption{\small{Evolution of two different bound objects in the simulation. 
Each row shows the evolution of a single object at $a=1$, $a=2$, $a=5$, $a=10$ 
and $a=100$. The particle positions are in physical coordinates. The box sizes 
have been scaled to the size of the mass concentrations and therefore differ 
amongst each other. Top row: intermediate mass bound object, 
$M$$\sim5.6\times10^{14}h^{-1}$M$_{\odot}$ ($a=1$). Boxsize is 
$6h^{-1}\hbox{Mpc}$ . Bottom row: least massive bound object, with a 
present-day mass of $\sim 2 \times10^{14}h^{-1}$M$_{\odot}$ (boxsize at $a=1$: 
$4h^{-1}\hbox{Mpc}$, boxsize at $a=100$: $3h^{-1}\hbox{Mpc}$). The evolution of 
these medium and low mass objects should be compared to that of the the massive
supercluster in Fig.~\ref{fig:cumulos_a0}.}}
\label{fig:cumulos_a1}
\end{figure*}

\subsection{Case Studies}
To get an impression of the internal and external evolution of the bound 
objects in our simulation, we focus on a few specific objects. By following 
qualitatively the ``internal'' evolution of three bound objects we intend to 
set the scene for the further analysis in this study.

Fig.~\ref{fig:cumulos_a0} displays the evolution of one of the most massive 
objects in our sample. At $a=1$, it has a substantial degree of substructure 
(Fig.~\ref{fig:cumulos_a1}, lefthand). The centre of the supercluster is 
dominated by the massive central cluster that is the virialized object from 
which we constructed the remainder of the supercluster. It forms the centre of 
a huge complex, connecting the surrounding matter distribution via prominent 
filamentary extensions. These form the transport channels along which mass 
flows into the central supercluster region. Noteworthy is the large number and 
variety of subclumps along the filaments and around the centre of the 
supercluster. This high mass supercluster undergoes a radical change towards 
the future. 

By $a=2$ we see that most of the surrounding material has fallen into the 
central core, with a dramatic decrease in the number of surrounding subclumps. 
While at $a=5$ we still see a significant number of small clumps around the 
central supercluster, by $a=10$ only a few individuals seem to have survived. 
Comparing the 1st and 2nd panel with the panels corresponding to $a=5$ and 
$a=10$ also seems to suggest that the infalling subclumps at later epochs have 
a lower mass. 

With respect to the morphological character of the surrounding mass 
distribution, we note that at $a=2$ we can still discern the vague remnants of 
the salient filamentary patterns at the present epoch. Nonetheless, most of the
formerly richly patterned weblike structure has disappeared and seems to have 
resolved itself as its mass accreted onto the supercluster. The supercluster 
has also assumed a more smooth and roundish appearance, even though at $a=2$ we
can still recognize the original geometry, both in terms of its elongated shape
and its orientation along the same direction. At even later epochs, the trend 
towards a highly centralized and regular mass concentration with a nearly 
perfect spherical shape continues inexorably. At $a=10$ we may still just 
recognize some faint matter extensions at the edge of the supercluster. 
However, these do no longer bear the mark of its original orientation and 
shape. At the final timestep, $a=100$, the object has reached the ultimate 
configuration of a perfectly regular, nearly spherical, and centrally 
concentrated and largely virialized dark matter halo. All surrounding 
substructure within the binding radius has fallen and has been entirely 
absorbed while the supercluster attained a perfectly virialized configuration: 
the big mass concentration has become a true island universe.

The evolution of the massive supercluster is compared with that of two more 
moderate bound mass clumps in Fig.~\ref{fig:cumulos_a1}. The top row shows the 
evolution of a medium mass bound object, with a mass of 
$M\sim5.6\times10^{14}h^{-1}$M$_{\odot}$ (at $a=1$), while the bottom row 
depicts the development of the least massive bound object. The present-day mass
of the latter is $M\sim2\times10^{14}h^{-1}$M$_{\odot}$. Even though the medium 
mass object (central row) shares a similar trend towards a centrally 
concentrated virialized clump, we also notice that its influence over the 
surroundings is considerably less pronounced and extends over a considerably 
smaller region. At $a=1$ we recognize some relatively large subclumps in its 
surroundings, most of which by $a=2$ have fallen in. The subclumps and the more
diffusely distributed surrounding matter do not seem to display a pronounced 
spatial pattern and they do not appear to be organized in weblike filamentary 
extensions. Even though to some extent this may be a consequence of the limited
resolution of our simulation, it undoubtedly pertains also to the substantially
lower dynamical (tidal) influence of the clump over its surroundings \citep[see
e.g.][]{bondweb1996}. After $a\approx 2-3$, the infall of matter proceeds 
mostly through quiescent accretion, resulting in a gradual contraction of the 
object  into a moderately elongated ellipsoidal halo. 

Even less outstanding is the evolution of the low-mass bound object (bottom 
row). It hardly shows any substructure and seems to consist only of a central 
region and a few particles within the binding radius. Between an expansion 
factor of $a=2$ and $a=100$ the changes in appearance are only marginal in 
comparison to those seen in the more massive bound structures. 

\section{The spatial distribution of bound objects and superclusters}
\label{sec:spatial}
\begin{figure*}
\vskip 0.0truecm
\includegraphics[width=0.33\textwidth]{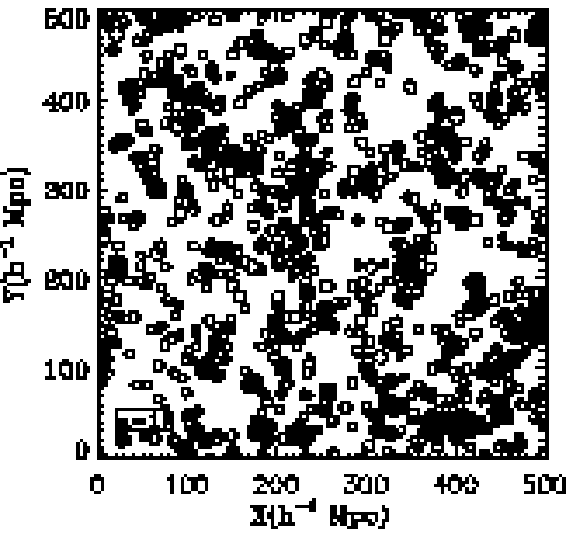}
\includegraphics[width=0.33\textwidth]{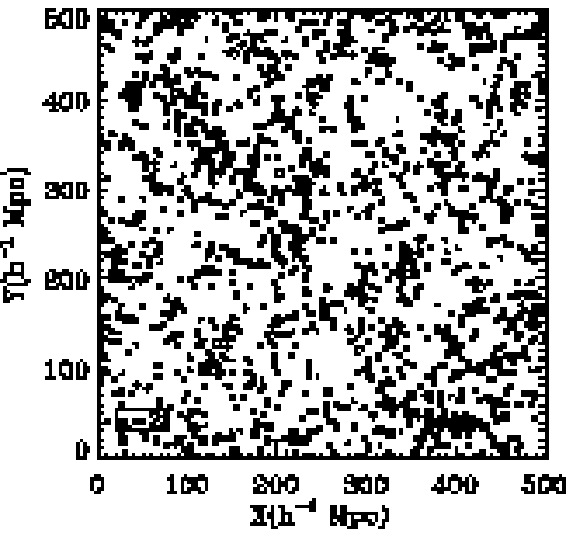}
\includegraphics[width=0.33\textwidth]{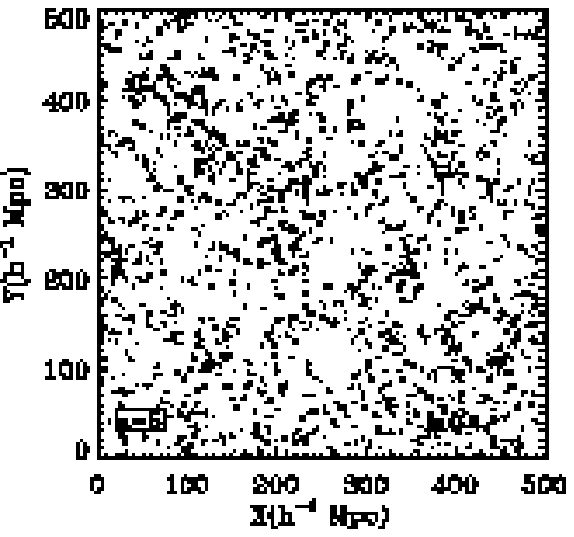}
\includegraphics[width=0.33\textwidth]{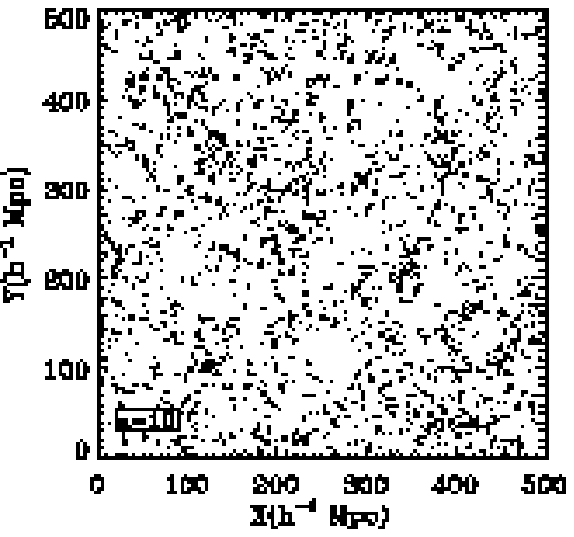}
\includegraphics[width=0.33\textwidth]{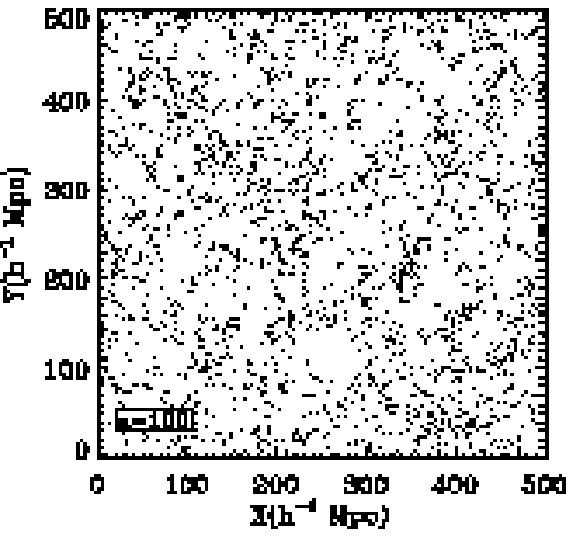}
\vskip -0.25truecm
\caption{\small{Spatial distribution of bound structures at five different 
expansion factors: $a=1$ (top left panel), $a=2$ (top center panel), $a=5$ (top
right panel), $a=10$ (bottom left panel) and $a=100$ (bottom right panel). The 
panels shows the distribution in comoving coordinates in a 100$h^{1-}$Mpc thick
slice projected along the $z$-axis. The bound structures are presented as 
circles with a size equal to their comoving radius (note that the comoving size
of the bound objects shrinks as the Universe evolves into the future). Amongst 
the bound objects, the dark circles indicate the superclusters 
(M$\geq 10^{15}h^{-1}$M$_{\odot}$).}}
\label{fig:bound_struct_dist}
\end{figure*}

The evolving spatial distribution of the bound objects in our simulation is 
shown in Fig.~\ref{fig:bound_struct_dist}. From this direct visual inspection, 
we see that the spatial distribution of the bound structures hardly changes 
between $a=1$ and $a=100$.

Immediately striking is the fact that the bound object or supercluster 
``filling factor'' $\bf{f}$ dramatically declines as time proceeds. Defining 
the filling factor as the fractional volume occupied by bound objects and 
superclusters,
\begin{equation}
\bf{f}\,\equiv\,\sum_j{V_j}/V_{box}\,,
\end{equation}
we may appreciate the strength of the effect by inspecting 
Table~\ref{tab:tabla_filling}. At the onset, ie. the present time, the bound 
objects and superclusters take up a sizeable fraction of the cosmic volume: 
$\sim 3.3\%$ for the bound objects and $\sim 1.2\%$ in the case of the 
superclusters. Even while towards later times the number of bound objects with 
viralized cores is increasing (unlike that for the superclusters), the volume 
fraction is rapidly declining. After $a=10$ they occupy a negligible fraction 
of the Universe, superclusters only a fractional volume in the order of 
$\sim 10^{-8}$. The tremendous exponential expansion of the Universe as a 
result of the cosmological constant clearly renders each bound cluster and 
supercluster an ever more isolated and lonely island in the Universe ! 

The other important observation is that the spatial distribution of bound 
clumps hardly changes from $a=1$ to $a=100$. The spatial patterns visible in 
the object distribution at $a=1$ are very similar to the ones at a much later 
epoch. Clustering does not seem to get any more pronounced after the present 
epoch $a=1$. 

\begin{figure*}
\vskip -7.0truecm
\mbox{\hskip -0.0truecm\includegraphics[width=18.5truecm]{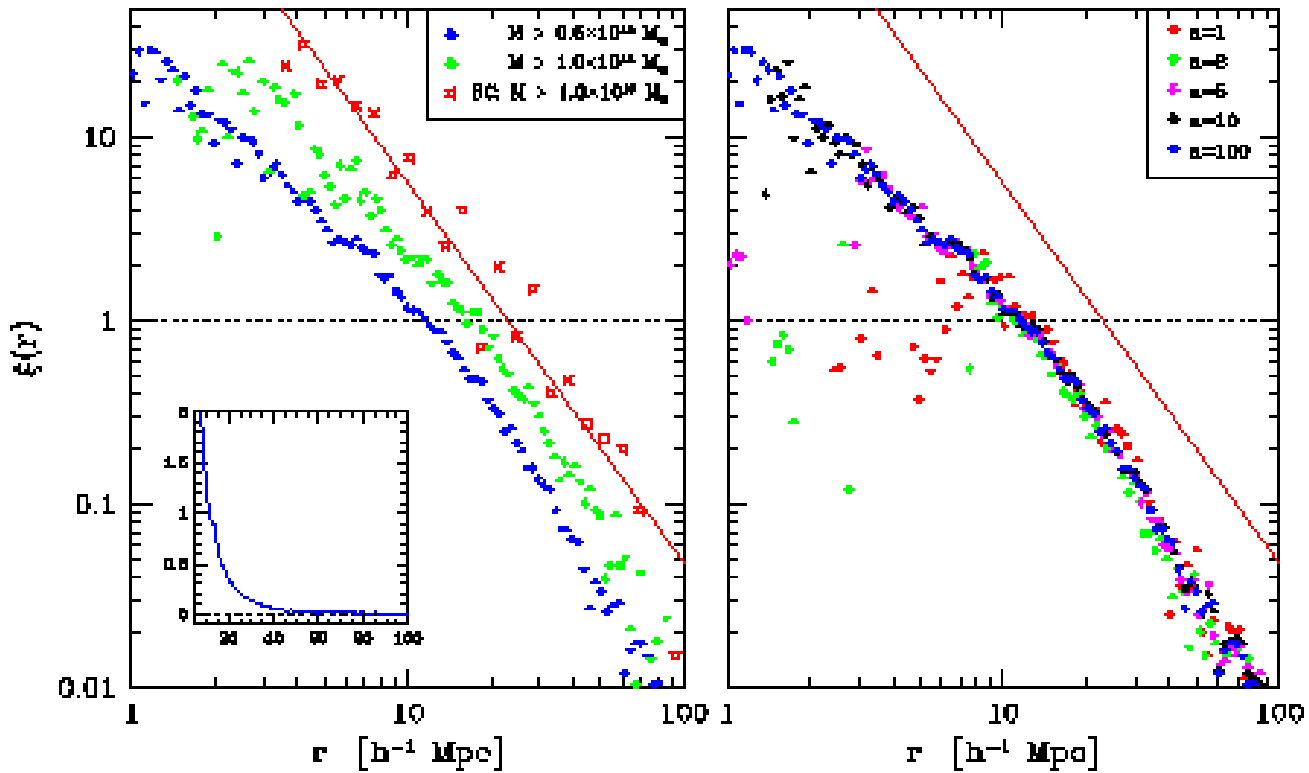}}
\vskip -1.25truecm
\caption{Two-point correlation function $\xi(r)$ of bound objects and of 
superclusters. The correlation function is shown in a log-log plot. Left: 
Two-point correlation function at $a=100$ for three different samples: all 
bound objects in simulation sample (blue dots), bound objects with mass 
M$>10^{14}$M$_{\odot}$ (green dots), superclusters (red stars). The red line is 
a power-law fit to $\xi(r)$ of the superclusters. The insert is the lin-lin 
plot of $\xi(r)$ for the whole object sample. Right: the evolution of $\xi(r)$.
The two-point correlation function of the whole objects sample at five 
different epochs, $a=1$, 2, 5, 10 and 100. The corresponding colours are 
indicated in the legenda the top-right corner of the frame. For guidance the 
red line shows the power-law fit to the supercluster $\xi(r)$ at $a=100$.} 
\label{fig:correlation}
\end{figure*}

\subsection{Correlation Analysis}
To quantify the visual impression of the spatial distribution of bound objects 
and superclusters in our sample, we have determined their spatial two-point 
correlation function $\xi(r)$ \citep[see e.g.][]{peebles80} at the different 
epochs. 

The left frame of Fig.~\ref{fig:correlation} presents the two-point correlation
function $\xi(r)$ at $a=100$ for three different samples: the complete 
simulation sample of bound objects, the subset of bound objects with mass 
M$>10^{14}$M$_{\odot}$ and the superclusters. We assess $\xi(r)$ in terms of 
comoving distances, in physical coordinates one should evidently take into 
account the dilution factor due to the expansion of the Universe. In our 
correlation analysis we take $a=100$ as reference point, as we will see that 
the situation at other epochs is more or less comparable. In all three cases we
observe the strong clustering of the populations. Over a large range, almost 
extending out to a distance of $r \approx 100h^{-1}$Mpc, the bound objects and 
superclusters do show a significant level of clustering. This can be inferred 
from the inspection of the lin-lin plot of $\xi(r)$ in the insert in 
Fig.~\ref{fig:correlation} (lefthand frame). 

\begin{table}
 \caption{\emph{Filling factor of bound objects (left) and superclusters 
(right) at 5 expansion factors: a=1, a=2, a=5, a=10 and a=100. }}
 \begin{minipage}{0.95\linewidth}
   \centering
   \begin {tabular}{|l||c|c|}
     \hline
     & & \\
     & Bound Structures & Superclusters \\
     & & \\
     \hline
     \hline
     & & \\
     $a=1$   & 3.34$\times10^{-2}$ & 1.24$\times10^{-2}$ \\
     $a=2$   & 4.34$\times10^{-3}$ & 1.37$\times10^{-3}$ \\
     $a=5$   & 3.31$\times10^{-4}$ & 8.64$\times10^{-5}$ \\
     $a=10$  & 4.24$\times10^{-5}$ & 1.01$\times10^{-5}$ \\
     $a=100$ & 4.14$\times10^{-8}$ & 0.91$\times10^{-8}$ \\
     & & \\
     \hline
   \end {tabular}
 \end{minipage}
 \label{tab:tabla_filling}
\end {table}

To first approximation, $\xi(r)$ behaves like a power-law function, 
\begin{equation}
\xi(r)\,=\,\left({\displaystyle r_0 \over \displaystyle r}\right)^{\gamma}\,. 
\end{equation}
This is particularly evident for the sample of superclusters, which at $a=100$ 
has a correlation length $r_0=23 \pm 5h^{-1}$Mpc and a slope 
$\gamma\approx 2.1\pm 0.1$. Within the error bounds the numbers at the other 
timesteps are equivalent. While the overall 
behaviour of the correlation function of all bound objects, or that of the ones
with M$>10^{14}$M$_{\odot}$, does resemble that of a power-law we also find 
marked deviations. At the small distances where the clustering is strongest, 
$r < 5h^{-1}$Mpc, we find that $\xi(r)$ has a distinctly lower slope, 
$\gamma \approx 1.4-1.5$. In the large-scale regime, i.e. beyond the clustering
length $r_0 \approx 11.5h^{-1}$Mpc, the slope steepens considerably and attains
a value close to $\gamma \approx 2.0$ at $a=100$. 

We find that more massive objects are more strongly clustered, confirming the 
impression obtained from Fig.~\ref{fig:bound_struct_dist} (see e.g. top-left 
frame). Comparison between the clustering of the whole sample with that of the 
subsample of objects with mass M$>10^{14}$M$_{\odot}$,  and in particular that 
of superclusters, shows that the more massive samples have a stronger 
correlation function over the whole range of distances. In addition to this, we
may also observe that the clustering of superclusters seems to extend out to 
larger distances than that of all objects. That is, for superclusters 
$\xi(r)>0$ out to distances larger than $r \approx 100h^{-1}$Mpc. 

This clustering scaling behaviour is a telling illustration of the (clustering)
bias of more massive objects with respect to the global average, and has to 
some extent be related to the properties of the peaks in the Gaussian 
primordial density field \citep{kaiser1984}. Our findings relate directly to 
the known clustering behaviour of clusters. Since \cite{bahcallson83} found 
that the clustering strength of clusters is an increasing function of the 
cluster richness, a large variety of studies have uncovered the systematics of 
this clustering bias \citep[see e.g.][]{szalayschramm1985,peacockwest1992,
einasto2002}. In particular enticing are the recent results obtained for 
cluster samples extracted from the SDSS survey \citep{bahcall2003,estrada2009}. 

In the righthand frame of Fig.~\ref{fig:correlation} we assess the evolution of
the two-point correlation function of the complete sample of bound objects, 
from $a=1$ to $a=100$. With the exception of some evolution at short distances,
$r<5h^{-1}$Mpc, there is not any significant evolution of the two-point 
correlation function over the full range of distances. This is a manifest 
illustration of the stagnation of structure evolution in an accelerating 
Universe on scales larger than several Megaparsec. The situation appears to be 
exactly the same for the stronger correlation function of superclusters: from 
$a=1$ to $a=100$ superclusters keep the same level of clustering. Of course 
this is a clear manifestation of the expected end of structure growth after the
Universe went into acceleration at $z\approx 0.7$. Note that for reasons of 
clarity, in the righthand frame of Fig.~\ref{fig:correlation} we restrict 
ourselves to merely plotting the power-law fit to the supercluster correlation 
function. 

\section{Global Evolution: Mass Functions}
\label{sec:ps_mf}
With the identification of superclusters we may ask how much mass is contained 
in them at the present epoch and at $a=100$. At $a=1$, the entire sample of 
bound objects in our simulation box amounts to  
$2.73\times10^{18}h^{-1}$M$_{\odot}$. At $a=100$ this has grown to a total mass 
of $2.83\times10^{18}h^{-1}$M$_{\odot}$. This represents $26\%$, respectively 
$27\%$, of the total mass in our simulated Universe 
($1.04\times10^{19}h^{-1}$M$_{\odot}$). 

We may make two immediate observations. First, given that bound objects 
tend to lose around $28\%$ of their mass towards $a=100$ (see below), it 
must mean that the number of bound objects fulfilling our criterion - 
of containing one or more virialized cores - is still growing 
from $a=1$ towards $a=100$. It may also indicate a problem in applying 
a purely spherical density criterion: at $a=1$ mass concentrations 
are more aspherical and inhomogeneous, while at $a=100$ they are 
nearly spherical concentrated mass concentrations (see sec.~\ref{sec:shapes}). 
Perhaps even more tantalizing is the fact that apparently more than $70\%$ of 
mass in the Universe will remain outside of the supercluster islands and will 
keep on floating as a lonely population of low-mass objects in a vast cosmic 
void. 

Perhaps the most outstanding repercussion of the slowdown of large-scale 
structure formation in hierarchical cosmological scenarios is the fact that the
condensaton of new objects out of the density field will gradually come to a 
halt. This should be reflected in the mass spectrum of the objects that were 
just on the verge of formation around the time of the cosmological transition. 
Here we investigate the mass distribution of superclusters, i.e. bound but not 
yet fully virialized structures. For comparison we also investigate a sample of
virialized halos.

\subsection{Theoretical mass functions}
\begin{figure}
\vskip -0.25truecm
\mbox{\hskip -0.0truecm\includegraphics[width=0.45\textwidth]{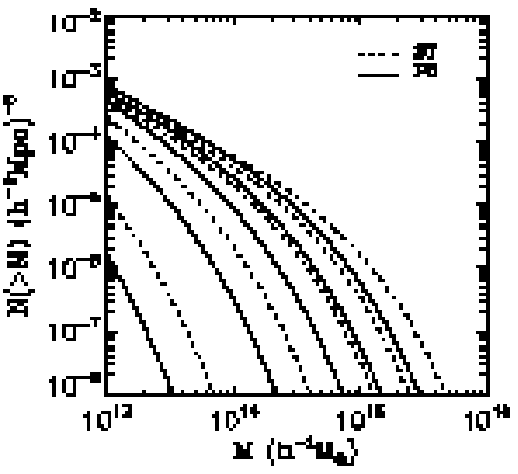}}
\vskip -0.25truecm
\mbox{\hskip -0.0truecm\includegraphics[width=0.45\textwidth]{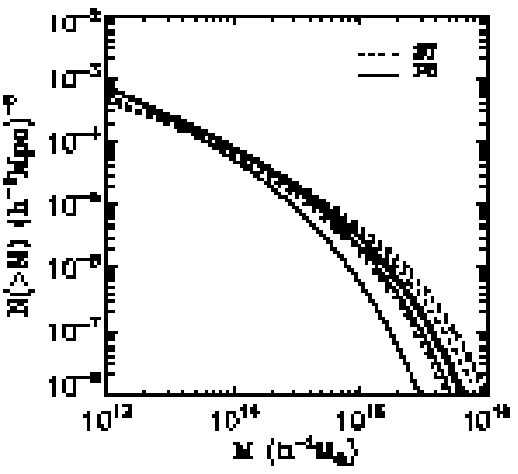}}
\caption{Theoretical mass functions for virialized objects in 
a $\Lambda$CDM Universe ($\Omega_{m,0}=0.3,\Omega_{\Lambda,0}=0.7,h=0.7$). 
Shown are the Press-Schechter mass functions (dashed lines) and the 
Sheth-Tormen mass functions (solid lines). 
Top: the mass functions at redshifts $z=4,2,1,0.5$ and $0$ (from left to 
right). Bottom: the mass functions at expansion factors $a=1,2,4,10$ and $100$ 
(from left to right).} 
\label{fig:massfuncps}
\end{figure}

Fig. \ref{fig:massfuncps} shows the evolution of the mass function of virialized
objects predicted by the Press-Schechter formalism \citep{press74} and 
the Sheth-Tormen excursion set prescription \citep{sheth99}, for the current 
cosmology. The Sheth-Tormen expression takes into account the anisotropic 
collapse of dark halos. We also compare our mass functions to the heuristic 
simulation-based mass function suggested by \cite{jenkins01}. We refer to 
appendix~\ref{app:app_ps} for a listing of the expressions for these mass 
functions. 

The top panel shows the strong evolution in the past, by depicting the mass 
functions at $z=4,2,1,0.5$ and $0$. It shows that structure grows in mass and 
number while the Universe expands. The future evolution is a lot less strong, 
as evidenced by the mass functions at expansion factors $a=1,2,4,10$ and $100$ 
in the bottom panel. After $a=1$ the number of low mass objects does not change
substantially, while after $a=4$ evolution comes to a complete halt. This may 
be best appreciated from the fact that the curves for $a=10$ and $a=100$ 
overlap completely. 

We see that the Sheth-Tormen approximation predicts a higher number of massive 
clusters than the Press-Schechter formalism. As anisotropic collapse speeds up 
the contraction along the minor axis of an object, there is a higher number of 
regions reaching a sufficiently large overdensity before dark energy prevents 
any further evolution. Implicitly, this lowers the number of low-mass objects 
as more get absorbed into the high mass superclusters. 

\begin{figure*}
\includegraphics[width=1.\textwidth]{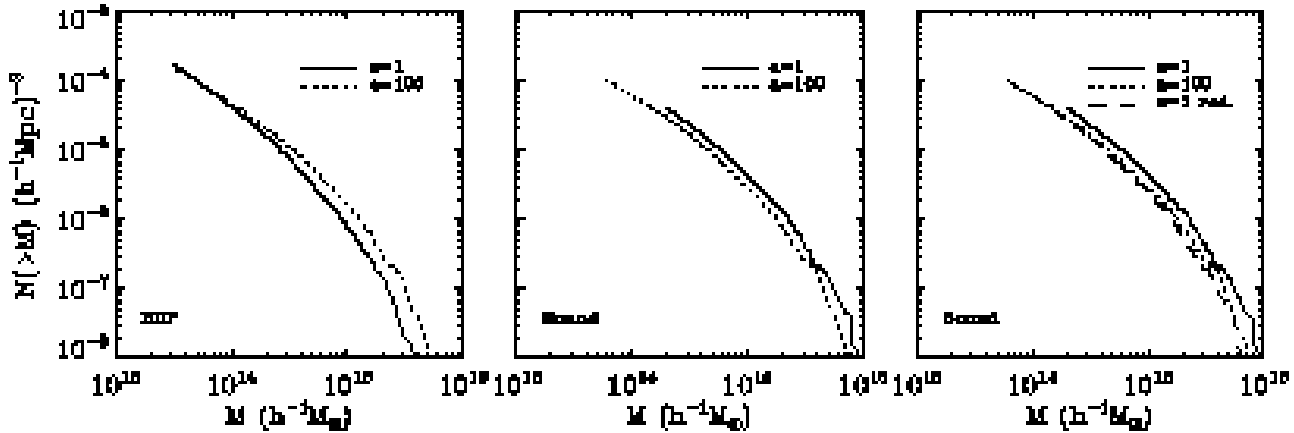}
\caption{Mass functions of virialized halos and of bound objects. 
Left: integrated mass function N($>$M) for virialized HOP objects in the 
simulation, at $a=1$ (solid line) and $a=100$ (dotted line). 
Centre: integrated mass function N($>$M) for bound objects in the simulation, 
at $a=1$ (solid line) and $a=100$ (dotted line). Right: the integrated 
mass function N($>$M) for bound objects compared with the {\it reduced} 
$a=1$ mass function (dashed line). See text for explanation.}
\label{fig:hop_bound_a1_comp}
\end{figure*}
\begin{figure*}
\centering
\includegraphics[width=1.\textwidth]{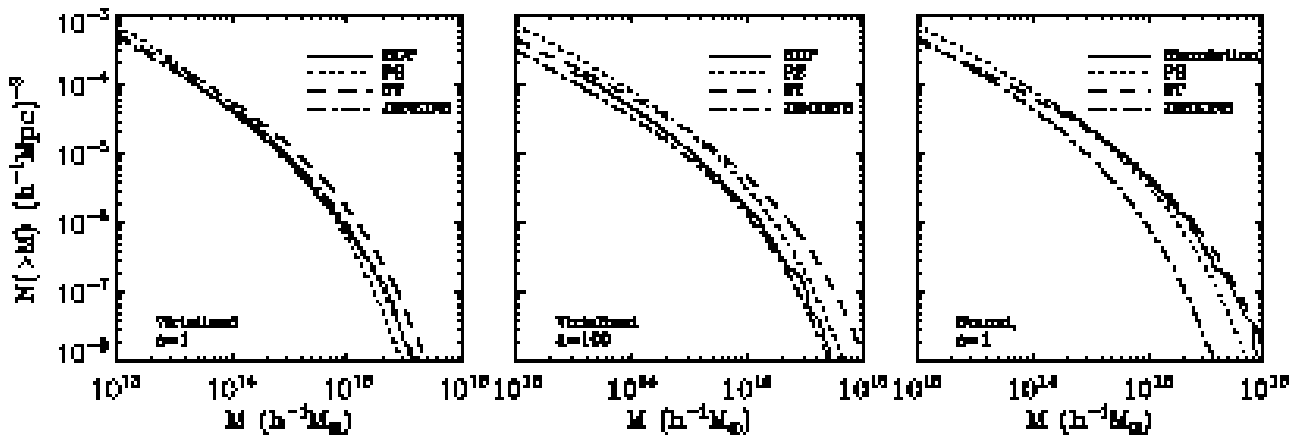}
\caption{Mass functions of virialized halos and of bound objects compared 
with three theoretical mass functions. These are the Press-Schechter 
mass function (dotted line), the Sheth-Tormen function (dashed line) and 
the Jenkins function (dot-dashed line). For the virialized HOP halos the 
critical overdensity $\delta_c$ for the PS and ST functions is the 
one for collapse, for the bound objects the value $\delta_b$ for assuring 
a bound object (D\"unner et al. 2006, this study). Left panel: 
the $a=1$ integrated mass function N($>$M) of HOP halos. Centre: the 
$a=100$ integrated mass function N($>$M) of HOP halos. Righthand: 
the $a=100$ integrated mass function N($>$M) of bound objects.} 
\label{fig:comparisontheo}
\end{figure*}

\subsection{Simulation mass functions}
\label{sec:scmassfunc}
We assess the mass function of the virialized halos, the objects identified by 
HOP on the basis of the prescription in sec.~\ref{sec:virhalo}, and that of 
the bound (supercluster) objects that were identified according to the 
criterion specified in section~\ref{sec:boundid}. Given that the we do not 
expect a radical change in mass functions between the present epoch and 
$a=100$, we restrict the comparison to the simulation mass functions at $a=1$ 
and $a=100$. 

Fig. \ref{fig:hop_bound_a1_comp} shows the mass functions of the virialized 
halos found by HOP (left panel) and of the bound objects (central panel), at 
$a=1$ and $a=100$. As expected, the number of massive virialized halos 
increases as we go from $a=1$ to $a=100$. The increase is only minor, yet 
significant, and a manifestation of the freezing of structure formation and 
hence that of the corresponding mass functions \citep[see also][]{nagamine03}. 
The mass functions at $a=1$ and $a=100$ also reflect the continuing 
hierarchical evolution within the realm of the bound supercluster regions. 
There is a definitive increase in the number of the most massive clumps, going 
along with a decrease at the low-mass side of the mass function. 

When turning towards the mass function for the bound (supercluster) regions 
(central panel) we find the same mass function at $a=1$ and $a=100$, except 
for a slight decrease in the mass of the objects over the whole mass range. 
There is a loss of mass, amounting to some $28 \pm 13\%$ of the mass enclosed 
within the critical radius of the superclusters at $a=1$ \cite{dunner06}. This 
is substantially more than the mere $1\%$ mass gain as a result of accretion of
mass in between $a=1$ and $a=100$. The loss of mass has to be ascribed to the 
virialization process of and within the bound object. A major factor in this is
the abundant substructure in the supercluster at the present epoch as opposed 
to the smoothened mass distribution within the ultimate supercluster island at 
$a=100$. In order to correct for this ``reduced'' mass we renormalize the 
supercluster mass function into a {\it reduced} supercluster mass function, 
simply by multiplying the masses by a factor $0.72$. As may be observed in the 
righthand panel of Fig.~\ref{fig:hop_bound_a1_comp} we find an almost perfect 
overlap between the mass function at $a=100$ and its reduced equivalent at 
$a=1$. 

\subsection{Comparison of simulated and theoretical mass functions}
\label{sec:theormass}
Fig. \ref{fig:comparisontheo} shows the cumulative mass function of the 
virialized objects found by HOP at $a=1$ and at $a=100$, and compares them 
with three theoretical mass functions (see app.~\ref{app:app_ps}). 

\subsubsection{Mass functions of bound objects and superclusters}
To compare the mass function of the bound (supercluster) regions with the 
theoretical Press-Schechter and Sheth-Tormen functions we need to specify a 
critical overdensity corresponding to the bound (supercluster) regions in our 
sample (see app.~\ref{app:app_ps}). Because in this study these are assembled 
on the presumption that they are marginally bound, we use the corresponding 
value of the linear extrapolated density excess, $\delta_{b}=1.17$, derived in 
section~\ref{sec:deltab} (see Eqn.~\ref{eq:deltabnd0}). 

At $a=1$, the Sheth-Tormen function seems to provide a better fit than the 
Press-Schechter function to the bound object mass spectrum, in particular for 
the tail of massive superclusters. It demonstrates the importance of 
morphological and tidal influences on the mass spectrum of these generically 
nonspherical objects (see next section~\ref{sec:shapes}). At $a=100$ the 
Press-Schechter function provides a substantially better fit to the bound 
object mass function. This may be related to the fact that the PS formalism 
implicitly assumes pure spherical collapse, which we will see in 
sect.~\ref{sec:shapes} agrees quite well with the shape of superclusters at 
$a=100$. Also, we find that in the far future, when most of our superclusters 
have become isolated and largely virialized islands, it is appropriate to 
compare with a Press-Schechter or Sheth-Tormen mass function based on another 
critical density value. In between $a=1$ and $a=100$ most {\it cosmic islands} 
will reside in a dynamical phase somewhere between marginally bound and full 
collapse, implying a critical value in between $\delta_b=1.17$  
(Eqn.~\ref{eq:deltabnd0}) and the collapse threshold $\delta_c=1.675$ 
(Eqn.~\ref{eq:deltacrit}). 

It is also clear that the Jenkins function does not provide a suitable fit to 
the supercluster mass function. This may not be surprising given the fact that 
it is a numerical approximation of the mass function of collapsed and 
virialized halos in N-body simulations and as such does not explicitly include 
an adjustable density threshold $\delta_c$. 

\subsubsection{Cluster mass function}
It is more straightforward to compare the mass function of the clusters, or HOP
halos, in our simulation with that of the three theoretical mass functions for 
virialized objects. For these fits we use the critical collapse density value 
$\delta_c=1.675$, the value for our cosmology according to 
Eqn.~\ref{eq:deltacrit}. The Jenkins approximation is of course independent of 
the value of $\delta_{c}$.

At $a=1$ the Jenkins mass function \citep{jenkins01} is the one that fits best 
(dot-dashed line), perhaps not entirely surprising given its N-body simulation 
background. The Press-Schechter mass function represents a good fit at the 
lower mass end, although it underestimates the number of high mass clusters. 
\cite{governato99} claim that a critical density value $\delta_{c}=1.775$ would
provide a better fit and indeed it would lead to a small yet significant 
improvement of the Sheth-Tormen mass function. 

\begin{figure*}
\mbox{\includegraphics[width=0.33\textwidth]{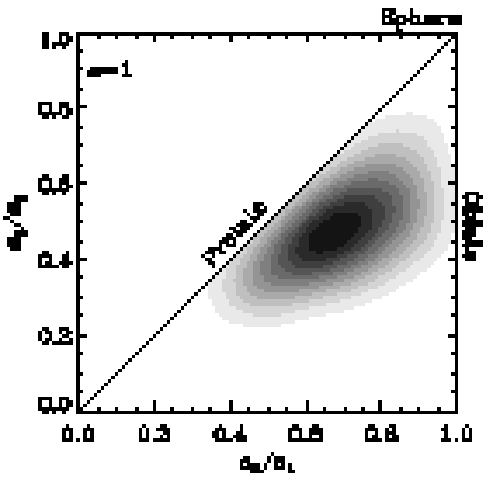}}
\mbox{\includegraphics[width=0.33\textwidth]{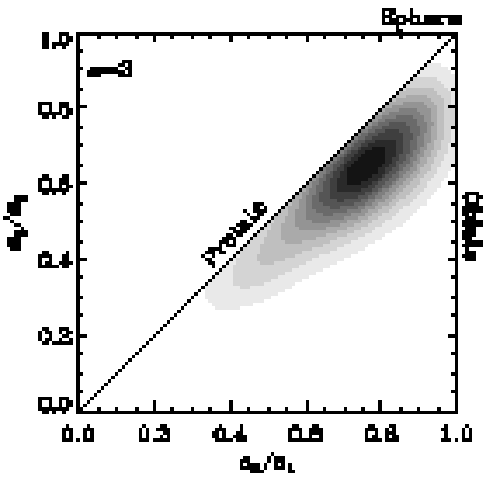}}
\mbox{\includegraphics[width=0.33\textwidth]{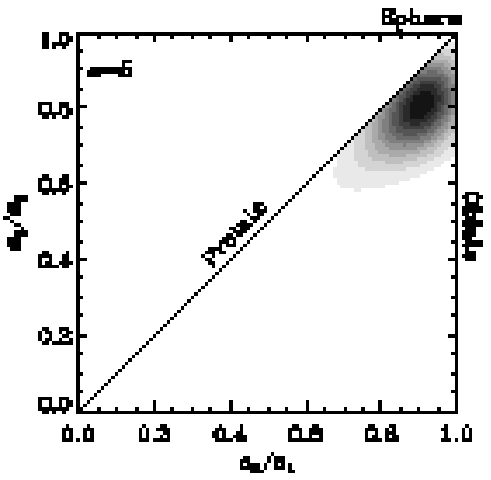}}
\mbox{\includegraphics[width=0.33\textwidth]{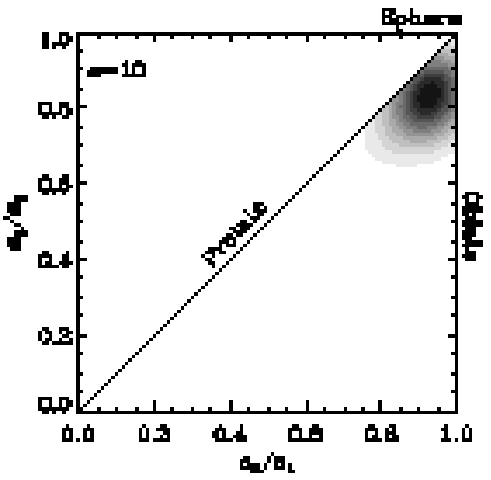}}
\mbox{\includegraphics[width=0.33\textwidth]{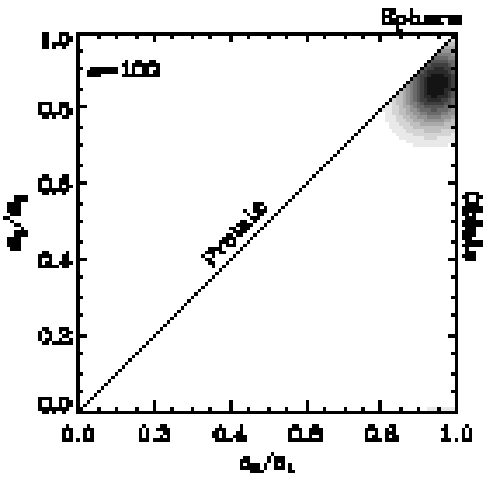}}
\caption{\small{Distribution of axis ratios for bound objects at five different
expansion factors. Top row: $a=1$, 2 and 5; bottom row: $a=10$ and 
$a=100$. Each diagram shows the probability density distribution in the plane 
of $s_{2}/s_{1}$ versus $s_{3}/s_{1}$ values, with principal axes $s_3<s_2<s_1$. 
Plotted is the axis ratio probability distribution in iso-probability 
greyscales values, with percentile steps of $10\%$. Dark colors correspond to 
high probability, decreasing to lower probability as colours fade to light (in 
steps of $5\%$).}}
\label{fig:comp_a1}
\end{figure*}

For $a=100$ we adjust the parameters of the Jenkins function, using the fitting
parameters for $\Omega_{m}=0$ listed in \cite{evrard02}. With these parameters,
it agrees very well with the HOP mass function, although it slightly 
overestimates the number of lower mass objects. However, for Jenkins' original 
parameter values, it would lead to a significant overabundance of objects with 
respect to the ones found in the simulations. 

Neither the pure Press-Schechter nor the Sheth-Tormen function manage to fit 
the mass spectrum at $a=100$ over the entire mass range. Press-Schechter does 
agree at the high mass end while Sheth-Tormen results in a better agreement at 
lower masses. This may be an indication for the more substantial role of 
external tidal forces on the evolution of the low mass halos. Such external 
influences are entirely ignored by the Press-Schechter formalism, while they 
are succesfully modelled by the Sheth-Tormen fits \citep{sheth99}.

\section{Shapes of Bound Structures}
\label{sec:shapes}
While the formation and evolution of structure on large scales comes to a
halt once the Universe starts to accelerate, the internal evolution of
overdense patches continues. One of the most telling manifestations of 
the internal evolution of these collapsing objects is their changing shape. 
The substructures that are within the bound radius will merge with each other 
into an increasingly smooth and concentrated clump that will gradually assume 
a more and more spherical configuration. 

Using a variety of definitions for superclusters, their shape has been studied 
both using real data \citep[e.g.][]{plionis92,sathya1998,basilakos01,sheth2003,
einasto2007c} and in N-Body simulations \citep[e.g.][]{sathya1998,shandarin2004,
basilakos06,wray06,einasto2007c}. Most studies agree that the dominant shape of 
superclusters at the present time is prolate, most evident in the presence of 
elongated filaments. These predominantly anisotropic shapes are a clear 
indication for the quasi-linear dynamical stage at which we find the 
present-day superclusters. 

\subsection{Definitions}
In order to determine the shape, we calculate the inertia tensor using all 
particles inside the spheres defined by Eqn.\ref{eq:ratio} with respect to 
their center of mass:
\begin{equation}
I_{ij}=\sum x_{i}x_{j}m\,.
\end{equation}
Since the matrix is symmetric, it is possible to find a coordinate system such 
that it is diagonal, yielding the eigenvalues $a_{1}$, $a_{2}$ and $a_{3}$. 
These give a quantitative measure of the degree of symmetry of the 
distribution. With major axis $s_{1}$, medium axis $s_{2}$ and minor axis 
$s_{3}$, ie. $s_{3}<s_{2}<s_{1}$, the two axis ratios $s_{2}/s_{1}$ and 
$s_{3}/s_{1}$ are given by 
\begin{equation}
\frac{s_2}{s_1}=\sqrt{\frac{a_{2}}{a_{1}}},\hspace{1cm}\frac{s_3}{s_1}=
\sqrt{\frac{a_{3}}{a_{1}}}\,,
\end{equation}
where $a_{1}>a_{2}>a_{3}$. 

The object is almost spherical if both ratios $s_{2}/s_{1}$ and $s_{3}/s_{1}$ 
are close to one. Oblate objects have axis ratios $s_{3} \ll s_{2} \sim s_{1}$, 
prolate objects $ s_{3} \sim s_{2} \ll s_{1}$. 

\begin{figure*}
\includegraphics[width=0.95\textwidth]{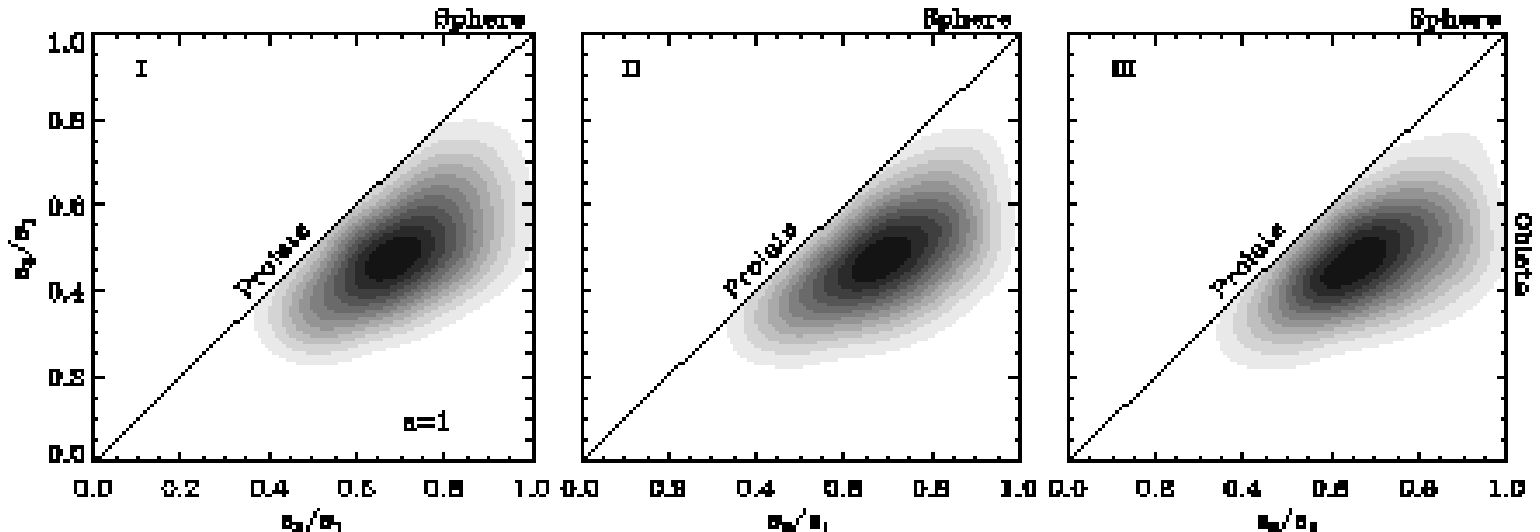}
\vskip 0.5truecm
\includegraphics[width=0.95\textwidth]{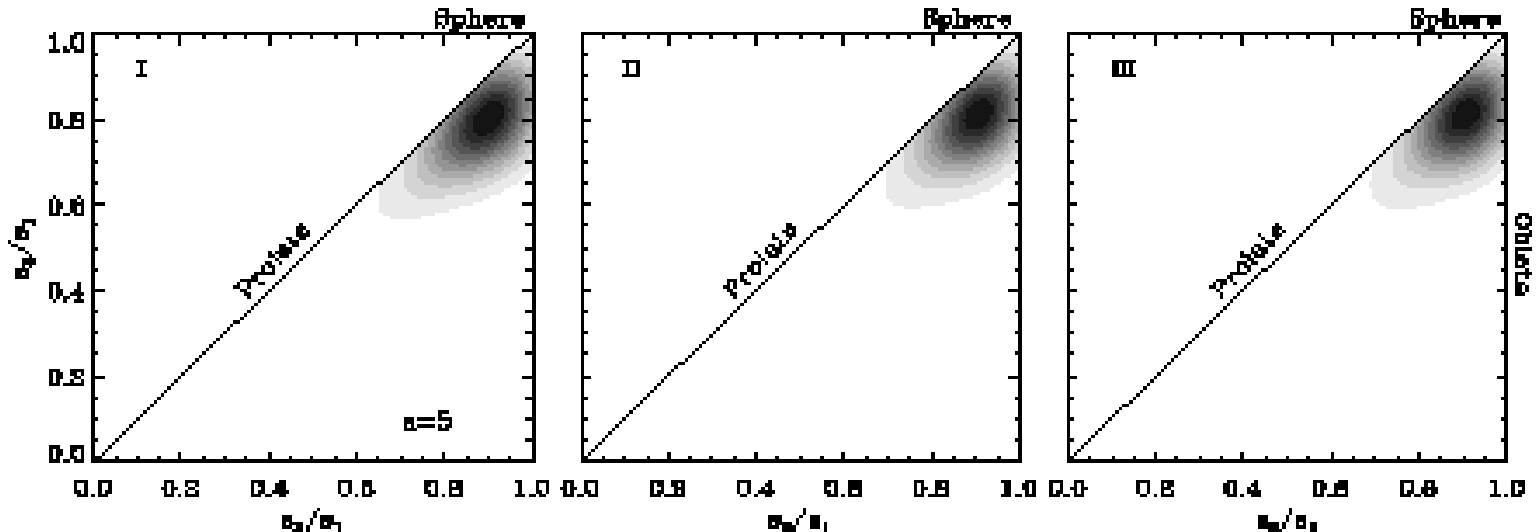}
\vskip 0.5truecm
\includegraphics[width=0.95\textwidth]{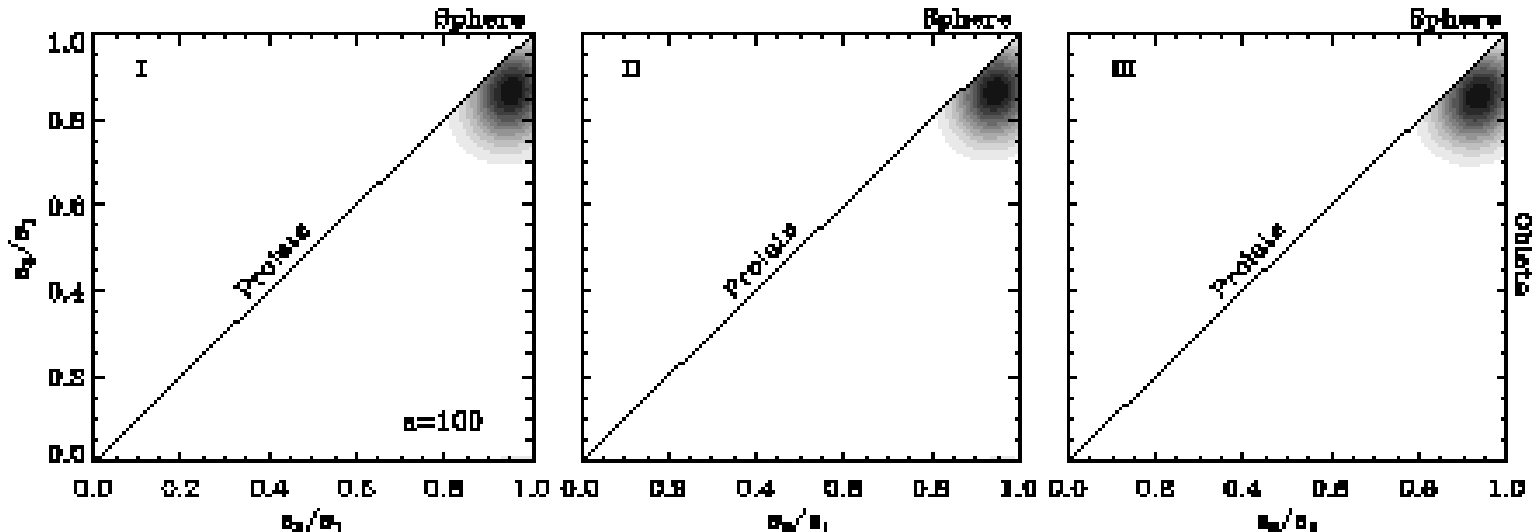}
\vskip 0.5truecm
\caption{Distribution of axis ratios for three mass ranges. 
I: high-mass bound objects with M$>4.8\times10^{14}h^{-1}$M$_{\odot}$. II: 
medium mass range bound objects, 
$2.8\times10^{14}h^{-1}$M$_{\odot}< M\le 4.8\times10^{14}h^{-1}$M$_{\odot}$. 
III: low mass bound objects, 
$2\times10^{14}h^{-1}$M$_{\odot}< M\le 2.8\times10^{14}h^{-1}$M$_{\odot}$. Each 
diagram shows the probability density distribution in the plane of 
$s_{2}/s_{1}$ versus $s_{3}/s_{1}$ values, with principal object axes 
$s_3<s_2<s_1$. Dark colors correspond to high probability, decreasing to lower 
probability as colours fade to light, with level steps corresponding to 
percentile differences of $10\%$. Top row: axis ratio distribution at the 
current epoch, $a=1$. Middle row: axis ratio distribution at $a=5$. Bottom row:
axis ratio distribution at $a=100$.}
\label{fig:dist_axis}
\end{figure*}
\subsection{Shape evolution}
Fig.~\ref{fig:comp_a1} reveals the overall evolving shape distribution in 
from the present epoch until $a=100$. The shape distribution is shown 
in terms of a plot of the axis ratios $s_{2}/s_{1}$ versus $s_{3}/s_{1}$ 
of the objects. The corresponding shape distribution function is shown by means
of greyscale iso-probability contour maps.   Because by definition 
$s_{3}<s_{2}<s_{1}$, only the righthand triangle of the diagram is populated. 
Spherical groups are located at (1,1), oblate groups tend towards the line 
$s_{2}/s_{1}=1$ while prolate groups are found near the diagonal 
$s_{2}/s_{1}=s_{3}/s_{1}$. 

\begin{table}
  \caption{Average values of the axis ratios $s_{2}/s_{1}$ and $s_{3}/s_{1}$, 
with their standard deviation, at five expansion expansion factor: $a=1$, 
$a=2$, $a=5$, $a=10$ and $a=100$.}
  \begin{minipage}{0.95\linewidth}
    \centering
    \begin {tabular}{|l||c|c|c|c|}
      \hline
      & & & &\\
      & $\langle s_{2}/s_{1} \rangle$ & $\langle s_{3}/s_{1} \rangle $ & 
      $\sigma_{s_{2}/s_{1}}$ & $\sigma_{s_{3}/s_{1}}$ \\ 
      & & & &\\
      \hline
      \hline
      & & & & \\
      $a=1$   & 0.69 & 0.48 & 0.13 & 0.11  \\
      $a=2$   & 0.73 & 0.61 & 0.14 & 0.13  \\
      $a=5$   & 0.87 & 0.77 & 0.10 & 0.10  \\
      $a=10$  & 0.90 & 0.81 & 0.08 & 0.08  \\
      $a=100$ & 0.94 & 0.85 & 0.03 & 0.05  \\
      & & & & \\
      \hline
    \end {tabular}
  \end{minipage}
  \label{tab:tabla_axis}
\end {table}

At the first two timesteps, $a=1$ and $a=2$, we find a clear dominance of 
nonspherical objects. Most objects are distinctly anisotropic, which agrees 
with the observations of individual objects in sect.~\ref{sec:evolution}. At 
$a=1$ the objects occupy a wide range of mostly triaxial shapes, with a 
mean value ($\langle s_{2}/s_{1} \rangle$, $\langle s_{3}/s_{1} \rangle $)=
(0.69,0.48) and a standard deviation of ($\sigma_{s_{2}/s_{1}}$,
$\sigma_{s_{3}/s_{1}}$)=(0.13,0.11) (see table~\ref{tab:tabla_axis}). There are 
hardly any thin pancake-shaped structures, as low values of $s_{3}/s_{1}<0.25$ 
seem to absent. To some extent this may be a reflection of our bound group 
identification procedure, given its bias towards spherical configurations (see 
sect.~\ref{sec:join}). Starting at $a=2$ we observe 
a stretching of the shape distribution (Fig.~\ref{fig:comp_a1}), along the 
direction of the prolate configurations, $s_3/s_1=s_2/s_1$, and towards less 
anisotropic and more spherical morphologies. A sizeable fraction of the objects
shows a tendency towards a prolate shape. 

The fact that there are almost no spherical, or even nearly spherical objects, 
at the earlier epochs is hardly surprising given the relative youth of these 
supercluster objects. It is a reflection of the fact that the primordial 
density field does not contain spherical peaks \citep{bbks} and that the first 
stages of gravitational contraction and collapse proceed via strongly flattened
and elongated geometries \citep{zeldovich1970,icke73}. 

\begin{table*}
 \caption{Number of objects and average values of the axis ratios $s_{2}/s_{1}$
and $s_{3}/s_{1}$, with their standard deviation, at $a=1$, $a=5$ and $a=100$ 
for three mass intervals (with mass range set at $a=1$).}
 \begin{minipage}{0.95\linewidth}
\centering
    \begin {tabular}{|c|l|l||c|c|c|c|}
      \hline
       & & & & & &\\
       & & & $\langle s_{2}/s_{1} \rangle$ & $\langle s_{3}/s_{1} \rangle $ & 
      $\sigma_{s_{2}/s_{1}}$ & $\sigma_{s_{3}/s_{1}}$ \\ 
       & & & & & &\\
      \hline
      \hline
       & & & & & & \\
      I & $M>4.8\times10^{14}h^{-1}$M$_{\odot}$ & $a=1$ & 0.70 & 0.49 & 0.13 & 
      0.11  \\
      & & $a=2$  & 0.72 & 0.60 & 0.13 & 0.12  \\
      & & $a=5$  & 0.86 & 0.76 & 0.10 & 0.10  \\
      & & $a=10$ & 0.90 & 0.81 & 0.07 & 0.08  \\
      & & $a=100$& 0.94 & 0.85 & 0.03 & 0.05  \\
       & & & & & & \\
      II &$2.8\times10^{14}h^{-1}$M$_{\odot}< M\le 4.8\times10^{14}h^{-1}$M$_{\odot}$ 
      &   $a=1$ & 0.70 & 0.48 & 0.13 & 0.11  \\
      & & $a=2$ & 0.74 & 0.63 & 0.14 & 0.13  \\
      & & $a=5$ & 0.87 & 0.77 & 0.09 & 0.10  \\
      & & $a=10$ & 0.91 & 0.81 & 0.07 & 0.08  \\
      & & $a=100$ & 0.94 & 0.85 & 0.03 & 0.05  \\
       & & & & & & \\
      III & $2\times10^{14}h^{-1}$M$_{\odot}< M\le 2.8\times10^{14}h^{-1}$M$_{\odot}$
      & $a=1$ & 0.68 & 0.47 & 0.14 & 0.11  \\
      & & $a=2$   & 0.73 & 0.62 & 0.15 & 0.14  \\
      & & $a=5$   & 0.86 & 0.76 & 0.11 & 0.11  \\
      & & $a=10$  & 0.90 & 0.80 & 0.08 & 0.08  \\
      & & $a=100$ & 0.93 & 0.85 & 0.04 & 0.05  \\
      & & & & & & \\
       \hline
    \end {tabular}
    \end{minipage}
\label{tab:tabla}
\end {table*}

After $a=2$ there is a rapid evolution of the vast majority of objects towards 
a nearly spherical shape. At $a=5$ nearly all clumps are found in the corner 
with axis ratio $s_2/s_1=s_3/s_1=1$. Also after $a=5$ the width of the shape 
distribution continues to shrink. Not only do the bound clumps and 
superclusters on average attain a more spherical shape as the expansion of the 
Universe accelerates: it is the entire population of objects that appears to 
follow this trend. This systematic change of shape can be most clearly 
appreciated from inspection of table~\ref{tab:tabla_axis}. 

In all, the contrast between the shape distribution at $a=1$ and the one at 
$a=100$ reveals a manifest and even radical internal evolution of the 
supercluster complexes, at the same time when evolution on larger scales has 
been virtually frozen. 

\subsection{Mass dependence}
One potentially relevant issue concerns the possible dependence of shape 
on the mass of bound structures. In order to investigate this, we divide our 
sample in three mass ranges, all approximately including the same number 
of bound structures. Class I are the most massive third of objects, which at 
$a=1$ have a mass $M>4.8\times10^{14}h^{-1}$M$_{\odot}$. The medium mass 
class II includes objects with masses at $a=1$ of 
$2.8\times10^{14}h^{-1}$M$_{\odot}<M\leq 4.8\times10^{14}h^{-1}$M$_{\odot}$ and 
the low mass class III  are those with masses at $a=1$ of 
$2\times10^{14}h^{-1}$M$_{\odot}<M\leq 2.8\times10^{14}h^{-1}$M$_{\odot}$. 

We found that over the entire interval of $a=1$ to $a=100$ there is hardly any 
distinction between the shape distribution in the different mass ranges (see 
Fig.~\ref{fig:dist_axis}). The three mass ranges have similar mean axis ratios 
at all five expansion factors (see Table~\ref{tab:tabla}).  This must be 
related to the fact that even though of different mass, the objects have been 
selected on the basis of similar (over)density values. The latter is an 
indication for the evolutionary state of the objects. In turn, as we have seen,
this is reflected in their shape. 

We conclude that irrespective of their mass, all objects evolve into single, 
virialized and spherical objects as they grow in complete isolation after the 
Universe assumed an accelerated expansion and after all surrounding clumps and 
substructure within their binding radius has been accreted.

\subsubsection{Cluster vs. Supercluster Shapes}
Keeping in mind that the objects in our sample are bound but perhaps not 
virialized, it is instructive to contrast them to virialized galaxy clusters. 
Although their masses may be comparable, the bound group radii are much larger 
as they have substantially lower densities than clusters. Within this larger 
region there is a considerably more pronounced substructure.  

\begin{figure*}
\mbox{\hskip -0.75truecm\includegraphics[width=0.285\textwidth]{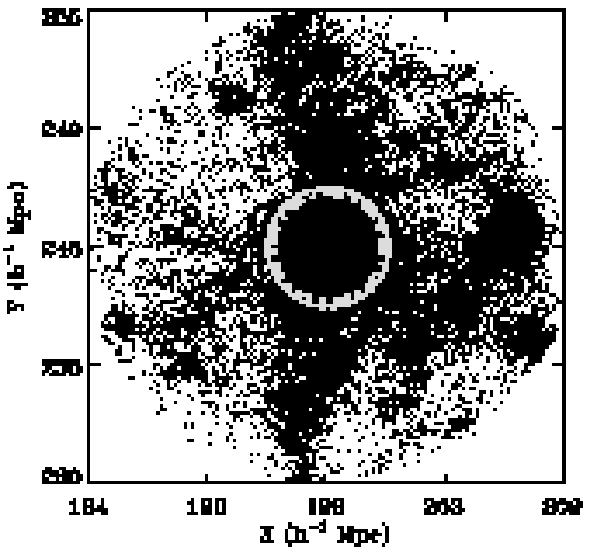}}
\mbox{\hskip -0.45truecm\includegraphics[width=0.285\textwidth]{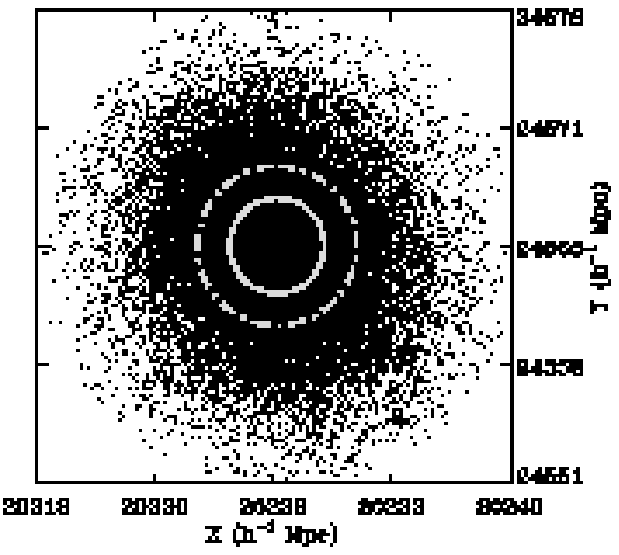}}
\mbox{\hskip -0.1truecm\includegraphics[width=0.49\textwidth]{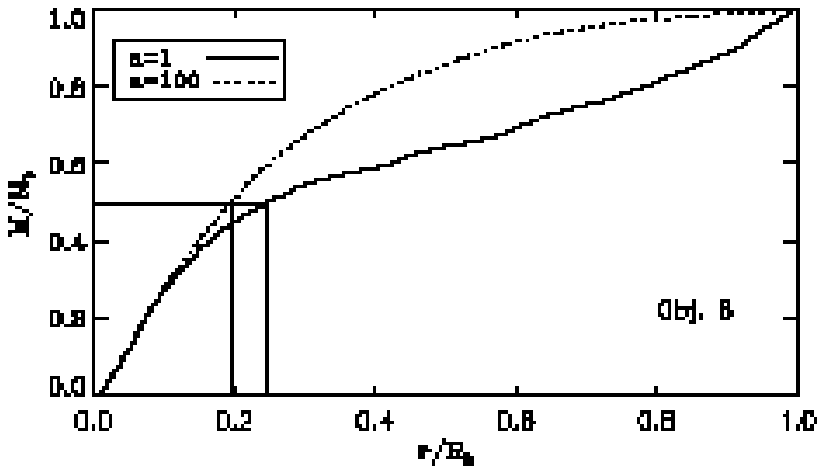}}
\vskip 0.1truecm
\mbox{\hskip -0.75truecm\includegraphics[width=0.285\textwidth]{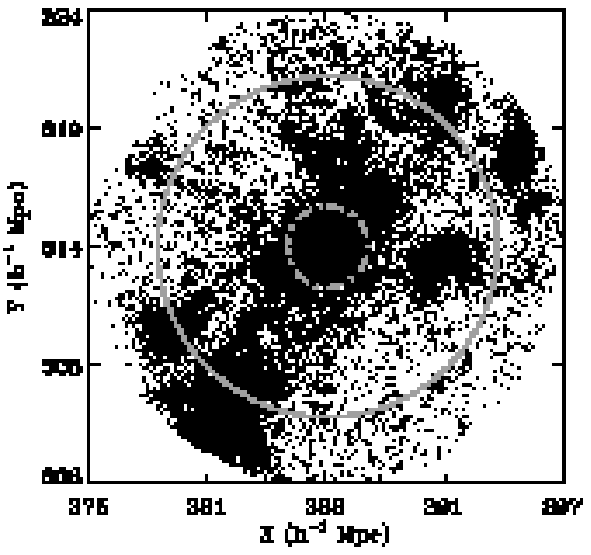}}
\mbox{\hskip -0.45truecm\includegraphics[width=0.285\textwidth]{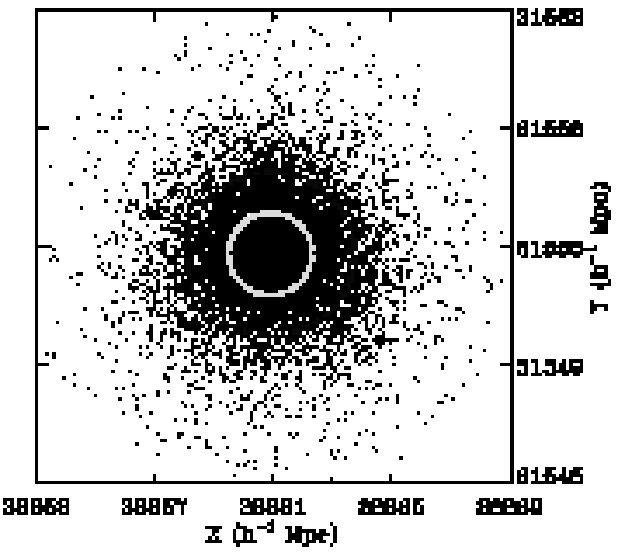}}
\mbox{\hskip -0.1truecm\includegraphics[width=0.49\textwidth]{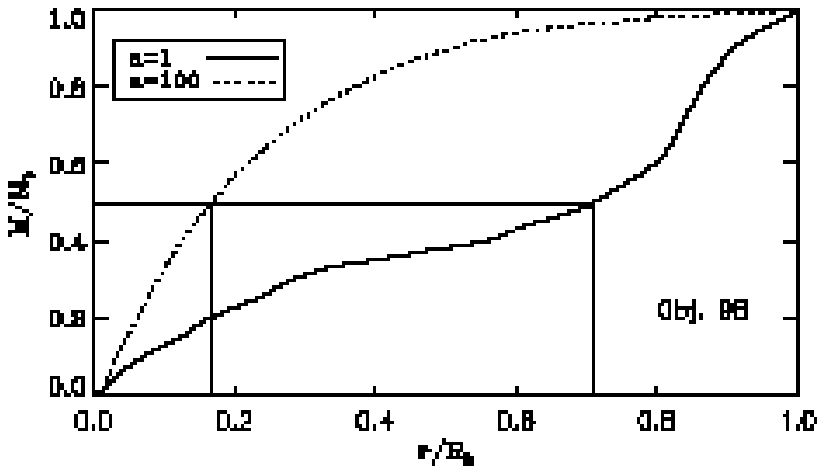}}
\caption{Top: supercluster \# 8 ($M\sim5.4\times10^{15}h^{-1}$M$_{\odot}$). 
Bottom: supercluster \# 98 ($M\sim3.6\times10^{15}h^{-1}$M$_{\odot}$). Dark 
matter/particle distribution at $a=1$ (left) and $a=100$ (centre) of object 
\# 8, in physical coordinates. For comparison the half mass radius (solid 
circle) and the virial radius (dashed-dot circle) are superimposed on the 
matter distribution. Righthand panel: Cumulative mass distribution of the 
object as a function of its (normalized) radius $r/r_b$, at $a=1$ (solid line) 
and at $a=100$ (dotted line). The cumulative mass M/M$_b$ is normalized with 
respect to the corresponding final (bound) mass M$_b$. The solid vertical lines
indicate the value of the half-mass radius (for which M/M$_b$=0.5, the 
horizontal solid line), at both epochs $a=1$ and $a=100$.}
\label{fig:clus8_98}
\caption{\small{Radial density profiles of Object 8 (top panel) and Object 98 
(bottom panel). In a log-log diagram each of the panels shows the radial density
profile $\rho(r)$, in units of the critical density $\rho_{\rm crit}$, as a 
function of the normalized radius $r/r_{vir}$ (normalized wrt. the supercluster 
virial radius). Solid line: profile at $a=1$. Dotted line: profile at $a=100$. 
For comparison in both panels we include a short line with an isothermal slope 
$-2$.}}
\vskip -0.5truecm
\label{fig:densprof}
\end{figure*}

At $a=1$, the clusters in our simulation are triaxial, with average axis ratios 
($\langle s_{2}/s_{1} \rangle$,$\langle s_{3}/s_{1} \rangle$)=($0.83$,$0.71$)
and standard deviation ($\sigma_{s_{2}/s_{1}}$,$\sigma_{s_{3}/s_{1}}$)=
($0.09$,$0.09$). These values are somewhat more pronounced than those quoted in
other studies \citep[e.g.,][]{dubinski91, katz91, haarlem93, jing02, kasun05, 
paz06, allgood06}. All agree that they tend to be more prolate as the halo mass 
increases. \cite{dubinski91} found that halos are ``strongly triaxial and very 
flat'', with mean axis ratios of $\langle s_{2}/s_{1} \rangle$=0.71 and 
$\langle s_{3}/s_{1} \rangle$=0.50. For simulations of isolated halos with 
different power spectra indices \cite{katz91} found values ranging from 
$s_{2}/s_{1}\approx 0.84-0.93$ and $s_{3}/s_{1}\approx 0.43-0.71$, while for 
massive clusters \cite{kasun05} found peak values of 
$(s_{2}/s_{1},s_{3}/s_{1})=(0.76,0.64)$.

Even though the radii of the bound structures in our sample are considerably 
larger than the virial radii of clusters, they do have similar triaxial shapes. 
Nonetheless, we do notice a distinct tendency of the superclusters to be more 
anisotropic than that of the virialized clusters. This must be a reflection of 
their different dynamical states.

\section{Internal Mass Distribution and Density Profiles}
\label{sec:dens_prof}
We have seen that superclusters are gradually decoupling from the global 
cosmic expansion, and towards the far future evolve in isolation as 
genuine {\it cosmic islands}. As these contract and collapse, and finally 
even virialize, they will develop into a much more compact object. 
To understand their internal evolution it is important to look into 
the development of their internal mass distribution, and hence 
their density profile. 

\subsection{Supercluster Case Studies}
Before trying to draw some general conclusions we look in more 
detail at the evolving mass distribution inside and around two 
representative individual superclusters in our sample. These are objects 
\# 8 and \# 98. The first one is a massive supercluster with a mass of 
$\sim5.4\times10^{15}h^{-1}$M$_{\odot}$, while the second one 
has a mass of $\sim3.6\times10^{15}h^{-1}$M$_{\odot}$. Supercluster 
\# 8 is one of the most massive objects identified at $a=1$ and 
also ends up as such at $a=100$. 

\begin{figure}
\centering
\includegraphics[width=0.49\textwidth]{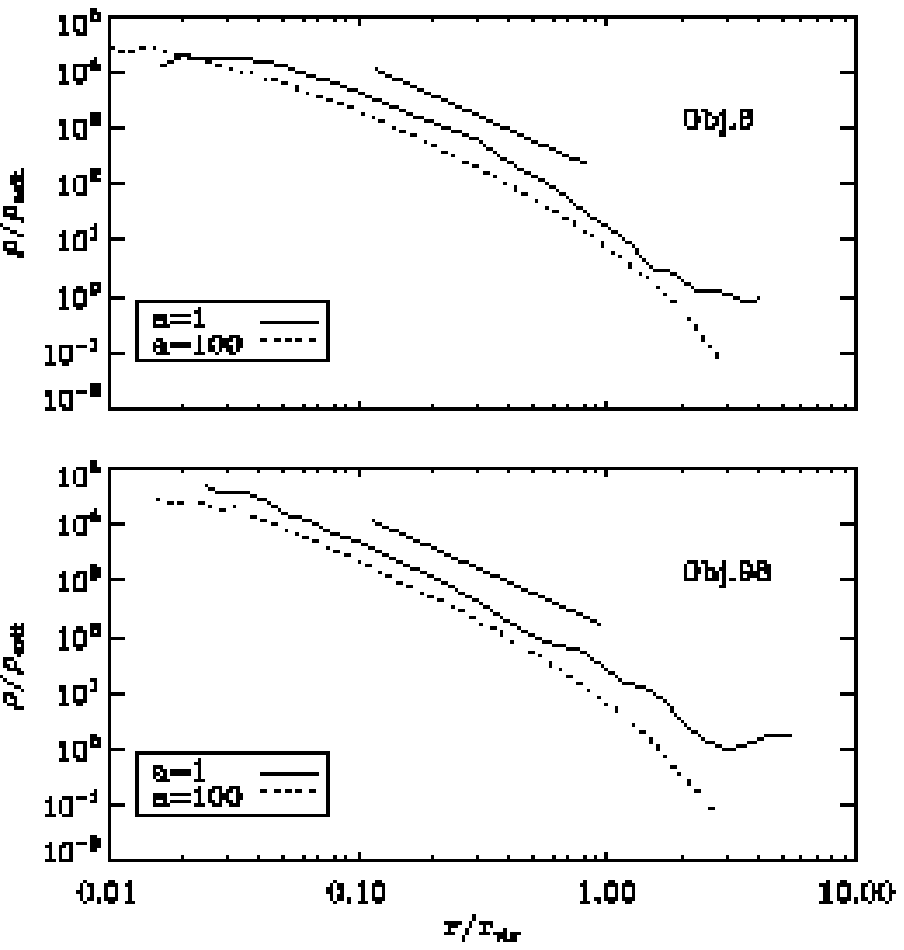}
\vskip -0.5truecm
\end{figure}

\begin{figure*}
\mbox{\hskip -0.5truecm\includegraphics[width=1.1\textwidth]{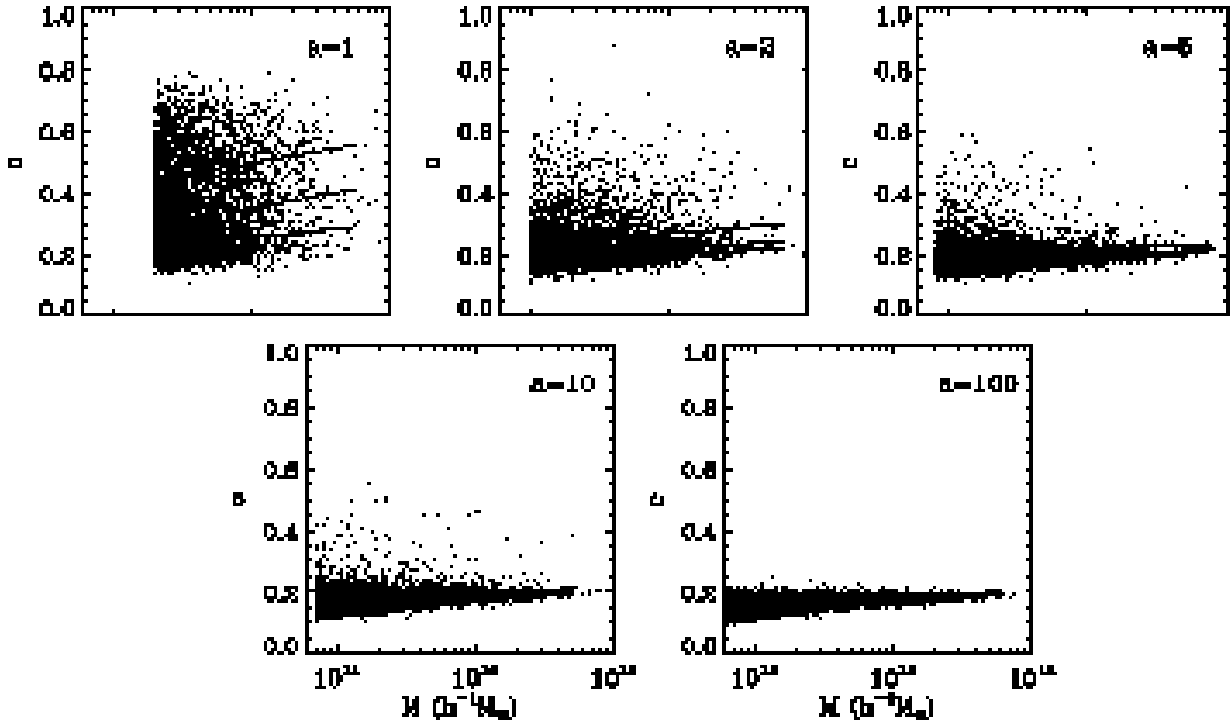}}
\caption{\small{Concentration parameter $c$ of each bound object in the bound 
object sample as a function its mass M. The concentration parameter is the 
ratio of the half-mass radius of the object to its total binding radius $r_b$, 
$c \equiv r_{hm}/r_b$. Superimposed on each of the plots is the line indicating
the median in mass bins, along with the corresponding 25 and 75 percentile 
lines. Lefthand panel: $a=1$. Righthand panel: $a=100$. At $a=100$ nearly all 
objects have a highly concentrated mass distribution, ie. $c < 0.2$. Note that 
the bound object sample is different at each expansion factor (see 
sec.~\ref{sec:objectsample}).}}
\vskip -0.25truecm
\label{fig:conc}
\end{figure*}

The evolution of the mass distribution in and around superclusters 8 and 98 
from a weblike irregular and structured pattern at $a=1$ into  
smooth and highly concentrated and nearly spherical mass clumps at 
$a=100$ can be observed from the changing particle distributions in the 
lefthand and central frames of Fig.~\ref{fig:clus8_98}. For guidance, 
on all four panels we superimposed circles centered at the 
supercluster's core. These are the ``half-mass sphere'' (solid 
circle), enclosing half of the total mass of the supercluster, 
and the ``virial sphere'', enclosing the central virialized core of 
the supercluster (dashed-dotted circle). 

For the structure of the virial core we turn to the log-log diagrams of the 
radial density profiles $\rho(r)$ in Fig.~\ref{fig:densprof}.  All profiles 
have the same basic shape, a high density central core embedded within an 
isothermal power-law region with slope $\sim -2$. While there may still be some
deviations from this slope at $a=1$, comparison with the inserted short line of
slope $-2$ shows that at $a=100$ the object cores are almost perfectly 
isothermal. The profiles confirm the impression from Fig.~\ref{fig:clus8_98} of
the growth of the virial core, given the smooth near power-law profiles at 
$a=100$ as opposed to the more irregular behaviour at the outer edges in the 
$a=1$ profiles. 

In addition, we note a radical change of the cosmic surroundings 
of both superclusters. At $a=1$ both superclusters are still solidly integrated 
and embedded within the Megaparsec Cosmic Web. Their central cores, indicated 
by the ``virial spheres'', are connected to the surroundings via filamentary 
tentacles along which we find a large variety of mass clumps. Note that these
outer structures at $a=1$ are actually bound to the supercluster core and is 
will fall in and merge with the central cluster as time proceeds. At 
$a=100$ the resulting supercluster concentrations have turned into isolated 
islands. 

It is telling that at $a=100$ both superclusters have attained 
an almost equivalent internal mass distribution, even though at $a=1$ 
their morphology was quite different. At the present epoch, supercluster 
\# 8 is already a centrally concentrated object. Given that the 
half-mass and virial spheres nearly overlap, we see that at $a=1$ 
its virialized core contains nearly half of its total mass. This impression 
is underlined by the cumulative mass distribution in the righthand 
panel of Fig.~\ref{fig:clus8_98}. Comparison between the 
cumulative mass distribution at $a=1$ (solid line) and $a=100$ 
(dashed line) shows that there is only a small increase in mass in 
the inner region of the supercluster, immediately around the virialized core. 
Also, the fact that the half-mass radius moves only slightly inward implies 
a moderate change in the inner mass distribution. It is the outer half 
of the supercluster's mass which rearranges itself more strongly: as it 
falls in towards the supercluster's interior, the mass distribution 
becomes more concentrated and more regular. 

The changing mass distribution is more pronounced for 
supercluster \# 98. At the current epoch its mass distribution 
is more extended: its half-mass radius is located at an outward 
position. The central core is not nearly as 
prominent as that in supercluster \# 8. The supercluster's 
mass increases rather slowly until the half-mass radius. 
Beyond this radius there is an abrupt rise until the outer  
supercluster radius. This is related to the presence of another 
major mass clump near the outer boundary. When we would have observed 
this supercluster in the observational reality, we would find it to 
be dominated by two very rich clusters.  

Also, we note that the superclusters do hardly gain mass from beyond 
their (binding) radius $R_b$. This is reflected in the flattening of 
the cumulative mass curves at $a=100$. 

\subsection{Supercluster Mass Concentration}
From the discussion in the previous subsection we have learned 
that the superclusters are turning into bodies with a highly 
concentrated mass distribution at $a=100$. It is only towards 
these later cosmic epochs that the superclusters have turned 
into highly nonlinear and concentrated regions. It would 
not be appropriate to seek to fit a theoretical density profile 
to their radial mass distribution in order to determine their 
concentration. Instead of seeking to fit an Einasto profile 
\citep{einasto1965}, or the profusely popular universal NFW profile 
\citep{nfw97}, we therefore prefer to define a concentration 
parameter that is independent of assumptions about the 
dynamical state of the mass concentration, 
\begin{equation}
c=\frac{r_{hm}}{r_{b}}\,,
\end{equation}
where $r_{hm}$ is the radius that encloses half of the mass. Note 
that with this definition a mass distribution with a ``delta peak'' 
in the centre would have $c=0$, while a perfectly uniform distribution 
would have $c\approx 0.8$. An isothermal distribution would 
correspond to $c=0.5$. 

Fig.~\ref{fig:conc} shows the distribution of $c$ as a function of mass, for 
the object population at $a=1$ until $a=100$. At the present epoch there still 
is a considerable spread of the concentration parameter ($\bar{c}=0.35$, 
$\sigma_c=0.14$). There are even a few objects that get close to the $c=0.8$ 
value corresponding to uniform mass profiles. Although there is a slight 
tendency towards higher concentrations, in general the concentration parameter 
at $a=1$ reflects the irregular and prominent outer mass distribution. Low-mass
bound objects appear to be more strongly concentrated than the 
$M > 10^{15}h^{-1}$M$_{\odot}$ superclusters. 

The additional panels to reveal the expected development. After $a=1$, there is
a relatively rapid evolution towards much more concentrated configurations. 
While at $a=2$ there are still a sizeable number of bound objects and 
superclusters with a rather diffuse or uniform internal mass distribution, at 
$a=5$ nearly all objects have attained a high level of concentration of 
$c<0.4$, with most having a concentration index $c\sim 0.2$. This development 
continues towards a radical conclusion at around $a=10$. Over the whole mass 
range objects are highly concentrated from $a=10$ onward. Also note that the 
trend of low-mass objects to be more concentrated than high-mass superclusters 
persist, even though it gradually weakens towards later epochs.  
At the final time of $a=100$, nearly all objects have $c<0.2$, with an average 
concentration index $\bar{c}=0.16$ and $\sigma_c=0.02$. 

It is clear that this evolution ties in strongly with the tendency to assume a 
near perfect spherical shape and, as we will see in the next section, with a 
rapidly decreasing level of substructure and multiplicity of the superclusters. 

\section{Supercluster Substructure and Multiplicity Function}
\label{sec:multi}
The third aspect of the internal evolution of superclusters is that of their 
substructure. Because superclusters are usually identified via their cluster 
content we focus on their \emph{multiplicity} N$_{\rm SC}$, ie. the number of 
clusters they contain. Clusters are taken to be virialized subclumps with a 
mass higher than $3\times10^{13}h^{-1}$M$_{\odot}$, the low-mass threshold 
for the virialized groups at $a=1$. We should take into account that the 
definition of the multiplicity N$_{\rm SC}$ depends on the cluster mass 
threshold in our sample. 

\begin{figure}
\includegraphics[width=0.45\textwidth]{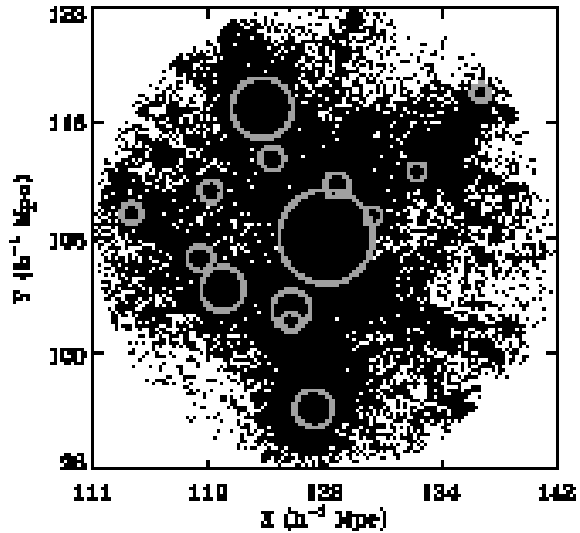}
\caption{Supercluster multiplicity. The (dark) mater distribution within the 
binding radius of a supercluster, at $a=1$. The clusters within the 
supercluster realm are indicated by circles. The size of the circle 
reflects the (virial) size of each cluster halo.} 
\label{fig:scmulti}
\end{figure}

Fig.~\ref{fig:scmulti} gives an impression of the substructure of one of the 
superclusters in our sample, at $a=1$. The cluster population within the 
supercluster is indicated by circles. The supercluster area is still a rather 
polymorphic assembly of matter, connected by means of filamentary extensions. 
The most prominent concentrations, the clusters, roughly follow these weblike 
structures. 

Interestingly, we find that the mean mass of all the clusters in our 
simulation volume is M$_{cl}=9.4\times10^{13}h^{-1}$M$_{\odot}$ while those 
residing within the realm of superclusters have an average mass 
M$_{cl}=3.6\times10^{14}h^{-1}$M$_{\odot}$. The fact that superclusters 
contain more massive clusters is partly a result of the stronger clustering 
of higher mass clusters \citep{kaiser1984,bahcall1988}. An additional 
factor is the active dynamical environment inside superclusters. Because 
of the high concentration of subclumps, these are continuously merging and 
falling into ever more massive clumps. Clusters will be centres of action and 
thus grow rapidly in mass. Meanwhile, lower mass and lower density clumps are 
more liable to lose mass or even to get gradually dismantled by the prevailing 
strong tidal forces in and near the superclusters. 

\subsection{Multiplicity Evolution}
\label{sec:multievol}
\begin{figure}
\includegraphics[height=0.235\textwidth]{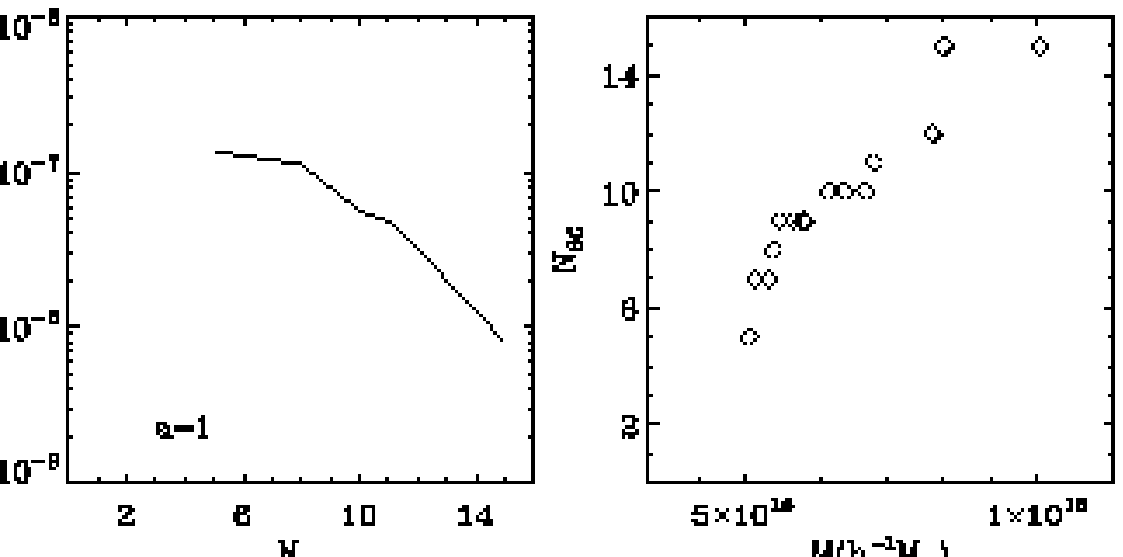}
\includegraphics[height=0.235\textwidth]{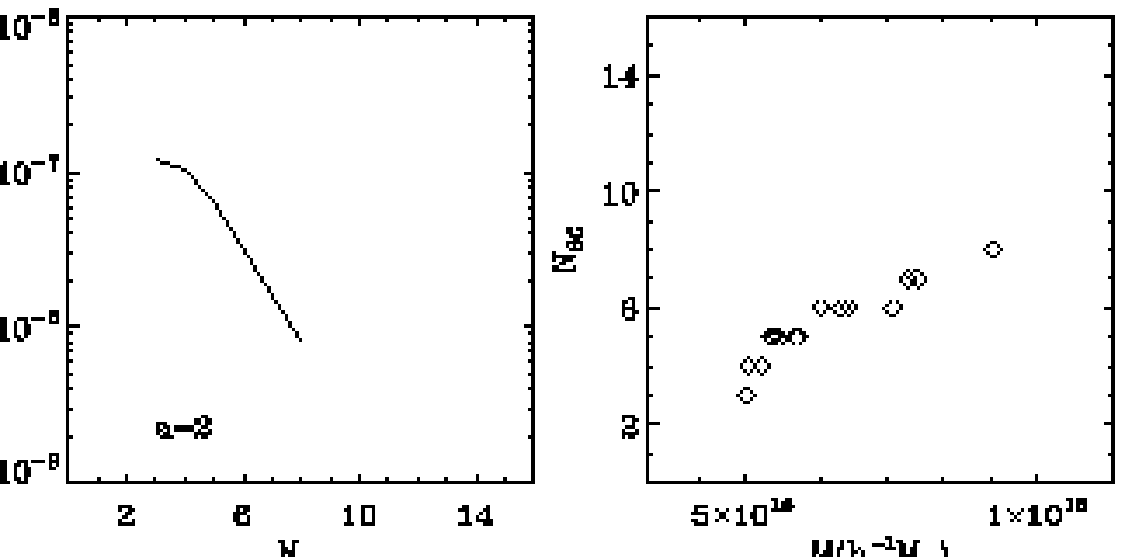}
\includegraphics[height=0.235\textwidth]{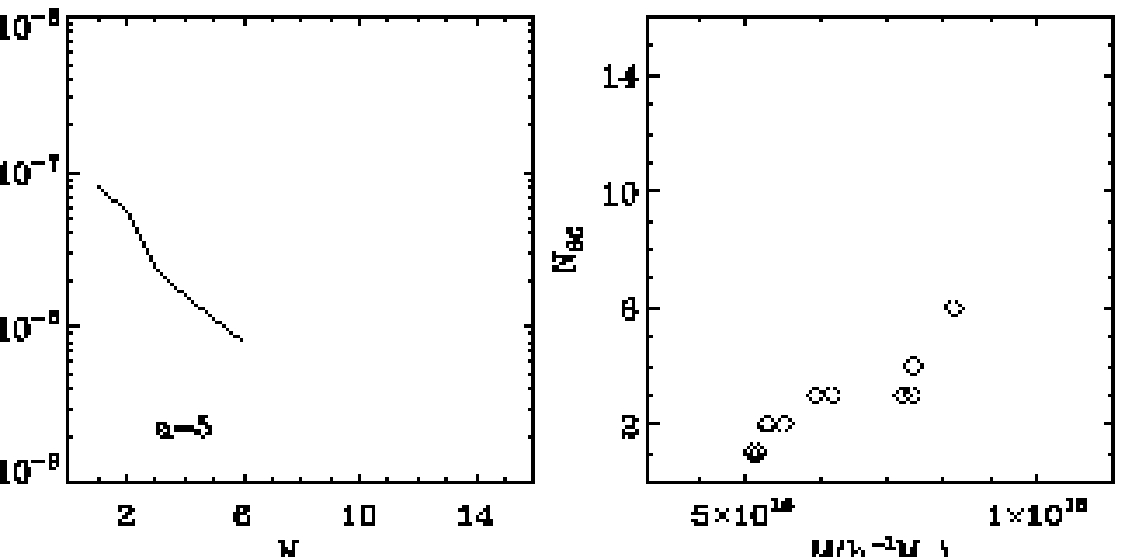}
\includegraphics[height=0.235\textwidth]{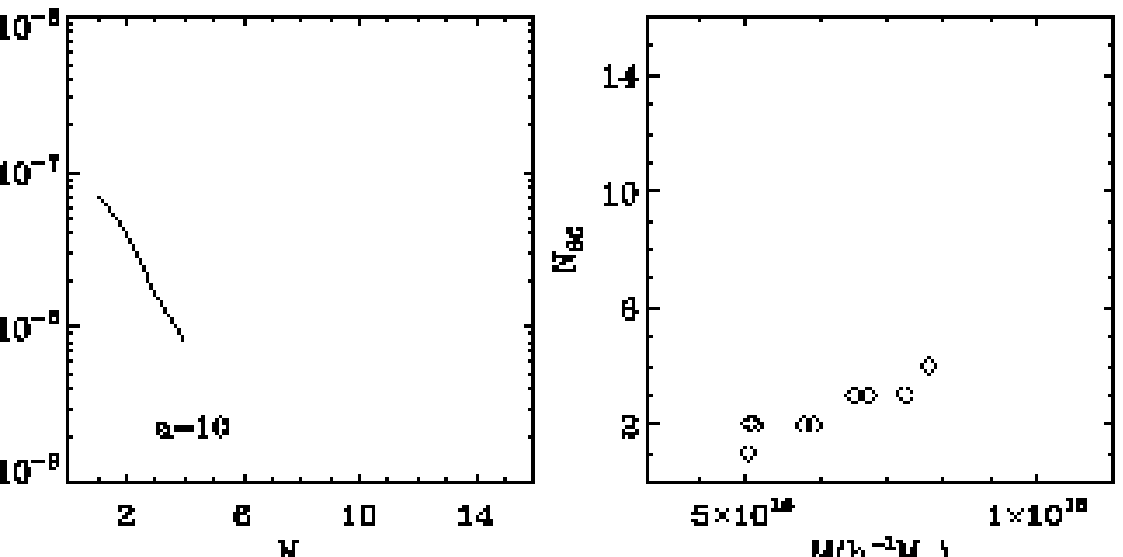}
\caption{Supercluster multiplicity. Left panel: the cumulative multiplicity 
distribution $n_{\rm SC}($N$_{\rm SC}$) of superclusters containing 
N$_{\rm SC}$ or more clusters at four different expansion factors: $a=1$, $a=2$,
$a=5$ and $a=10$. At each timestep the distribution concerns the most massive 
superclusters in the simulation, with M$>10^{15}h^{-1}$M$_{\odot}$. Righthand: 
Supercluster multiplicity N$_{\rm SC}$ as a function of the mass M of the 
supercluster.} 
\label{fig:multiplicity}
\end{figure}

One of the principal findings of our study is that, without exception, at 
$a=100$ all superclusters in our sample have attained a multiplicity one. By 
that time, they all have evolved into compact and smooth density concentrations,
akin to the one seen in Fig.~\ref{fig:cumulos_a0}. The hierarchical development
of the supercluster involves the gradual merging of its constituent subclumps 
into one condensed object. As a result, we see that superclusters which at 
the present epoch contain several to dozens of clusters ultimately end up as 
an object of unit multiplicity. Fig.~\ref{fig:multiplicity} reveals the 
systematic evolution towards this configuration. 

Turning to the present epoch, we assess the multiplicity function for the 17 
most massive superclusters, those whose mass is in excess of 
M$=5\times10^{15}h^{-1}$M$_{\odot}$. When evaluating the (cumulative) 
multiplicity distribution, the number density of superclusters with more than 
N$_{\rm SC}$ clusters (Fig.~\ref{fig:multiplicity}, top-left row), we see that 
half of the superclusters have 10 or more members. Another important trend, 
not entirely unexpected, is that larger and more massive superclusters 
contain a higher number of cluster members. One can immediately infer this when 
looking at the multiplicity N$_{\rm SC}$ against supercluster mass M$_{\rm SC}$, 
plotted in the righthand row of Fig.~\ref{fig:multiplicity}. This trend is 
particularly strong for superclusters with N$_{\rm SC}<10$. For a reason which 
we do not entirely understand, the multiplicity seems to level off for more 
massive and larger supercluster complexes. 

At first gradually, at $a=2$, and later more radically we see a strong change 
in the multiplicity N$_{\rm SC}$ of superclusters as more and more clusters 
within their realm merge and mix with the central matter concentration. The 
lefthand column of panels shows the systematic decline of high multiplicity 
objects. At $a=2$ we can no longer find superclusters with more than 
10 cluster members. At that epoch 2 of the 17 superclusters 
have collapsed into a single object of multiplicity one. Towards the 
later epochs the multiplicity of superclusters quickly declines further, and 
at $a=10$ there are no superclusters around with more than 5 clusters. 

\begin{figure}
\includegraphics[height=0.32\textwidth]{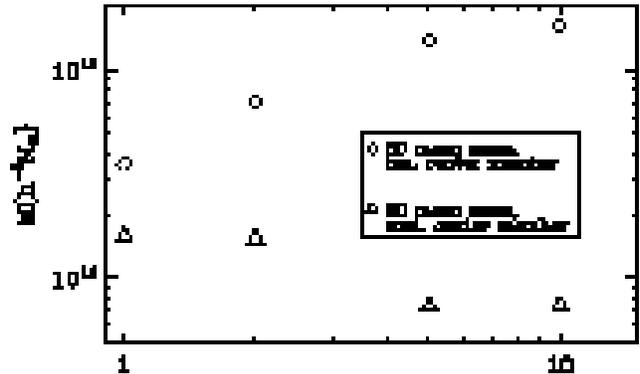}
\caption{Mean mass of the cluster members of superclusters in our 
sample, including (diamonds) and excluding (triangles) the central 
mass concentration.}
\label{fig:meanmass}
\end{figure} 

When looking at the corresponding frames for the multiplicity as a function of 
supercluster mass, we see that the number of superclusters of unit multiplicity 
is continuously increasing. Meanwhile, the superclusters that still have more 
than one cluster member appear to have less and less clusters within their 
realm. 

An additional interesting issue is that of the masses of the clusters that 
populate superclusters. In Fig.~\ref{fig:meanmass} we can see that the 
mean mass of clusters in superclusters is continuously growing, from 
M$_{cl}=7\times10^{14}h^{-1}$M$_{\odot}$ at $a=2$ to 
M$_{cl}=1.6\times10^{15}h^{-1}$M$_{\odot}$ at $a=10$. However, this reflects 
the continuous growth of the central mass concentration, accompanied by the 
decreasing supercluster multiplicity: the mean mass is becoming more 
and more dominated by that of the central supercluster core. Meanwhile, 
the mass of the remaining clusters outside the core is decreasing. In other 
words, the first clusters to merge with the supercluster core tend to be its 
most massive companions. The ones that accrete at a later epoch have a 
considerably lower mass. This is in accordance with our observation in 
sect.~\ref{sec:evolution} (see Fig.~\ref{fig:cumulos_a0}).

\subsection{Multiplicity Criterion}
Because we have used a physical criterion for the identification of clusters 
and superclusters, we assume the inferred multiplicities to be close to the 
one found in the observed reality. To some extent supercluster multiplicity 
estimates will depend on the supercluster identification procedure. There is 
certainly a dependence on percolation radius when identified on the basis of a 
percolation criterion \citep{zeldovich1982,shandzeld89}. 

This may be one of the reasons why our measured supercluster multiplicities 
differ from those obtained in other studies (in addition to the dependence on 
cluster mass threshold). For example, \cite{wray06} found superclusters with 
more than 30 members. This certainly relates to the choice of linking length 
for defining the supercluster: dependent on the linking length they found 
maximum supercluster sizes ranging from $\sim 30h^{-1}$Mpc to 
$\sim 150h^{-1}$Mpc. The latter are much larger than the superclusters we find 
according to our definition based on binding-density.

\section{Shapley-like Superclusters}
\label{sec:shapley}

The Shapley concentration, first noted by \cite{shapley1930}, is one of 
the most outstanding supercluster complexes out to $z=0.12$ 
\citep[see][]{raychaud1989,ettori1997,quintana2000,proust2006}. It amasses at 
least 30 rich Abell galaxy clusters in a core region of 
$\sim 25h^{-1}\hbox{Mpc}$ and is located at a distance of 
$\sim 140h^{-1}\hbox{Mpc}$. Its total mass is estimated to be 
$\sim5\times10^{16}h^{-1}$M$_{\odot}$ within a radius of 
$\sim 30h^{-1}\hbox{Mpc}$ \citep[see e.g.][]{proust2006,munoz08}. 
Almost as massive is another similar assembly of massive clusters in the local 
Universe, the Horologium-Reticulum supercluster 
\citep[see e.g.][]{fleenor2005}. Both may have a major influence on the motion 
of the Local Group with respect to the background Universe 
\citep[see e.g.][]{plionis1991,kocevski2006}.

We investigated in how far we can find back the equivalents of the Shapley and 
Horologium superclusters in our simulation. On the basis of the mass functions 
determined in sect.~\ref{sec:scmassfunc}, we estimate that the most massive 
supercluster present in a volume akin to the local Universe ($z<0.1$) may have 
a mass of $\sim 8\times10^{15}h^{-1}$M$_{\odot}$. This mass is slightly larger 
than the one in the bound region of the Shapley concentration as determined by 
\cite{dunner08}. Extrapolating this conclusion, we find that one may typically 
find two Shapley-like superclusters out to $z \approx 0.1$, a volume of the 
size of the Local Universe. 

When turning towards the multiplicity of the detected simulation superclusters,
we find from Fig.~\ref{fig:multiplicity} that the largest supercluster in the 
Local Universe would have 15 members. A Shapley-like supercluster would have a 
radius of $\sim14h^{-1}$Mpc and host between 10 to 15 members, close to the 
number found in the bound region of the Shapley supercluster \citep{dunner08}, 
which contains $\sim 1/3$ of the clusters traditionally assigned to this 
structure \citep[e.g.][]{proust2006}. In the observational reality of our Local
Universe ($z<0.1$), we find 5 superclusters with 10 or more members, the 
largest one containing 12 members \citep[see e.g.][]{einasto1994}.

This brings us to the issue of the extent to which the supercluster population 
in our LCDM simulation at $a=1$ resembles the one seen in our nearby Universe. 
To this end we are in the process of translating our theoretical supercluster 
criterion into one that would be able to identify structures along the same 
lines in large magnitude-limited galaxy redshift surveys. By applying this to 
the 6dF galaxy sample \citep{6dfdr3} and the SDSS DR7 galaxy sample 
\citep{sdssdr7}, we plan a detailed comparison of morphology, size and spatial 
distribution of the identified superclusters with those seen in our simulation.
First indications from similar comparison between superclusters 
in the 2dFGRS sample \citep{colless2003} and the Millennium simulation 
\citep{springmillen2005}, based on a more conventional supercluster 
identification process, seems to indicate that there are no discernable 
discrepancies \citep{einasto2007b}.

\section{Conclusions}
\label{sec:conclusions}
In this work, we have followed the evolution of bound objects from the present 
epoch up to a time in the far future of the Universe, at $a=100$, in a standard
$\Lambda$CDM ($\Omega_{m,0}=0.3$, $\Omega_{\Lambda,0}=0.7$ and $h=0.7$) 
Universe. We contrasted the external global evolution of the population of 
bound objects with their vigorous internal evolution, starting from the 
contention that in a dark energy dominated Universe they have the character of 
{\it island universes}. Within such a Universe we expect them to become 
increasingly isolated objects in which cosmic evolution proceeds to the 
ultimate equilibrium configuration of a smooth, spherical, virialized and 
highly concentrated mass clumps. We identify the most massive of these objects 
with superclusters. 

For the external evolution, we investigate the spatial distribution and 
clustering of bound objects and superclusters, along with the weak change of 
their mass functions. To assess the internal structure, we have looked into 
their rapidly changing shape, their evolving density profile and mass 
concentration and the level of substructure of the superclusters in terms of 
their multiplicity, i.e. the number of clusters within their realm. 

We defined the bound structures by the density criterion derived in Paper I 
(see Eqn.~\ref{eq:ratio}), and identified them from a 500$h^{-1}$ Mpc 
cosmological box with $512^3$ dark matter particles in a $\Lambda$CDM 
($\Omega_{m,0}=0.3$, $\Omega_{\Lambda,0}=0.7$ and $h=0.7$) Universe. We ran the 
simulation up to $a=100$, which is a time where structures have stopped 
forming. We used the HOP halo identifier in order to identify independently 
virialized structures, at each of the five timesteps we have analyzed in 
detail: $a=1$, 2, 5, 10 and 100. 

The main results of the present study can be summarized as follows:

\begin{itemize}

\item While the large-scale evolution of bound objects and superclusters comes 
to a halt as a result of the cosmic acceleration, their internal evolution 
continues vigorously until they have evolved into single, isolated, almost 
perfectly spherical, highly concentrated, virialized mass clumps. This 
development is very strong between $a=1$ and 10, and continues up to $a=100$. 

\item The marginally bound objects that we study resemble the superclusters in
the observed Universe. While clusters of galaxies are the most massive, fully 
collapsed and virialized objects in the Universe, superclusters are the largest
bound --but not yet collapsed-- structures in the Universe.

\item The superclusters are true \emph{island universes}: as a result of the
accelerating expansion of the Universe, no other, more massive and larger, 
structures will be able to form. 

\item While the superclusters collapse between $a=1$ and $a=100$, their 
surroundings change radically. While at the present epoch they are 
solidly embedded within the Cosmic Web, by $a=100$ they have turned into 
isolated cosmic islands.

\item The large scale distribution of bound objects and superclusters 
(in comoving space) does not show any significant evolution in between $a=1$ 
and $a=100$. The cluster and supercluster correlation functions do not 
change over this time interval, and retain their near power-law 
behaviour. Superclusters remain significantly stronger clustered than 
the average bound object, with a supercluster correlation length of 
$23 \pm 5h^{-1}$Mpc compared to $r_0\approx 11.5h^{-1}$Mpc for 
the full bound object distribution in our simulation sample. 

\item The mass functions of bound objects and superclusters hardly change 
from $a=1$ to $a=100$, as we expect on theoretical grounds. 
The mass functions in the simulations are generally in good agreement 
with the theoretical predictions of he Press-Schechter, Sheth-Tormen, and 
Jenkins mass functions. At $a=1$, the Sheth-Tormen prescription provides a 
better fit. At $a=100$, the pure Press-Schechter function seems to be 
marginally better. This may tie in with the more anisotropic shape of 
superclusters at $a=1$ in comparison to their peers at $a=100$.

\item The change in the internal mass distribution and that in the surroundings 
is directly reflected by the radial density profile. Without exception towards 
$a=100$ all objects attain a highly concentrated internal matter distribution, 
with a concentration index $c=0.2$. In general, the vast majority of objects 
has evolved into a highly concentrated mass clump after $a=10$. 

\item The mass profile in the outer realms of the supercluster changes
radically from $a=1$ to $a=100$. At $a=1$ it is rather irregular, while there 
are large differences between the individual objects. This is a reflection of 
the surrounding inhomogeneous mass distribution of the Cosmic Web. In between 
$a=5$ and $a=10$, nearly all superclusters have developed a smooth, regular and 
steadily declining mass profile. 

\item The inner density profile steepens substantially when the inner region of
the supercluster is still contracting. On the other hand, when at $a=1$ it has 
already developed a substantial virialized core, the inner density profile 
hardly changes.

\item As a result of their collapse, the shapes of the bound objects 
systematically change from the original triaxial shape at $a=1$ into an almost 
perfectly spherical configuration at $a=100$. For example, at $a=1$ their mean 
axis ratios are ($\langle s_{2}/s_{1} \rangle$, $\langle s_{3}/s_{1} \rangle $)=
(0.69,0.48). At $a=100$, they have mean axis ratios of 
($\langle s_{2}/s_{1} \rangle$,$\langle s_{3}/s_{1} \rangle $)= (0.94,0.85). 

\item At the current epoch the superclusters still contain a substantial amount
of substructure. Particularly interesting is the amount of cluster-mass 
virialized objects within its realm, expressed in the so called 
\emph{multiplicity function}. Restricting ourselves to superclusters with a 
mass larger than $5\times10^{15}h^{-1}$M$_{\odot}$, of which we have 17 in our 
simulation sample, we find a multiplicity of 5 to 15 at the current epoch. As 
time proceeds there is a systematic evolution towards unit multiplicity at 
$a=100$, following the accretion and merging of all clusters within the 
supercluster's realms.

\item In a volume comparable to the Local Universe ($z<0.1$) we find that the 
most massive supercluster would have a mass of 
$\sim8\times10^{15}h^{-1}$M$_{\odot}$. This is slightly more massive than
the mass of the Shapley Supercluster given in \cite{dunner08}. When turning 
towards the multiplicity, we find that the largest superclusters in the Local 
Universe would host between 10 and 15 members, close to the number found in the 
bound region of the Shapley supercluster \citep{dunner08} (which contains 
$\sim 1/3$ of the clusters traditionally assigned to this structure, e.g., 
\cite{proust2006}).
\end{itemize}

While in this study we have addressed a large number of issues, our study 
leaves many related studies for further investigation. One of the most pressing 
issues concerns an assessment of the nature of our supercluster objects. This 
involves a comparison with other supercluster definitions, in particular in how
far our density-based definition fares at the earlier epochs when most objects 
of similar mass will have distinct anisotropic shapes. 

Also, while here we follow the evolution of superclusters in the standard 
$\Lambda$CDM Universe, in an accompanying publication we will systematically 
address the influence of dark matter and dark energy on the emerging 
supercluster. There we found that dark matter is totally dominant in 
determining the supercluster's evolution. For a preliminary, detailed report on
the analysis of the role of the cosmological constant in the formation and 
evolution of structures, we refer to \cite{arayamelo08}.

\section*{Acknowledgments}
The authors gratefully acknowledge the very useful comments and recommendations 
by the referee, responsible for a substantial improvement of the paper. 
PA and RvdW are grateful to Bernard Jones for valuable and incisive comments 
and many useful discussions and Thijs van der Hulst for continuous and 
unceasing encouragement. HQ is grateful to the FONDAP Centro de Astrof\'isica 
for partial support, AR acknowledges support by FONDECYT Regular Grant 1060644 
and in addition AR and HQ thank Proyecto Basal PFB-06/2007. AM was supported by
the Fondo Gemini-Conicyt through grant no. 32070013. RD acknowledges support
from a CONICYT Doctoral Fellowship. PA acknowledges support by a NOVA visitor 
grant during the completion phase of the manuscript.

\appendix

\section{Linear density excess of critical shell}
\label{app:rdelta}
\noindent In order to find the linear density excess $\delta_0$ for the 
critically bound shell, we evaluate its evolution at early epochs ($a\ll1$). 
At these early times -- when density perturbations are still very small, 
$\delta \ll1$ -- the linearly extrapolated density excess 
(Eqn.~\ref{eq:denlin}),
\begin{equation}
\delta(a)\,=\,D(a)\,\delta_0
\end{equation}
\noindent represents a good approximation for the (real) density 
of the object (Eqn.~\ref{eq:ratio}).

At early times the Universe is very close to an Einstein-de Sitter Universe, 
expanding according to
\begin{equation}
a(t)=\left(\frac{t}{t_{\ast}}\right)^{2/3}\,,
\end{equation}
where $t_{\ast}$ is the characteristic expansion time. 

Following this cosmic evolution, 
\begin{equation}
1+\delta\,\equiv\,\frac{\rho}{\rho_{u}}=\frac{2\Omega_{\Lambda 0}}{\Omega_{m,0}}
\left(\frac{a}{\tilde{r}}\right)^3\,.
\label{eq:deltamasuno}
\end{equation}
to first order yields 
\begin{equation}
\frac{t(\tilde{r})}{\tilde{r}^{3/2}}\,\approx\,{t_{\ast}}\,\left(1+\frac{\delta}{2}\right)
\left(\frac{\Omega_{m,0}}{2\Omega_{\Lambda,0}}\right)^{1/2}\,.
\label{eq:andreasway}
\end{equation}
This approximation, neglecting contributions of order $\delta^2$ and higher, is 
reasonably accurate for density perturbations $\delta \ll 1$. The evolution of 
the critical shell's radius $\tilde r(t)$, the solution to the energy equation 
eq.~\ref{eq:energy} for ${\tilde E}={\tilde E}^{\ast}=-\frac{3}{2}$, is given
 by the integral expression
\begin{equation}
\tilde{t}=\int_{0}^{\tilde{r}}\frac{\sqrt{r}dr}
{(1-r)\sqrt{r+2}}\,.
\label{eq:integraltiempo}
\end{equation}
Retaining the two lowest-order terms of this equation,
\begin{equation}
t(\tilde{r})\approx\sqrt{\frac{2}{3\Lambda}}\tilde{r}^{3/2}+\frac{3}{10}
\sqrt{\frac{3}{2\Lambda}}\tilde{r}^{5/2}\,,
\label{eq:rt12}
\end{equation}
we find that
\begin{equation}
\frac{t}{{\tilde r}^{3/2}}\,\approx\,\sqrt{\frac{2}{3\Lambda}}\,\left(1\,+\,\frac{9}{20}{\tilde r}\right)\,.
\end{equation}
The front factor on the righthand side of this equation should be equal to that
of Eqn.~(\ref{eq:andreasway}), so that the characteristic time $t_{\ast}$ is 
found to be  
\begin{equation}
t_{\ast}=\left[\frac{4\Omega_{\Lambda,0}}{3\Lambda\Omega_{m,0}}\right]^{1/2}\,.
\label{eq:tast}
\end{equation}
Combining expression~(\ref{eq:rt12}) with that of Eqn.~(\ref{eq:andreasway}) we find that 
\begin{equation}
t_{\ast}\,\left(1+\frac{\delta}{2}\right)\left(\frac{\Omega_{m,0}}{2\Omega_{\Lambda,0}}\right)^{1/2}\,
\approx\,\sqrt{\frac{2}{3\Lambda}}\,\left(1\,+\,\frac{9}{20}{\tilde r}\right)
\label{eq:tdeltar123}
\end{equation}
resulting in the following relation between the early density excess of the bound sphere 
and its dimensionless radius ${\tilde r}$:
\begin{equation}
{\delta}(t)\,=\,\frac{9}{10}\,{\tilde r}(t)\,.
\end{equation}

\section{Press-Schechter modelling of superclusters}
\label{app:app_ps}
According to the Press-Schechter formalism 
\citep{press74,peacheav1990,bond91,sheth1998}, the comoving number density of 
halos of mass $M$ at redshift $z$, in a cosmic background of density $\rho_u$, 
is given by the expression
\begin{equation}
\label{eq:corcho}
\frac{\mathrm{d}n}{\mathrm{d}M}=\sqrt{\frac{2}{\pi}}\frac{\rho_{u}}{M^{2}}\frac{\delta_{c}}
{\sigma(M,z)}\left|\frac{\mathrm{d}\ln{\sigma(M,z)}}{\mathrm{d}\ln{M}}
\right|e^{\frac{-\delta_{c}^{2}}{2\sigma^{2}(M,z)}}\,,
\end{equation}
in which $\delta_{c}/\sigma$ quantifies the relative critical overdensity 
$\delta_c$ of collapse with respect to the variance of density fluctuations 
$\sigma(M,z)$ on a mass scale $M$. For a scenario with a power spectrum $P(k)$ 
of the linear density field,
\begin{equation}
\label{eq:varsigma}
\sigma^{2}(M)=4\pi\int^{\infty}_{0} P(k)\omega(kr)k^{2}\mathrm{d}k\,,
\end{equation}
where $\omega(kr)$ is the Fourier-space representation of a real-space top-hat 
filter enclosing a mass $M$ in a radius $r$ at the mean density of the 
Universe, which is given by
\begin{equation}
\omega(kr)=3\left[\frac{\sin(kr)}{(kr)^{3}}-\frac{\cos(kr)}{(kr)^2}\right]\,.
\end{equation}

\vskip 0.5cm
\noindent The critical spherical collapse overdensity value $\delta_c$  
has a weak dependence on the cosmological background 
\citep{gunn72,lacey93,eke1996, kitayama1996}. Useful fitting formulae for the 
linear spherical model collapse value $\delta_{c}$ were obtained by 
\cite{nfw97} for $\Omega_{\Lambda}=0$ FRW Universes and for flat Universes,
\begin{equation}
\delta_{c}(\Omega_{m})\,=\,\left\{\begin{array}{ccc}
0.15(12\pi)^{2/3}\Omega_{m}^{0.0185} & & \Omega_{m}<1, \Omega_{\Lambda}=0,\\
\ \\
0.15(12\pi)^{2/3}\Omega_{m}^{0.0055} & & \Omega_{m}+\
\Omega_{\Lambda}=1.
\end{array}\right.
\label{eq:deltacrit}
\end{equation}
According to this formula $\delta_{c}=1.68$ at $a=1$ (for $\Omega_m=0.3$), 
while at $a=100$, when $\Omega_m=0.43\times01^{-6}$, one finds $\delta_c=1.56$. 
A similar expression for the critical virial density has been 
given by \cite{bryan98}, 
\begin{equation} 
\Delta_{vir}\,=\,\left\{\begin{array}{ccc}
18\pi^2\,+\,82(\Omega_m-1)\,-39(\Omega_m-1)^2&&\\
\qquad\qquad\qquad\qquad\Omega_{m}+\Omega_{\Lambda}=1&&\\
\ \\ 
18\pi^2\,+\,60(\Omega_m-1)\,-32(\Omega_m-1)^2&&\\
\qquad\qquad\qquad\qquad\Omega_{m}<1, \Omega_{\Lambda}=0&&\\
\end{array}\right.
\end{equation}

While most applications of the (extended) PS formalism assume perfectly 
spherical collapse, we know that generic gravitational collapse of primordial 
density peaks proceeds anisotropically. \cite{sheth99} improved the PS 
formalism by taking into account the anisotropic collapse implied by the 
anisotropic primordial shape of density peaks and the anisotropic tidal 
stresses imparted by external mass concentrations. Modelling this by means of 
the ellipsoidal collapse model \citep[e.g.,][]{lynbell1964,icke73, 
whitesilk1979,eistloeb1995,bondmyers1996,desjacques2008} they showed this 
translates into a more fuzzy {\it moving collapse density barrier}. The 
resulting mass function, 
\begin{equation}
\begin{split}
\frac{\mathrm{d}n_{ST}}{\mathrm{d}M}=A\sqrt{\frac{2a}{\pi}}\left[1+
\left(\frac{\sigma(M)^{2}}{a\delta_{c}^{2}}\right)^{p}\right]
\frac{\rho_{u}}{M^{2}}\frac{\delta_{c}}{\sigma(M)}\times\\
\left|\frac{\mathrm{d}\ln{\sigma(M)}}{\mathrm{d}\ln{M}}\right|
e^{\frac{-a\delta_{c}^{2}}{2\sigma^{2}(M)}}\,,
\end{split}
\end{equation}
with $a=0.707$, $p=0.3$ and $A\approx0.322$, gives a substantially better 
fit to the mass functions obtained in N-body simulations. In comparison with 
the standard PS mass function, ST predicts a higher abundance of massive 
objects and a smaller number of less massive ones. Later, \cite{jenkins01} 
reported a small disagreement with respect to N-body simulations: 
underpredictions for the massive halos and overpredictions for the less 
massive halos. They suggested the alternative expression:
\begin{equation}
\frac{\mathrm{d}n_{J}}{\mathrm{d}M}=A\frac{\rho_{u}}{M^{2}}
\frac{\mathrm{d}\ln{\sigma(M)}}{\mathrm{d}\ln{M}}
e^{(-|\ln\sigma^{-1}+B|)^{\epsilon}}\,.
\end{equation}
with $A=0.315$, $B=0.61$ and $\epsilon=3.8$. Note, however, that their 
expression does not depend explicitly on $\delta_{c}$. They showed that ``for a
range of CDM cosmologies and for a suitable halo definition, the simulated mass
function is almost independent of epoch, of cosmological parameters, and of 
initial power spectrum''.

\bsp

\label{lastpage}

\end{document}